\documentclass[twocolumn,twocolappendix]{aastex631}

\newcommand\feh{[Fe/H]}
\newcommand{\alpham}{[Mg/Fe]}

\usepackage{amsmath}
\usepackage{color,soul}
\usepackage [autostyle, english = american]{csquotes}
\MakeOuterQuote{"}

\shorttitle{Mapping the Milky Way with APOGEE}
\shortauthors{Imig et al. 2023}
\graphicspath{{./}{figures/}}

\begin{document}

\title{A Tale of Two Disks: Mapping the Milky Way with the Final Data Release of APOGEE}

\author[0000-0003-2025-3585]{Julie Imig}
\affiliation{Department of Astronomy, New Mexico State University, P.O.Box 30001, MSC 4500, Las Cruces, NM, 88033, USA}

\author[0009-0001-4527-6436]{Cathryn Price}
\affiliation{Department of Astronomy, New Mexico State University, P.O.Box 30001, MSC 4500, Las Cruces, NM, 88033, USA}

\author[0000-0002-9771-9622]{Jon A. Holtzman}
\affiliation{Department of Astronomy, New Mexico State University, P.O.Box 30001, MSC 4500, Las Cruces, NM, 88033, USA}

\author[0000-0003-4761-9305]{Alexander Stone-Martinez}
\affiliation{Department of Astronomy, New Mexico State University, P.O.Box 30001, MSC 4500, Las Cruces, NM, 88033, USA}

\author[0000-0003-2025-3147]{Steven R. Majewski}
\affiliation{Department of Astronomy, University of Virginia, Charlottesville, VA 22904, USA}

\author[0000-0001-7775-7261]{David H. Weinberg}
\affiliation{The Department of Astronomy and Center of Cosmology and Astro Particle Physics, The Ohio State University, Columbus, OH 43210, USA}
\affiliation{Institute for Advanced Study, Princeton, NJ 08540, USA}

\author[0000-0001-7258-1834]{Jennifer A. Johnson}
\affiliation{The Department of Astronomy and Center of Cosmology and Astro Particle Physics, The Ohio State University, Columbus, OH 43210, USA}

\author[0000-0002-0084-572X]{Carlos Allende Prieto}
\affiliation{Instituto de Astrof\'{i}sica de Canarias, E-38205 La Laguna, Tenerife, Spain}
\affiliation{Departamento de Astrof\'{i}sica, Universidad de La Laguna, E-38206 La Laguna, Tenerife, Spain}

\author[0000-0002-1691-8217]{Rachael L. Beaton}
\affiliation{Space Telescope Science Institute, 3700 San Martin Drive, Baltimore, MD 21218, USA
}
\affiliation{Department of Astrophysical Sciences, Princeton University, Princeton, NJ 08544, USA}
\affiliation{The Observatories of the Carnegie Institution for Science, 813 Santa Barbara St., Pasadena, CA~91101}

\author[0000-0003-4573-6233]{Timothy C. Beers}
\affiliation{Department of Physics and Astronomy and JINA Center for the Evolution of the Elements, University of Notre Dame, Notre Dame, IN 46556, USA}

\author[0000-0002-3601-133X]{Dmitry Bizyaev}
\affiliation{Apache Point Observatory and New Mexico State University, P.O. Box 59, Sunspot, NM, 88349-0059, USA}
\affiliation{Sternberg Astronomical Institute, Moscow State University, Moscow}

\author[0000-0003-1641-6222]{Michael R. Blanton}
\affiliation{Center for Cosmology and Particle Physics, Department of Physics, 726 Broadway, Room 1005, New York University, New York, NY 10003, USA}

\author[0000-0002-8725-1069]{Joel R. Brownstein}
\affiliation{Department of Physics and Astronomy, University of Utah, 115 S. 1400 E., Salt Lake City, UT 84112, USA}

\author[0000-0001-6476-0576]{Katia Cunha}
\affiliation{Steward Observatory, University of Arizona, Tucson, AZ 85721, USA}

\author[0000-0003-3526-5052]{Jos\'e G. Fern\'andez-Trincado}
\affiliation{Instituto de Astronom\'ia, Universidad Cat\'olica del Norte, Av. Angamos 0610, Antofagasta, Chile
}

\author[0000-0002-3101-5921]{Diane K. Feuillet}
\affiliation{Lund Observatory, Department of Astronomy and Theoretical Physics, Box 43, SE-221~00 Lund, Sweden}

\author[0000-0001-5388-0994]{Sten Hasselquist}
\affiliation{Space Telescope Science Institute, 3700 San Martin Drive, Baltimore, MD 21218, USA}

\author[0000-0003-2969-2445]{Christian R. Hayes}
\affiliation{NRC Herzberg Astronomy and Astrophysics Research Centre, 5071 West Saanich Road, Victoria, B.C., Canada, V9E 2E7}

\author[0000-0002-4912-8609]{Henrik J\"onsson}
\affil{Materials Science and Applied Mathematics, Malm\"o University, SE-205 06 Malm\"o, Sweden}

\author[0000-0003-1805-0316]{Richard R. Lane}
\affiliation{Centro de Investigación en Astronomía, Universidad Bernardo O'Higgins, Avenida Viel 1497, Santiago, Chile}

\author[0000-0001-5258-1466]{Jianhui Lian}
\affiliation{South-Western Institute for Astronomy Research, Yunnan University, Kunming, Yunnan 650091, People's Republic of China}

\author[0000-0001-8237-5209]{Szabolcs M{\'e}sz{\'a}ros}
\affiliation{ELTE E\"otv\"os Lor\'and University, Gothard Astrophysical Observatory, 9700 Szombathely, Szent Imre H. st. 112, Hungary}
\affiliation{MTA-ELTE Lend{\"u}let "Momentum" Milky Way Research Group, Hungary}
\affiliation{MTA-ELTE Exoplanet Research Group, Szombathely, Szent Imre h. u. 112., H-9700, Hungary}

\author[0000-0002-1793-3689]{David L. Nidever}
\affiliation{Department of Physics, Montana State University, P.O. Box 173840, Bozeman, MT 59717, USA}

\author[0000-0001-8654-9499]{Annie C. Robin}
\affiliation{Institut UTINAM - UMR 6213 - CNRS - University of Bourgogne Franche Comt{\'e}, France, OSU THETA, 41bis avenue de l'Observatoire, 25000, Besan\c{c}on, France}

\author[0000-0003-0509-2656]{Matthew Shetrone}
\affiliation{University of California Observatories, UC Santa Cruz, Santa Cruz, CA 95064}

\author[0000-0002-0134-2024]{Verne Smith}
\affiliation{NSF's National Optical-Infrared Astronomy Research Laboratory, 950 North Cherry Avenue, Tucson, AZ 85719, USA}

\author{John C. Wilson}
\affiliation{Astronomy Department, University of Virginia, Charlottesville, VA 22901, USA}




\begin{abstract}


We present new maps of the Milky Way disk showing the distribution of metallicity ({\feh}), $\alpha$-element abundances ({\alpham}), and stellar age, using a sample of \textcolor{black}{66,496} red giant stars from the final data release (DR17) of the Apache Point Observatory Galactic Evolution Experiment (APOGEE) survey. We measure radial and vertical gradients, quantify the distribution functions for age and metallicity, and explore chemical clock relations across the Milky Way for the low-$\alpha$ disk, high-$\alpha$ disk, and total population independently. The low-$\alpha$ disk exhibits a negative radial metallicity gradient of \textcolor{black}{$-0.06 \pm 0.001$} dex kpc$^{-1}$, which flattens with distance from the midplane. The high-$\alpha$ disk shows a flat radial gradient in metallicity and age across nearly all locations of the disk. The age and metallicity distribution functions shift from negatively skewed in the inner Galaxy to positively skewed at large radius. Significant bimodality in the {\alpham}-{\feh} plane and in the {\alpham}-age relation persist across the entire disk. The age estimates have typical uncertainties of $\sim0.15$ in $\log$(age) and may be subject to additional systematic errors, which impose limitations on conclusions drawn from this sample. Nevertheless, these results act as critical constraints on galactic evolution models, constraining which physical processes played a dominant role in the formation of the Milky Way disk. We discuss how radial migration predicts many of the observed trends near the solar neighborhood and in the outer disk, but an additional more dramatic evolution history, such as the multi-infall model or a merger event, is needed to explain the chemical and age bimodality elsewhere in the Galaxy.

\end{abstract}

\keywords{Milky Way Galaxy (1054) -- Milky Way disk (1050) -- Galactic abundances (2002) -- Stellar ages (1581) -- Galaxy stellar content (621) -- Galactic Archaeology (2178) -- Galaxy structure (622) -- Milky Way formation (1053) -- Milky Way evolution (1052)}


\section*{}




\textit{Accepted to ApJ July 21, 2023}





\section{Introduction} \label{sec:intro}

The positions, chemical compositions, and ages of individual stars in the Milky Way reflect the formation and evolution history of our Galaxy, with each individual star acting as a "fossil" containing the chemical fingerprint of the interstellar gas from which it formed. Our inside perspective in the Milky Way grants the ability to study it in greater detail than any other galaxy, placing strong observational constraints on formation models and simulations of disk galaxies. For this reason, constraining the chemical and dynamical properties of the stellar populations in the Milky Way disk remains a cornerstone of modern galactic astronomy. 

Understanding the present-day chemical structure of our Galaxy has been increasingly successful with the advent of large spectroscopic stellar surveys like APOGEE \citep{Majewski2017}, Gaia \citep{GaiaDR3_Brown2021}, Gaia-ESO \citep{Gilmore_GaiaESO}, LAMOST \citep{LAMOST_Luo2015}, GALAH \citep{GALAH_Buder2018}, RAVE  \citep{Steinmetz2020}, and SEGUE \mbox{\citep[e.g.,][]{Yanny_2009}}. These surveys obtain precise kinematic and chemical information for a combined millions of stars across the Milky Way, with increasing sample sizes and more complete spatial coverage with every generation of survey. When paired with precise distances and positions from Gaia astrometry \citep{GaiaDR2_Brown2018,GaiaDR3_Brown2021}, these large surveys can access the evolution history of a large fraction of the Galactic disk. Even stellar ages, notoriously difficult to infer as they previously could not be directly measured for individual stars, are now readily available through the precise measurements of the masses for thousands of red giant stars available through asteroseismology \citep[e.g.,][]{Pinsonneault_2018,Miglio_2021}. These asteroseismic data sets are additionally used as training sets for machine learning techniques, expanding the stellar sample with age estimates to hundreds of thousands of stars \citep[e.g.,][Stone-Martinez et al. 2023]{Ness2016,Leung_2018,Anders_2018,Mackereth2019,Wu_2019,Ciuc_2021}.

Despite this wealth of data, the debate remains heated around which physical processes played the largest roles in shaping the Milky Way's disk. The structural and chemical distribution of stars in the Milky Way has been well studied, leading to the discovery of two main stellar components, the "thin" and "thick" disk near the solar neighborhood \citep[e.g.,][]{Yoshii1982,Gilmore1983}. These components are distinct in their dynamic signature, with the thick disk characterized by kinematically hotter stellar orbits (larger vertical velocity dispersion), and a slower systemic rotational velocity than the thin disk \citep[e.g.,][]{Soubiran2003,Juric2008,Kordopatis2013,Robin_2017}. The thin disk is also generally accepted to be more radially extended, and as the name implies, has a smaller scale height than the thick disk \citep[e.g.,][]{Bensby_2011,Bovy2016,Mackereth2017,Lian2022_maps,Robin_2022}. The two disks also differ in their chemical fingerprints, with the thin disk generally containing younger metal-rich stars characterized by their lower $\alpha$-element\footnote{$\alpha$-elements are elements with an atomic number multiple of 4 (the mass of a Helium nucleus, an $\alpha$-particle), e.g., O, Mg, S, Ca} abundances relative to the older, more metal-poor thick disk \citep[e.g.,][]{Furhmann1998, Bensby2005, Reddy2006, Lee2011, Bovy2012, Bovy2016, Nidever2014,Kordopatis_2015_GaiaESO,Hayden2015, Mackereth2017, Vincenzo2021, Katz_2021}. The {\alpham} ratio reflects the relative iron enrichment by prompt, massive core-collapse supernovae compared to the longer timescale Type Ia supernovae. Because of this, the {\alpham} ratio is generally high in populations that formed during rapid and efficient starbursts, and approaches solar "$\alpha$-poor" values in populations that form steadily over long time periods \citep[e.g.,][]{mb1990,thomas2005}. Thus, the chemical differences between the thin and thick disk suggests that they formed via distinct pathways, leaving the evidence of their enrichment histories within the present-day chemical structure of the Galaxy. However, many of these studies are biased towards the solar neighborhood due to observation limitations, and there has been some debate on whether the two components are truly distinct at all \citep[e.g.,][]{Bensby_2007,Bovy_2012,Kawata_2016,Hayden_2017,Anders_2018}.


Nevertheless, different explanations for the origins of this chemical bimodality in the disk have been proposed, using a combination of physical processes such as star formation, gas accretion, quenching, galaxy mergers, and stellar radial migration to attempt to explain the observed trends. The different models can be generally categorized into three scenarios: the "two-infall" models where the thick disk forms first followed by the thin disk, the "superposition" models where the two disks form in different parts of the Galaxy and mix through stellar migration, and the "clumpy formation" models where the two disks form simultaneously but with different star formation efficiencies.


The "two-infall" class of models, originally of \cite{Chiappini1997} and \cite{Chiappini_2001}, describe a scenario wherein the Milky Way first forms from the collapse of primordial gas, creating the progenitor of the present-day thick disk in a fast burst of star formation. The gas reservoir of the Galaxy is then quenched, entering a quiescent period of little star formation until the Galaxy receives a second infall of pristine gas. This accretion of fresh material dilutes the metallicity of the interstellar medium before reigniting star formation that forms the thin disk. The second gas infall happens over a longer time scale, allowing for a period of more continuous star formation, resulting in the $\alpha$-poor nature of thin disk. \citet{Linden_2017} constrained the timing of the second infall to be between 7-8 Gyr ago based on the ages and chemistry of star clusters in APOGEE. \citet{Spitoni_2019,Spitoni_2020,Spitoni_2021} expand on this model, constraining the length of the delay between the two episodes of gas infall to be between 3 - 5.5 Gyr, and proposing the second gas infall corresponds to a merger event with a gas-rich dwarf galaxy around 8-11 Gyr ago. This may coincide with the Milky Way's accretion of the Gaia-Enceladus dwarf galaxy, estimated to have happened 10 Gyr ago \citep{Helmi_2018,Vincenzo_2019}. 

A number of \textit{three}-infall models have also recently been proposed, including the model of \cite{Spitoni_2022} constrained to Gaia data. Their most recent infall starts $\sim$2.7 Gyr ago and gives birth to the recently discovered young, low-$\alpha$ stars that are impoverished in some elements \citep{gaiamaps2022}. This latest infall may be linked with the Sagittarius dwarf spheroidal galaxy's most recent perigalactic passage through the Milky Way's disk \citep{RuizLara_2020,Laporte_2019,Antoja_2020}. A starburst 2-3 Gyr ago has been detected independently in \citet{Isern_2019} and \citet{Mor_2019}.

The works of \citet{Lian2020a,Lian2020b,Lian_2020c} and \citet{Lian2021} present a modified version of the two-infall model. In their version, an underlying continuous episode of gas accretion is interrupted by two rapidly quenched starbursts. The first starburst forms the high-$\alpha$ thick disk, and the second starburst forms the metal-poor end of the low-$\alpha$ sequence 6 Gyr later. The metal-rich low-$\alpha$ sequence is attributed to the secular evolution phase between the two bursts.

Another variation of the two-infall model without the inclusion of merger events has been supported by recent chemo-dynamical simulations from \cite{Khoperskov_2021}. As in previous models, the thick disk is formed early on in a burst of star formation in a turbulent, compact disk. Stellar feedback from the formation of the thick disk drives outflows that quench star formation, enrich the Galactic halo, and eventually, feed the gas back into the disk on a more sustained timescale, creating the thin disk with a "galactic fountain" \citep[e.g.,][]{Shapiro_1976,Bregman_1980,Marinacci_2011,Fraternali_2017}. The models of \cite{Haywood_2016,Haywood_2018,Haywood_2019} support this scenario, where the high-$\alpha$ population was formed early on in a turbulent gas-rich disk with strong feedback, and the leftover, diluted gas forms the low-$\alpha$ thin disk on longer timescales.



The "superposition" class of chemical evolution models, pioneered by \cite{Schonrich_2009a,Schonrich_2009b}, reproduce the observed disk dichotomy without the need for a violent merger history to heat the thick disk. In this scenario, the chemical locus of the thin disk is not an evolutionary track; it is a superposition of end points of evolutionary tracks from different Galactocentric radii \citep[e.g.,][]{Nidever2014,Kubryk_2015,Sharma_2021b}. Stars from these different tracks reach the solar neighborhood by radial migration, a natural consequence of the Galaxy's spiral structure \citep{Sellwood2002,Roskar2008}. 
Stars in the high-$\alpha$ thick disk formed early during an efficient phase of rapid star formation, primarily in the inner Galaxy, before migrating to their present day radial distribution.


This superposition model is expanded on in the works of \citet{Minchev_2013, Minchev_2014}, \citet{Minchev2017}, and \cite{Johnson_2021}, which also emphasize the importance of radial migration in the Milky Way's structure. In these models, gas inflows, outflows, and star formation rates vary with Galactic location, emphasizing the radial dependence of the disk's chemical evolution history. Stellar radial migration allows stars to move around the Galaxy as time progresses, and potentially enrich a different spatial zone than the one they were born in when they die. These models show that this radial migration is the key to reproducing many of the observed trends in the Milky Way, including the changes of [Mg/Fe] and [Fe/H] distributions with radius and height.



A third, qualitatively different scenario is proposed by the "clumpy formation" model of \cite{Clarke_2019}, motivated by results from hydrodynamical simulations \citep[e.g.,][]{Bournaud_2007} and observations of high-redshift galaxies \citep[e.g.,][]{Elmegreen2005}. In their picture, the low-$\alpha$ thin disk is a true evolutionary sequence corresponding to inefficient star formation in the disk, while the high-$\alpha$ population is formed mainly during rapid, clumpy bursts in the Galaxy's early gas rich phase. These clumps are comparable to those observed in high-redshift galaxies with the \textit{Hubble Space Telescope} \citep[e.g.,][]{Elmegreen2005}. In addition to the chemical bimodality, these models also reproduce the observed mass density structure of the Milky way, including the flared thin disk \citep{Silva_2020,Amarante_2020}.



The discrepancy between these different proposed explanations, which all reasonably reproduce the observed trends in the Milky Way's disk, can only be closed with more observational constraints. Detailed chemical maps that cover the entire span of the disk, robust measurements of the Milky Way's radial and vertical metallicity gradients, and the metallicity distribution function will help constrain which physical processes played an important role in the formation of the disk. Adding in the ages of stars can provide an important axis for interpreting these results, as they enable a direct temporal connection between the properties of individual stars and the evolutionary time scale of the Milky Way \citep[e.g.,][]{Mackereth2017,Feuillet2019,Vazquez_2022}.


In this paper, the final data release of the Apache Point Galactic Evolution Experiment \citep[APOGEE;][]{Majewski2017, SDSSdr17} is used to further explore the properties of the Milky Way, with a larger sample size and greater spatial coverage than previously available. 

The chemical cartography of the Milky Way has been extensively studied previously using a variety of different surveys including SEGUE \citep[e.g.,][]{Lee2011,Lee_2011b,Gomez_2012}, RAVE \citep[e.g,][]{Kordopatis2013,Robin_2017}, GALAH \citep[e.g.,][]{Lin_2019,Hayden_2020, Sharma_2021}, LAMOST \citep[e.g.,][]{Huang_2020,Vickers_2021,Hawkins_2022}, Gaia \citep[e.g.,][]{Lemasle2018,gaiamaps2022,Poggio2022}, Gaia-ESO \citep[e.g.,][]{Bergemann2014,Kordopatis_2015_GaiaESO,Magrini_2018,Vazquez_2022}, and previous data releases of APOGEE \citep[e.g.,][]{Nidever2014,Hayden2015,Weinberg_2019,Eilers2021,Katz_2021}. These works, and others, have impressively advanced the field of chemical cartography over the last decade, meaning that many of the results presented in this paper are not new. However, the final data release of APOGEE presents a larger and more detailed base data set than previously available. This allows us to cull a selected high-quality sample, minimizing systematics while still probing a large number of stars at different locations across the Galaxy. Additionally, APOGEE has the distinct advantage of working in the infrared, easily accessing the heavily dust-obscured regions like the Galactic center and mid-plane, which are often beyond the reach of optical surveys. 


Our study complements the DR17-based study of \cite{Weinberg2021}, which focused on abundance trends of [X/Mg] for many different elements. These trends, which are nearly universal throughout the disk, provide insights on nucleosynthetic processes, while the distribution of stars in [Mg/Fe], [Fe/H], and age across the disk provide constraints on Galactic history.

In this work, we build upon the decades of previous discoveries and explore the chemical trends in the Milky Way disk through the legacy of the APOGEE survey. A high-quality sample of \textcolor{black}{$66,496$} red giant stars and their precise measurements of metallicity ({\feh}), ages, and $\alpha$-element abundances ({\alpham}) are used to create maps, measure gradients, quantify distribution functions, and trace age-abundance relations across the Milky Way disk, and compare the observations with the most recent models. In short, we find evidence supporting all three classes of chemical evolution models; Radial migration is an important process in shaping the disk over time, but the observed bimodality in $\alpha$-element abundances and ages persists even in disk regions where radial migration is not expected to be as prevalent. This suggests a multi-phase star formation history, such as that presented in the two-infall or clumpy formation class of models, is at least partially responsible for the formation of the Milky Way as seen today.

Section \ref{sec:data} contains an overview of the APOGEE survey and supplementary data used in this study. Spatial maps, gradient measurements, distribution functions, and other results are presented in Section \ref{sec:results} and compared with previous literature. In Section \ref{sec:discussion}, we discuss our results in the context of chemical evolution models. The conclusions we draw from this study are presented in Section \ref{sec:conclusions}. 

\section{Data} \label{sec:data}

\subsection{APOGEE} \label{sec:data:apogee}

The Apache Point Observatory Galactic Evolution Experiment \citep[APOGEE;][]{Majewski2017} is a high-resolution ($R\sim22,500$) near-infrared ($1.51-1.70$ $\mu$m) spectroscopic survey containing observations of 657,135 unique stars released as part of the SDSS-IV survey \citep{Blanton_2017}. The spectra were obtained using the APOGEE spectrograph \citep{apogeespectrographs_Wilson2019} mounted on the $2.5$m SDSS telescope \citep{apo25m_Gunn2006} at Apache Point Observatory to observe the Northern Hemisphere (APOGEE-N), and expanded to include a second APOGEE spectrograph on the $2.5$ m Ir{\'e}n{\'e}e du Pont telescope \citep{lco25m_Bowen1973} at Las Campanas Observatory to observe the Southern Hemisphere (APOGEE-S). The final version of the APOGEE catalog was published in December 2021 as part of the 17th data release of the Sloan Digital Sky Survey \citep[DR17;][]{SDSSdr17} and is available publicly online through the SDSS Science Archive Server and Catalog Archive Server\footnote{Data Access Instructions: \url{https://www.sdss.org/dr17/irspec/spectro_data/}}.

The APOGEE data reduction pipeline is described in \cite{Nidever_2015_ApogeeDataReduction}. Stellar parameters and chemical abundances in APOGEE were derived within the APOGEE Stellar Parameters and Chemical Abundances Pipeline \citep[ASPCAP;][J.A. Holtzman et al. 2022 in prep.]{Holtzman_2015,aspcap,Holtzman_2018,Jonsson_2020}. ASPCAP derives stellar atmospheric parameters, radial velocities, and as many as 20 individual elemental abundances for each APOGEE spectrum by comparing each to a multi-dimensional grid of theoretical model spectra \citep{Meszaros2012,Zamora2015} and corresponding line lists \citep{Shetrone_2015,Smith_2021}, employing a $\chi^2$ minimization routine with the code \texttt{FERRE} \citep{AllendePrieto_2006} to derive the best-fit parameters for each spectrum. We highlight that several elements (notably {\alpham}) were updated in DR17 to include non-LTE effects in the stellar atmosphere. ASPCAP reports typical accuracy in metallicity measurements within 0.01 dex \citep{aspcap2018}. In this study, we adopt the calibrated values for surface gravity ($\log g$), metallicity ({\feh}), and $\alpha$-element abundances ({\alpham}) from ASPCAP. We adopt [Mg/Fe] for our $\alpha$-element abundance instead of the "total" [$\alpha$/M], because [Mg/Fe] is the most precisely measured abundance by ASPCAP, and this element ratio has been traditionally used to define the boundary between the chemical thin and thick disk.


\subsection{Sample Selection} \label{sec:data:selection}

\begin{figure}
    \centering
    \includegraphics[width=0.45\textwidth]{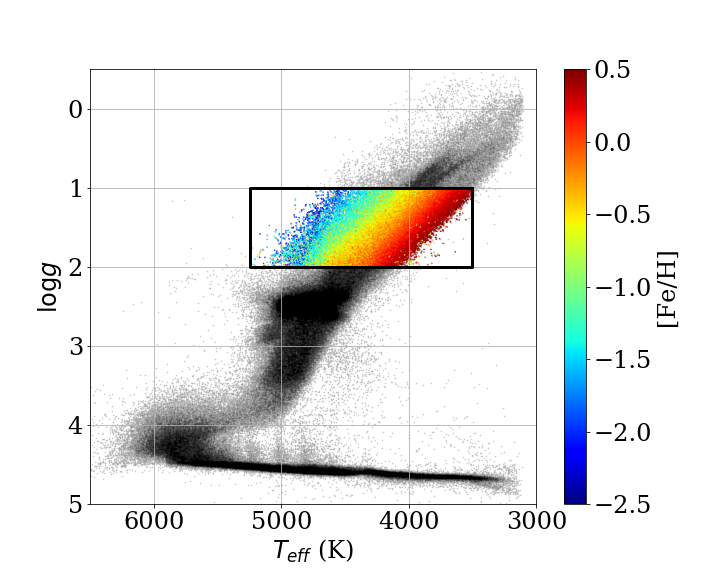}
    \caption{The T$_{\rm eff}$-$\log g$ distribution of stars in the sample described in Section \ref{sec:data}. Our adopted red giant sample is outlined in black and plotted by color (metallicity), while the full APOGEE sample is shown in gray in the background for reference.}
    \label{fig:sample_HRdiagram}
\end{figure}

\begin{figure}
    \centering
    \includegraphics[width=0.45\textwidth]{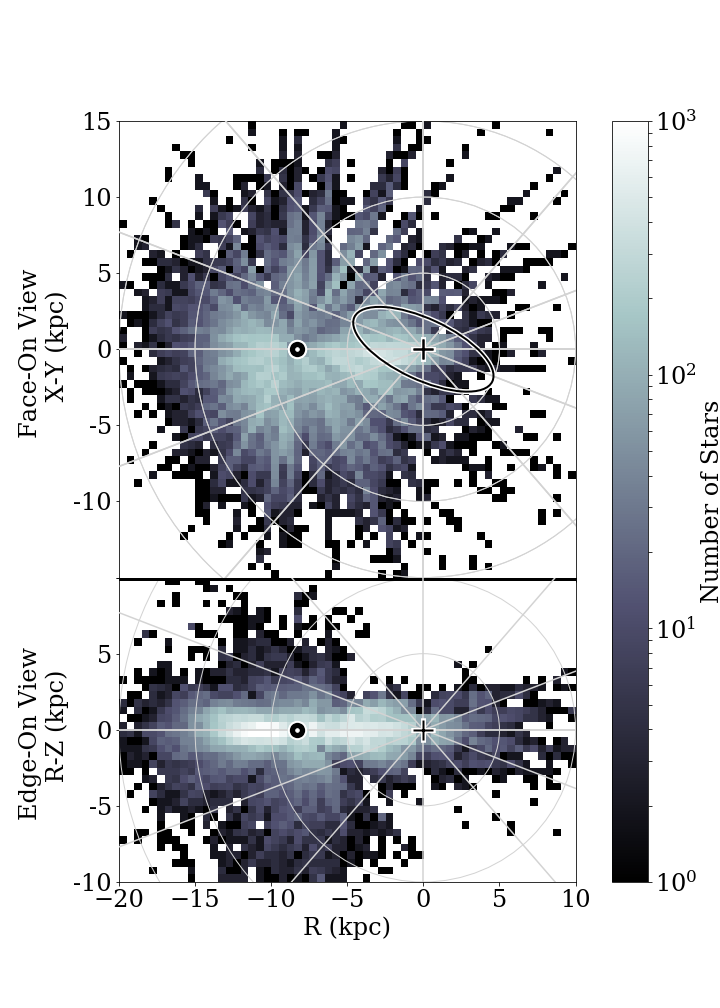}
    \caption{The spatial distribution of stars in our sample shown as a face-on view of the Galaxy ($X$-$Y$ plane; top panel), and an edge-on view ($R$-$Z$ plane; bottom panel). Each spatial bin of $\Delta X =  \Delta Y = \Delta Z$ = 0.5 kpc is colored by the number of stars in that location. The position of the Sun at $X = -8.3$ kpc is denoted by the solar symbol ($\odot$), and the Galactic center is marked with a plus (+). The ellipse around the Galactic center marks the approximate location of the bar, as an ellipse with major axis length 10 kpc, a 0.4 axis ratio, and rotated 25$^{\circ}$.}
    \label{fig:star_numbers}
\end{figure}

Several cuts were made to the full APOGEE catalog to refine our sample. First, only stars defined as APOGEE main survey targets (also sometimes called the "main red giant sample") were selected using the \texttt{EXTRATARG} flag. This removes any duplicate entries, as well as any ancillary science or other survey stars that were targeted for observation for a specific purpose (e.g., satellite or dwarf galaxy targets, star cluster member candidates, {\it Kepler} Objects of Interest). The main survey targets were randomly selected for observation from the 2MASS catalog, based on their $(J-K)$ color and $H$-band apparent magnitude. For more information on the targeting strategies of APOGEE, see \citet{Zasowski2013,Zasowski2017}, \citet{Beaton_2021}, and \citet{Santana_2021}.

Stars with noisy spectra ($S/N < 50$) or unreliable parameter estimates from ASPCAP were removed from our sample using the \verb|SN_BAD| and \verb|STAR_BAD| ASPCAP bits respectively. The latter is triggered when the derived parameters for a star are designated a bad fit by its high $\chi^2$ value, when the derived temperature does not match the star's observed color, when any individual stellar parameter measurement is flagged as bad, or when the derived parameters lie on an edge of the synthetic spectral grid.

The sample is further restricted to stars with surface gravity values between $1 \leq \log g \leq 2$. Limiting to a small range in $\log g$ minimizes potential systematic uncertainties in abundance measurements, which tend to present as a function across T$_{\rm eff}$ and $\log g$ in APOGEE \citep[e.g.,][]{aspcap2018,Eilers2021}. The higher luminosity of these giants helps probe larger distances, allowing for a wide range of positions to be sampled across the Galactic disk in our study. Fainter stars may be better sampled closer to the Sun, but to keep our sample consistent across all distances, we apply this $\log g$ cut to ensure the trends we are documenting are not attributed to any systematic bias. \citet{Eilers2021} presents an empirical correction for these systematics for those interested; this would mainly be a concern if the expected distribution of $\log g$ in observations varies significantly with distance, which we do not expect in our sample. The lower $\log g$ limit is also imposed by the availability of asteroseismic data, as no age estimates are available for stars with $\log g < 1$ in {\texttt{distmass}}. 

Figure \ref{fig:sample_HRdiagram} shows the T$_{\rm eff}$-$\log g$ distribution of the sample after these refinements. The final number of stars in our RGB sample is \textcolor{black}{66,496}. Due to the particulars of APOGEE field selection, there are more stars above the disk ($Z \geq 0; N=$ \textcolor{black}{38,031}) than below ($Z \leq 0; N=$ \textcolor{black}{28,465}), and more observations towards the Galactic center ($R \leq R_{\odot}; N= $ \textcolor{black}{36,317}) than outward ($R \geq R_{\odot}; N=$ \textcolor{black}{30,179}). The spatial distribution of our sample is shown in Figure \ref{fig:star_numbers}. The stellar distance estimates used for this Figure (and the remaining of the paper) are described in Section \ref{sec:data:ages}. 

In this work, we make no correction for the selection biases within the APOGEE survey. Stars close to the solar neighborhood will be over-represented in our sample. As shown in Figure \ref{fig:star_numbers}, as distance from the Sun increases, the number of stars available in the APOGEE sample decreases.\footnote{This explanation is an oversimplification of the APOGEE selection function. The actual selection function depends on more than just distance from the Sun, as targeting strategies may vary between fields, observing time availability and instrument specifications differed between the North and South, and the non-homogeneous dust distribution in the Milky Way plays a major role in what can be observed.}
There are certain limitations that this selection function imposes on this work and similar studies. Specifically, results should not be averaged over a large spatial range, as the relative number of observed stars will clearly weight the average towards the solar neighborhood. Additionally, nothing can be inferred from the relative number of stars between locations, or about the intrinsic density profile of stars in the disk, because the former is heavily influenced by the selection function. That said, the effect of the selection function should be negligible when confined within a small spatial zone and $\log g$ limit in the Galaxy, such that general abundance trends and \textit{normalized} number distributions should be consistent even without correcting for the APOGEE selection function \citep[e.g.,][Appendix A]{Hayden2015}. In this work, we consistently bin stars in different ranges of $R$ and $Z$ to avoid this bias and explore how chemical and age trends vary across the disk. A more complete prescription on how to account for the effects of the selection function in APOGEE has been published in \citet{Bovy2012,Bovy2016}, \citet{Mackereth2017} for previous data releases, and Imig et al. (in prep.) for DR17. Imig et al. (in prep.) will present the density distribution of mono-age mono-abundance stellar populations in APOGEE DR17 after correcting for the selection function.



\subsection{{\alpham} Subsamples} 
\label{sec:data:selection:alpha}

\begin{figure}
    \centering
    \includegraphics[width=0.45\textwidth]{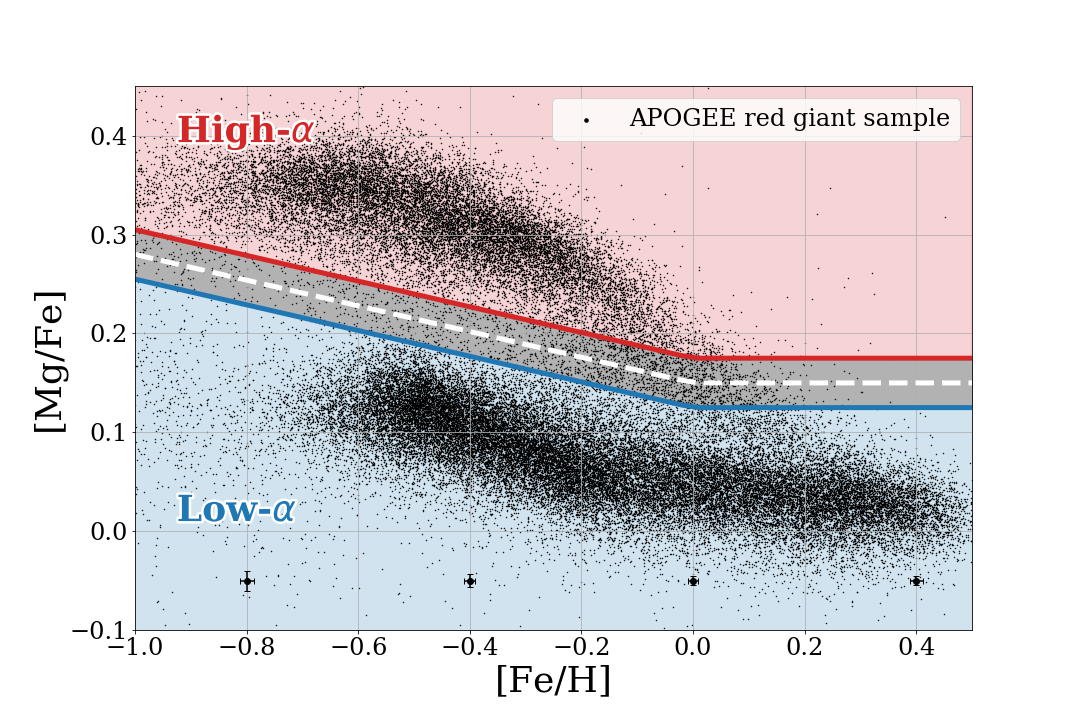}
    \caption{The {\alpham}-{\feh} plane for stars in our red giant sample (black points), demonstrating the adopted definition of the low-$\alpha$ (blue) and high-$\alpha$ (red) sequence defined in Equation \ref{eq:alpha_split}. The gray region is an added buffer zone of \feh=$\pm 0.025$ dex around the line to remove overlap between the two sequences due to abundance uncertainties. The typical uncertainties associated with each measurement of {\feh} and {\alpham} are shown in the error bars near the bottom of the plot (as the $\pm 1 \sigma$ value), for several selected metallicities.}
    \label{fig:alpha_samples}
\end{figure}

Figure {\ref{fig:alpha_samples}} shows the {\alpham}-{\feh} plane for our full sample, where the bimodal distribution in {\alpham} is obvious \citep[e.g.,][]{Furhmann1998,Bensby2005,Reddy2006,Lee2011,Kordopatis_2015_GaiaESO,Hayden2015,Katz_2021}. Although there is some debate on whether the two sequences are truly distinct (see Introduction), we use this figure to define two further subsamples in our data to investigate this question later. \citet{Vincenzo2021} demonstrate that the distribution in [Mg/Fe] at fixed [Fe/H] is genuinely bimodal when considering the full disk population at near-solar radii, after accounting for the APOGEE selection function. We define the $\alpha$-poor "thin disk" sequence and the $\alpha$-rich "thick disk" sequence by splitting the full sample into two groups defined by a line in {\alpham}-{\feh} space, shown in Figure \ref{fig:alpha_samples}. We adopt a similar limit as \cite{Weinberg_2019,Weinberg2021} parameterized by the equation:

\begin{equation}
    \label{eq:alpha_split}
    {\rm \alpham} =
    \begin{cases}
    0.15-0.13*{\rm \feh} & \text{if } {\rm \feh} \leq 0\\
    0.15 & \text{if } {\rm \feh} > 0\\
    \end{cases}
\end{equation}

Our equation differs from \cite{Weinberg_2019,Weinberg2021} by a small offset of [Mg/Fe] $=+0.03$ dex, correcting for abundance calibrations. 

This separation in $\alpha$-element abundances is shown in Figure \ref{fig:alpha_samples}. A conservative buffer zone within $\pm 0.025$ dex of the line is excluded to remove potential overlap between the two sequences. This value is larger than the typical uncertainties of APOGEE abundance measurements, shown as the $\pm 1 \sigma$ in the error bars across the bottom of the plot. 

\subsection{Age and Distance Estimates} \label{sec:data:ages}

Accurately mapping the Milky Way in three dimensions requires knowing precise distances to every star in our sample. Galactocentric positions were calculated for each star using the right ascension (RA) and declination (DEC) from APOGEE observations and distance estimates from the APOGEE {\texttt{distmass}} value added catalog (A. Stone-Martinez et al. 2023, submitted).
The {\texttt{distmass}} distances were obtained through a neural network that was trained to estimate a star's luminosity based on its ASPCAP parameters, using Gaia and cluster distances to provide the training labels. Distance estimates from the {\texttt{distmass}} catalog are typically precise within 10\%. For the purpose of calculating Galactocentric coordinates, we define the reference location of the Sun to be $R_{\odot} = 8.3$ kpc with a height of $z_{\odot} = 0.027$ kpc above the plane \citep{BlandHawthorn2016}.

For evolved red giant stars, carbon and nitrogen abundances provide mass information because of the mass-dependence of stellar mixing \citep[e.g.,][]{Iben_1965,Salaris_2005}, allowing the determination of stellar masses for stars without asteroseismology \citep[e.g.,][]{Martig_2016,Vincenzo_2021}. This fundamental property is used to derive stellar age estimates in the {\texttt{distmass}} catalog, wherein ages are derived by training a second neural network on the ASPCAP parameters of stars with asteroseismology masses from the APOGEE-{\it Kepler} overlap survey \citep[APOKASC; ][APOKASC3: Pinsonneault et al. 2023 in prep.]{Pinsonneault_2018}. The neural network learns the relations between the ASPCAP parameters and asteroseismic masses for stars from APOKASC, then it predicts the masses for all giant stars from DR17. Knowing the masses for evolved stars, ages can be derived through stellar evolution theory which predicts a star's location on an isochrone. For the {\texttt{distmass}} catalog, isochrones from \cite{Choi_2016} were adopted to make this conversion from derived mass to stellar age. The isochrones cover a range of ages ($5.0 \leq \mathrm{log}(\mathrm{age}) \leq 10.3$), metallicities ($-2.0\leq [{\rm{Z}}/{\rm{H}}]\leq 0.5$), and masses ($0.1\leq M/{M}_{\odot }\leq 300$). 

\begin{figure}
    \centering
    \includegraphics[width=0.45\textwidth]{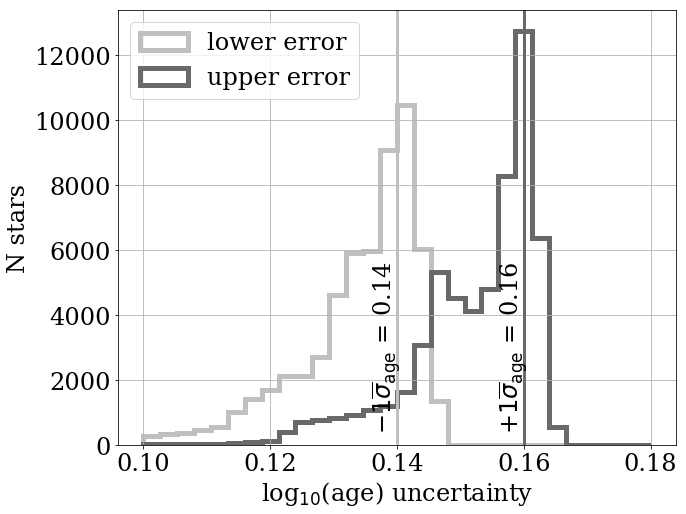}
    \caption{The reported $\pm 1\sigma$ uncertainty in stellar age estimates from the distmass value added catalog. The light gray histogram shows the lower uncertainty values for the sample and the dark gray histogram shows the upper uncertainties for the sample. The median of each distribution is plotted as a vertical line and labeled to highlight a "typical" uncertainty value.}
    \label{fig:age_errors}
\end{figure}

The reported uncertainties on the age estimates in {\texttt{distmass}} are shown as a histogram in Figure \ref{fig:age_errors} for our sample; the median lower uncertainty is $0.14$ $\log_{10}$(age) and the median upper uncertainty is $0.16$ $\log_{10}$(age). The uncertainties in stellar age are propagated from the uncertainties in stellar mass, which are predicted from the spread in mass from the neural network training set. The mass and age uncertainties have no strong dependence on metallicity or $\log g$ (A. Stone-Martinez et al. 2023, submitted).

To determine if these uncertainties are realistic, A. Stone-Martinez et al. (2023, submitted) performs an additional evaluation of the age estimates by comparing to previous literature. Compared to a small sample of cluster members with independent age estimates from main sequence turnoff fitting, they find that the {\texttt{distmass}} ages are accurate within $\pm1\sigma = 0.16 $ in $\log_{10}$(age) across 12 different star clusters with ages $ 9.2 \leq \log_{10}$(age) $\leq 9.7$ Gyr. Compared to a larger sample of field stars from the {\texttt{astroNN}} catalog \mbox{\citep{Leung_2018,Mackereth2019}}, the {\texttt{distmass}} ages show a typical spread of $\pm1\sigma = 0.11 $ in $\log_{10}$(age), although these ages are derived with a similar methodology. Both of these evaluations are consistent within the reported age uncertainties.


Precise stellar ages remain difficult to measure robustly for large samples of stars. Neural network derived ages like {\texttt{distmass}} are heavily dependent on their training set values, and any uncertainties in the training set labels will influence the neural network model. The {\texttt{distmass}} catalog trains on stellar masses derived from asteroseismology, which have their own systematics and different groups have derived different results. Appendix \ref{sec:app:ages} tests some of our results using different training sets and age catalogs, and motivates our choice of adopted ages. Even within {\texttt{distmass}}, there are six provided stellar age estimates for every star from neural networks trained on different asteroseismic mass estimates from three different research groups in APOKASC 3 (Pinsonneault et al. 2023 in prep.), and a set of "corrected" and "uncorrected" masses for each. The corrections are motivated by Gaia data, using parallaxes to calibrate the stellar radius derived from asteroseismology \citep[e.g.,][]{Zinn_2019}. These corrections may be less reliable at lower $\log g$ and produce results that are less consistent across different bins in $\log g$. For the remainder of this paper, we use the {\texttt{distmass}} results trained on the uncorrected SS ages from APOKASC 3 (Pinsonneault et al. 2023 in prep.), corresponding to the column named \verb|"AGE_UNCOR_SS"| in the {\texttt{distmass}} catalog.

Selecting different age catalogs does make some difference in our results, particularly among the oldest stars as shown in Appendix \ref{sec:app:ages}. Because this quantitative aspect can change considerably, we caution the reader against drawing strong conclusions from any of our age-related results without thoroughly understanding the related caveats outlined in  Appendix \ref{sec:app:ages}.

An additional quality flag from the {\texttt{distmass}} catalog is used to refine the sample when using the stellar age estimates. Namely, we remove stars that have bit 2 set, indicating stellar parameters T$_{\rm eff}$, {\feh}, and $\log g$ lie outside of the range covered by the APOKASC training set; this removes stars with potentially unreliable mass (and therefore age) estimates. For anything involving ages, the full RGB sample is additionally restricted to \textcolor{black}{57,756} stars with this cut. Notably, all stars with {\feh} $\leq -0.7$ are excluded by this criterion. Metal-poor stars have extra mixing that was not learned by the neural network because there were no metal poor stars in the training set. Because metal-poor stars tend to be older, this means that for age related figures, the oldest stars ($\rm{age}>10^{10}$ years) may not be well-represented in our sample, particularly at large radii in the Galaxy; the potential effects of this, and other age-related caveats, are explored more in Appendix \ref{sec:app:ages}.


\begin{figure*}
    \centering
    \includegraphics[width=\textwidth]{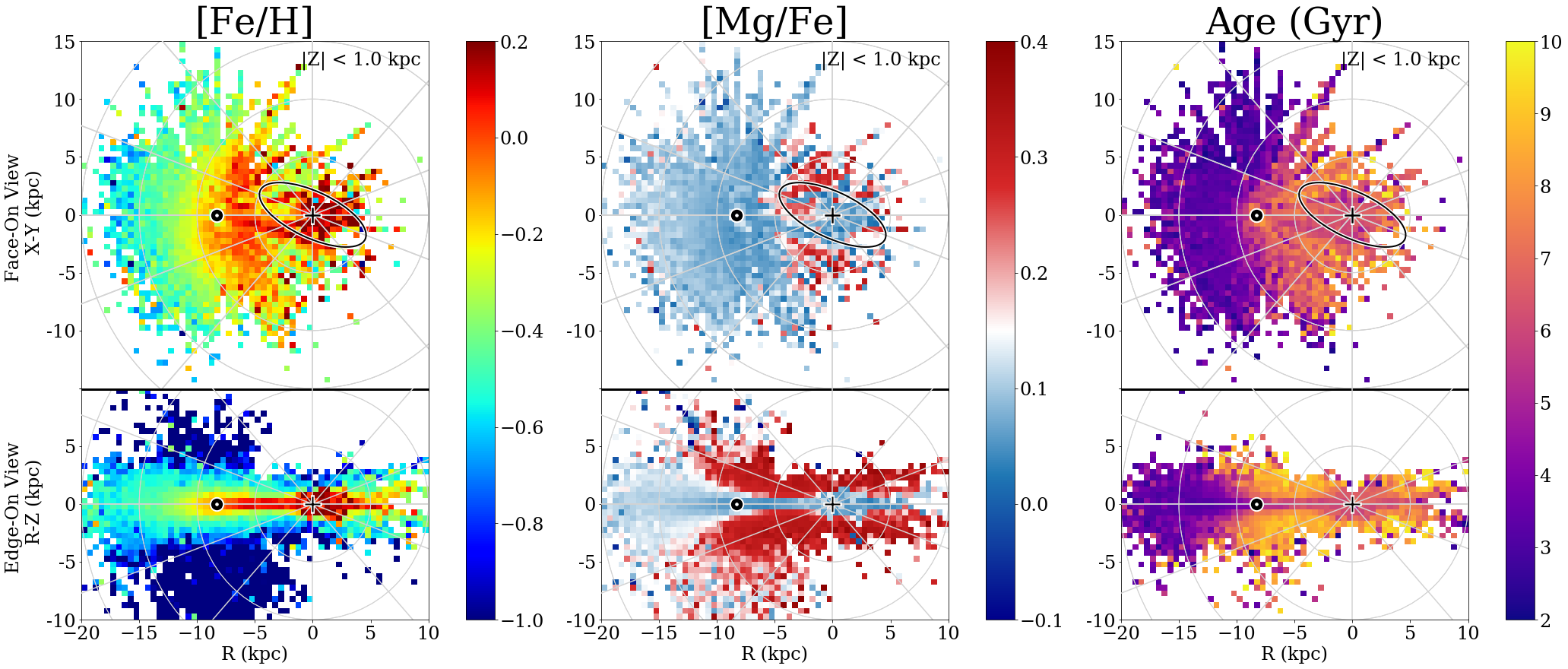}
    \caption{Global Maps of the Milky Way, showing the average distribution of {\feh} (left), {\alpham} (middle), and stellar age (right) across the Galaxy. The top row of panels shows a face-on view ($X-Y$ plane), integrated through the disk with $|Z|\leq 1.0$ kpc. The bottom row of panels shows an edge-on view ($R-Z$ plane, with $R$ preserving the sign of $X$ to show the opposite side of the Galaxy), integrated through the whole disk. Colors encode the median quantities in each or $X-Y$ or $R-Z$ pixel. In the face-on views, the age and metallicity gradients are visible, with the Galactic bar standing out as metal-rich and $\alpha$-poor. The location of the Sun at $X=-8.3$ kpc is denoted by the solar symbol ($\odot$), and the Galactic center is marked with a plus (+). The approximate location of the Galactic bar is also shown as an ellipse with major axis length 10 kpc, a 0.4 axis ratio, and rotated 25$^{\circ}$.}
    \label{fig:global_maps}
\end{figure*}

\newpage 

\section{Results} \label{sec:results}

\subsection{Cartography} \label{sec:results:cartography}

Maps of the Galactic disk as a function of {\feh}, {\alpham}, and stellar age are shown in Figure \ref{fig:global_maps} in face-on ($X-Y$ plane; top row) and edge-on ($R-Z$ plane; bottom row) perspectives. The median value of each parameter is calculated for different spatial bins sized $\Delta X = \Delta Y = 0.5$ kpc, and shown as the respective color on the figure. For the edge-on perspective, the sign of the $X$ coordinate is applied to the $R$ coordinate, to better highlight the spatial coverage of the observations on the opposite side of the Galaxy.

In median metallicity (left column), clear radial and vertical metallicity gradients are visible in the disk, with higher average metallicities near the Galactic center that decline towards outer radii. In median $\alpha$-element abundances (middle column), the bimodality in the disk shows low-$\alpha$ stars congregating in the "thin disk" near the Galactic midplane, and high-$\alpha$ stars populating the "thick disk" at higher $Z$ locations. At larger radii, the low-$\alpha$ stars extend farther above and below the plane. The high-$\alpha$ stars are more centrally concentrated. The right column, colored by median stellar age, contains fewer stars due to the additional cuts described in Section \ref{sec:data:ages} when dealing with ages from the {\texttt{distmass}} catalog. Once again, radial and vertical gradients appear in these maps, as well as younger stars extending farther above the plane in the outer Galaxy. The innermost structural features of the Galaxy, such as the bar (noted by the ellipse) and bulge stand out as metal-rich, $\alpha$-poor, and older-aged than stars at similar radii but different azimuthal angles, consistent with previous studies of the central regions of the Galaxy \citep{Wegg_2019,Zasowski_2019,Hasselquist_2020, Eilers2021,Lian2021,Queiroz_2021} for this metallicity range.

\begin{figure}
    \centering
    \includegraphics[trim={3cm 0 2cm 0},height=0.85\textheight]{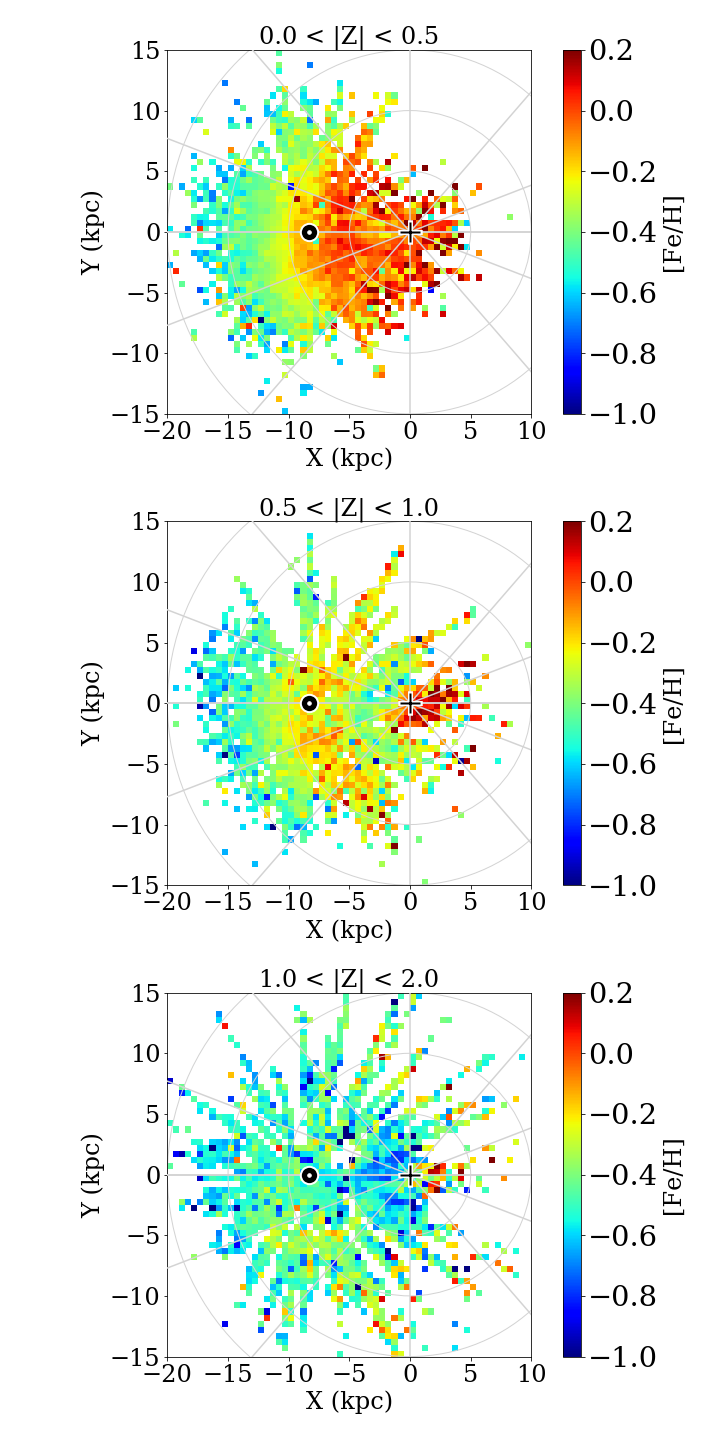}
    \caption{Face-on maps of Galactic disk, showing the median metallicity ({\feh}) of stars in spatial bins of $\Delta X = \Delta Y = 0.5$ kpc. The different panels are slices in vertical space, from closest to the Galactic plane (top panel; $|Z| < 0.5$ kpc) to increasing heights above the plane (bottom panel; $1 < |Z| < 2$ kpc). The Sun’s position is marked by the solar symbol ($\odot$), and the position of the Galactic center is indicated by a plus ($+$).}
    \label{fig:metal_map}
\end{figure}

\begin{figure}
    \centering
    \includegraphics[trim={3cm 0 2cm 0},height=0.85\textheight]{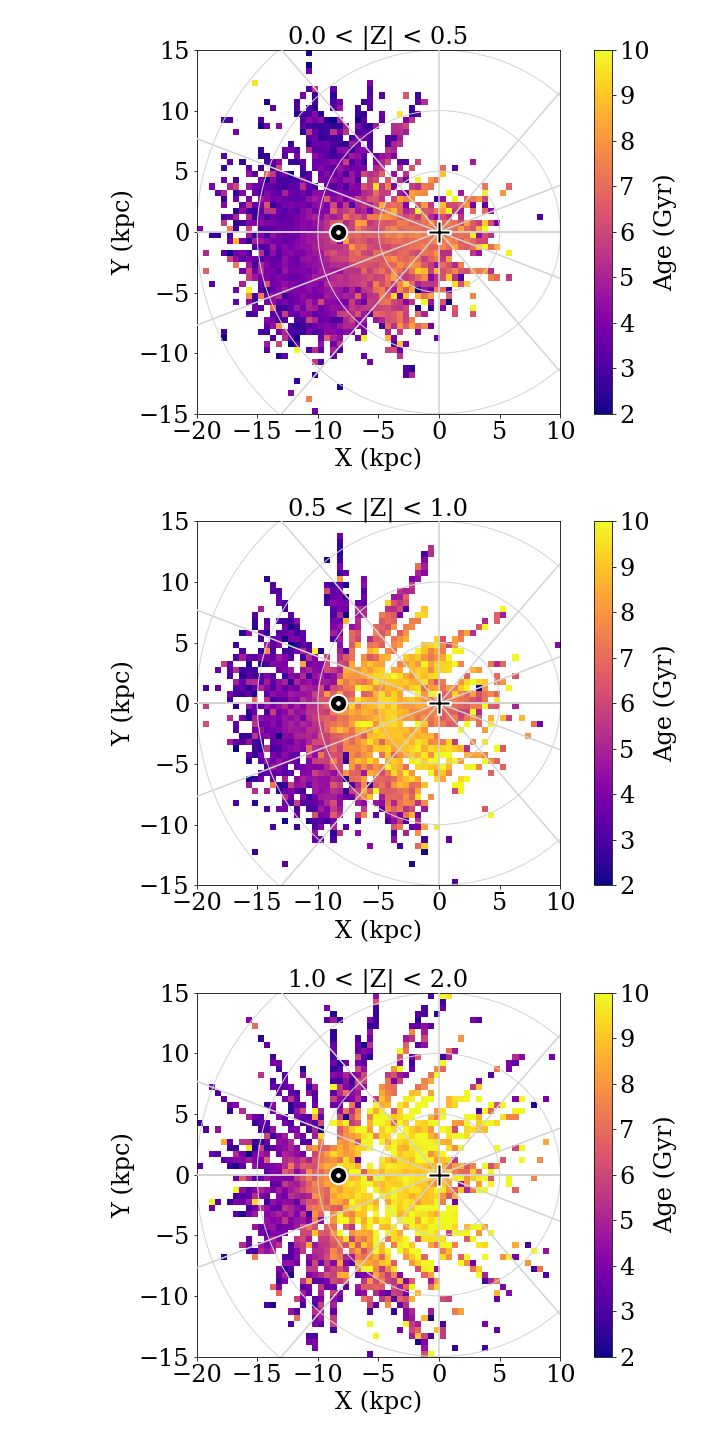}
    \caption{Same as Figure \ref{fig:metal_map}, but colored by the median age (in Gyr) of stars in each spatial bins of $\Delta X = \Delta Y = 0.5$ kpc.}
    \label{fig:age_map}
\end{figure}



Dividing the maps into vertical bins reveals more nuanced structure; Figure \ref{fig:metal_map} depicts face-on metallicity maps divided by height above the Galactic plane, from closest to the Galactic plane (top panel; $|Z| \leq 0.5$ kpc) to farthest away (bottom panel; $1 \leq |Z| \leq 2$ kpc). The metallicity gradient is strongest close to the Galactic plane, with locations near the Galactic center showing a higher median metallicity than those at larger radii, as expected. Farther above the mid plane, the trend becomes less apparent, with almost no obvious gradient present when $|Z| \geq 1$ kpc, and the stellar populations showing a lower median metallicity overall. The middle panel ($ 0.5 \leq |Z| \leq 1$ kpc) shows a peculiar trend where the median metallicity actually increases from $R=0$ until $R \sim 7$ kpc, and then decreases with a shallow metallicity gradient. This build-up of metal-rich stars in the center of the Galaxy is possibly the signature of the bulge.

The age distribution of the Galactic disk is shown in Figure \ref{fig:age_map}. Again, the age gradient is strong close to the Galactic plane, with older stars more common near the center and younger stars dominating in the outer Galaxy. Unlike in metallicity, there is still a radial age gradient above the Galactic plane ($|Z| > 1$ kpc), although in general the stars found above the plane are older than the stars found in the plane.



\subsection{{\alpham} Distribution} \label{sec:results:AlphaDF}

\begin{figure*}
    \centering
    \includegraphics[width=\textwidth]{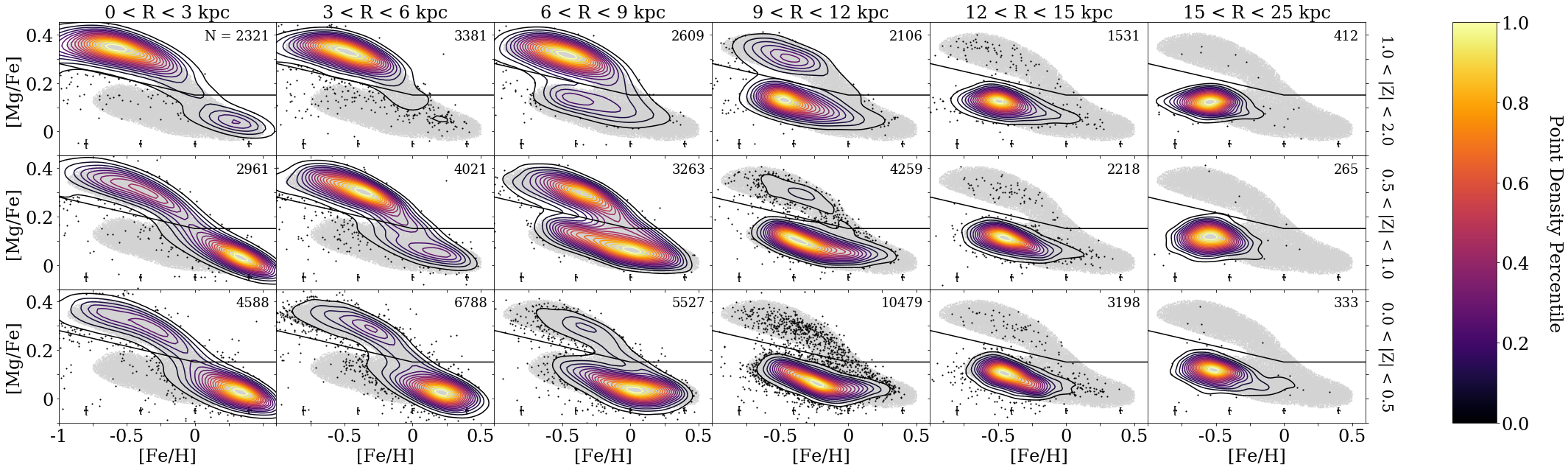}
    \caption{The distribution of stars in the {\alpham} vs. {\feh} plane as a function of $R$ and $|Z|$, as a contour map of point density. Spatial bins move from closest to the Galactic plane (bottom row, 0.0 $<|Z|<$ 0.5 kpc) to farthest above the Galactic plane (top row, 1.0 $<|Z|<$ 2.0 kpc), and from close to the Galactic center (left column, 0.0 $<|R|<$ 3.0 kpc) to farthest out in the disk (right column, 15.0 $<|R|<$ 25.0 kpc). The number in the top-right corner of each panel is the number of stars in our sample in that spatial bin. For reference, the gray background shape and black line is the same in each panel, to highlight how the sequence changes across location in the Galaxy. The black line is the boundary between high- and low-$\alpha$ populations defined in Equation \ref{eq:alpha_split}, and the gray shape is the contour containing 90\% of the points in the full sample. The typical uncertainties in abundance measurements as a function of metallicity are shown as a $\pm1\sigma$ value at the bottom of each panel for reference.}
    \label{fig:apogee_mhplots}
\end{figure*}

\begin{figure*}
    \centering
    \includegraphics[width=\textwidth]{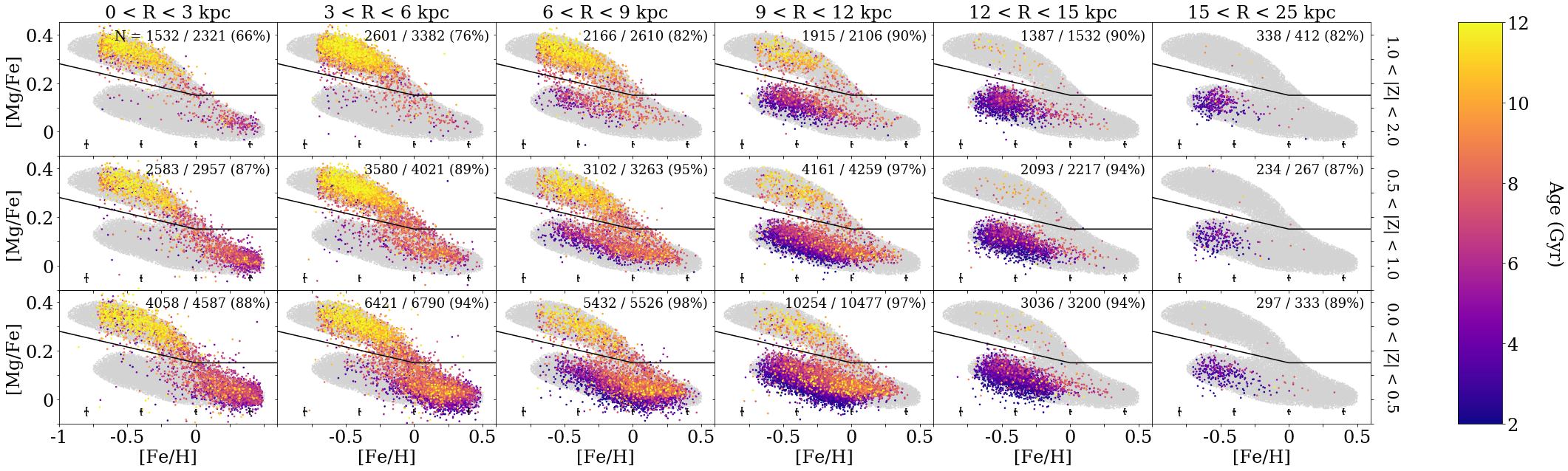}
    \caption{Same as Figure \ref{fig:apogee_mhplots}, but colored by stellar age. Points with [Fe/H]$\leq$ -0.7 have been excluded for potentially unreliable age estimates with the cuts described in Section \ref{sec:data:ages}. The percentage of stars ($f_{\rm distmass}$) in each bin which pass the distmass quality criterion is shown in parentheses in the upper right corner of each panel.}
    \label{fig:apogee_mhplots_age}
\end{figure*}

Figure \ref{fig:apogee_mhplots} shows the Galactic distribution of stars in the {\alpham}-{\feh} chemistry plane as a function of Galactic position. The different rows are the same vertical bins adopted in previous sections, with the bottom row closest to the Galactic plane ($0 < |Z| < 0.5$ kpc) and the top row farthest from the plane ($1.0 < |Z| < 2.0$ kpc). The columns are different radial bins, from closest to the disk center (left column) to farthest out (right column). Each panel shows the {\alpham}-{\feh} distribution of stars in its respective spatial zone, colored by stellar point density (in Figure \ref{fig:apogee_mhplots}) and stellar age (in Figure \ref{fig:apogee_mhplots_age}). Our adopted definition of the split between the high- and low-$\alpha$ sequences (equation \ref{eq:alpha_split}) is plotted in black. For reference, the gray background highlights the distribution of the full sample, indicating the contour within which 90\% of the sample is found.

Generally, the low-$\alpha$ sequence is concentrated close to the Galactic plane (bottom row), and the high-$\alpha$ sequence is more prominent outside the plane (top row) for $R < 12$ kpc. The location of the high-$\alpha$ sequence does not change based on location in the Galaxy. The low-$\alpha$ sequence is more metal rich near the center of the Galaxy (left column), and moves to more metal-poor with increasing radius (right column). Additionally, at large radii, the low-$\alpha$ sequence extends farther above the plane than it does close to the Galactic center. All of this has been well-documented in previous studies \citep[e.g.,][]{Bensby2005,Bensby_2011,Nidever2014,Kordopatis_2015_GaiaESO,Hayden2015,Katz_2021,Vincenzo2021,gaiamaps2022}.



Figure \ref{fig:apogee_mhplots_age} shows the spatial distribution of {\alpham} coded by stellar age. The high-$\alpha$ sequence is composed of older stars at all radial bins. The low-$\alpha$ sequence includes older stars close to the Galactic center ($R < 6$ kpc) and younger stars farther out in radius. At any radius, the lower {\alpham} stars within the low-$\alpha$ sequence have younger ages. Within the low-$\alpha$ sequence, stellar age correlates more with {\alpham} than it does with {\feh}, indicating that the low-$\alpha$ sequence is likely not a true single sequence.

\begin{figure}
    \centering
    \includegraphics[width=0.45\textwidth]{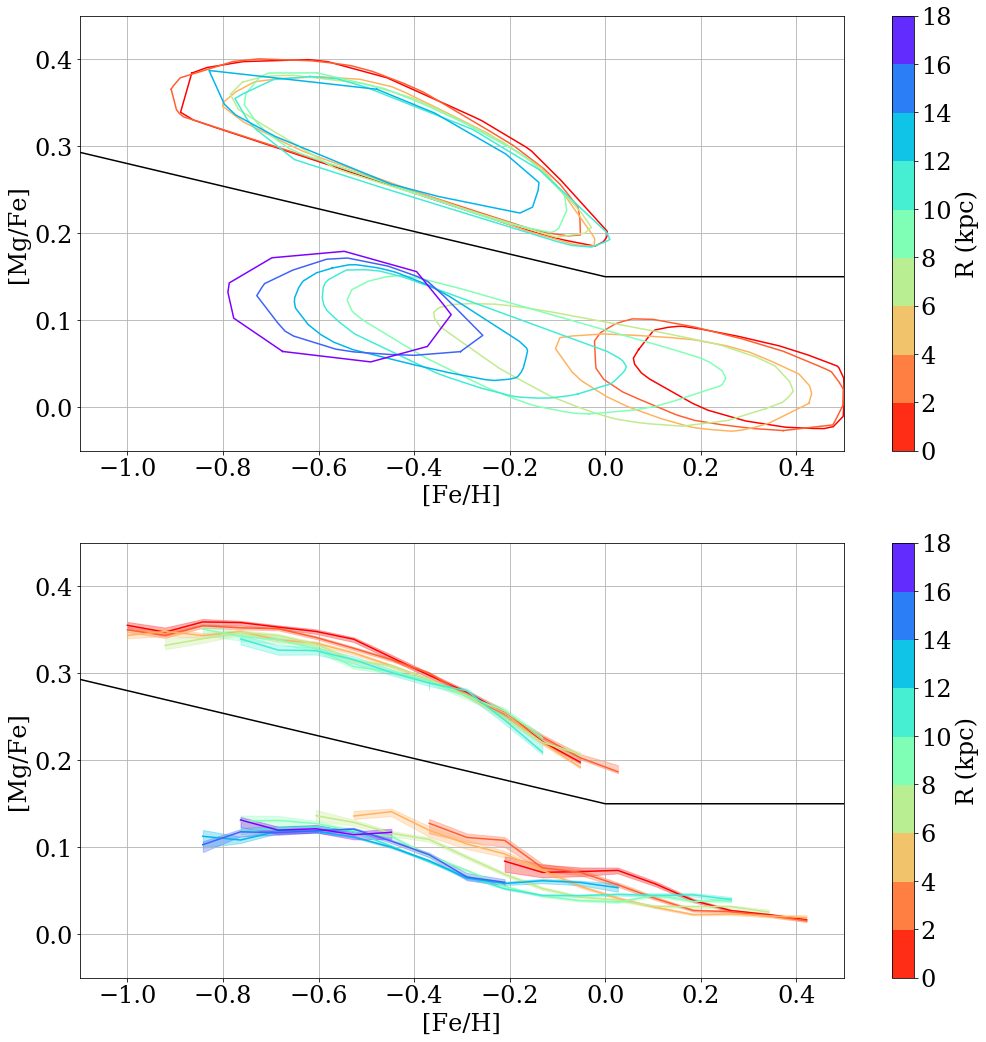}
    \caption{Variation in the {\alpham}-{\feh} plane as a function of Galactic radius (color) for stars within $|Z|<0.5$ kpc of the plane. \textbf{Top Panel:} The contour containing 75\% of points for both the low-$\alpha$ and high-$\alpha$ samples, calculated separately. \textbf{Bottom Panel:} The median {\alpham} as a function of {\feh} for both sequences. The shaded regions denotes the $\pm1\sigma$ uncertainty in the median. In both panels, the black line is our defined boundary between the low-$\alpha$ and high-$\alpha$ described in Equation \ref{eq:alpha_split}.}
    \label{fig:alpha_contours}
\end{figure}

To aid in the direct comparison between the radial and height bins, Figure \ref{fig:alpha_contours} shows the contours (top panel) and median (bottom panel) in the {\alpham}-{\feh} plane for both the low-$\alpha$ and high-$\alpha$ samples as a function of Galactic radius. In this Figure, the data are restricted to the plane ($|Z|<0.5$ kpc), equivalent to the bottom row in Figures \ref{fig:apogee_mhplots} and \ref{fig:apogee_mhplots_age}. The contour containing 90\% of points in both the low-$\alpha$ and high-$\alpha$ samples is shown in the top panel. The low-$\alpha$ sequence is more metal-rich near the center of the Galaxy, and shifts continuously to lower metallicities and higher-$\alpha$ moving outwards in radius. The high-$\alpha$ sequence's contour is generally the same shape and position at all radii, although close to the center of the Galaxy, the shape extends further to the metal-poor end. In the bottom panel, the median {\alpham} as a function of metallicity is shown for both samples. As before, the low-$\alpha$ sample shifts more metal-poor with increasing radius. The high-$\alpha$ sequence moves slightly downwards (towards more $\alpha$-poor) with increasing radius. This is also seen in \cite{Katz_2021} with APOGEE data, using the mode of the data, although they find a larger shift of $\sim 0.05$ dex between the inner and outer Galaxy, while ours is closer to half that at $\sim 0.025$ dex.

\subsection{Azimuthal Variance in Metallicity} \label{sec:results:azimuthal}

The degree to which trends in the Galactic disk are azimuthally symmetric has the potential to provide interesting insight into the history of the disk. The stellar distribution across the Galaxy is not uniform, with in-situ non-axisymmetric features such as the Galactic bar and spiral arms containing higher stellar density than surrounding populations, particularly for young stars \citep[e.g.,][]{BlandHawthorn2016,Reid2019,Khoperskov_2020}. Additionally, chemical enrichment is strongly dependent on local conditions, and spiral density fluctuations can lead to measurable differences in a galaxy's enrichment history across azimuth \citep{Spitoni_2019b}.

\begin{figure*}
    \centering
    \includegraphics[width=\textwidth]{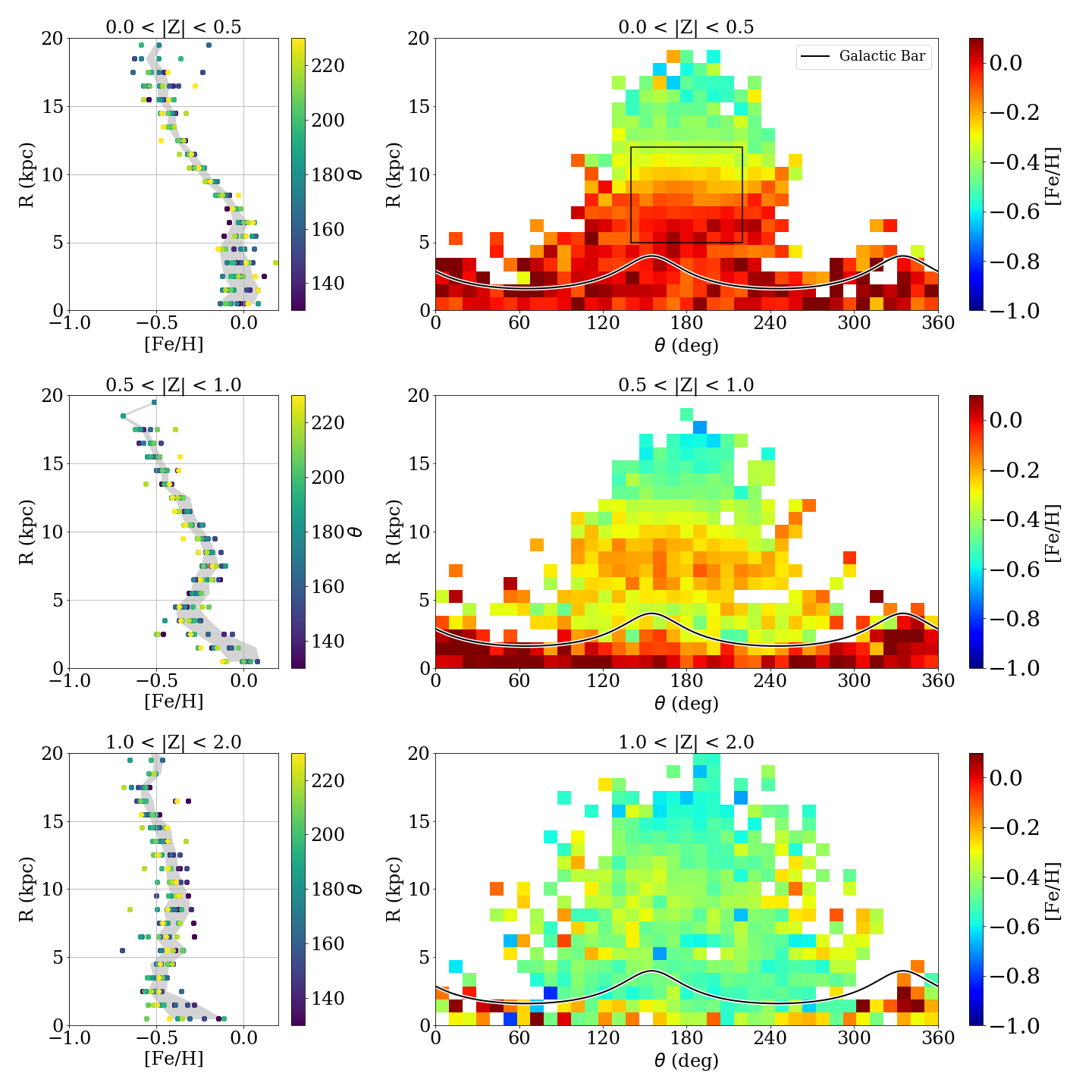}
   \caption{Maps of the median metallicity in the disk in polar coordinates, to highlight any non-axisymmetric features in the disk. The spatial bins are sized $\Delta R = 1$ kpc and $\Delta \theta = 10^{\circ}$. The rows are different slices in $Z$, moving from closest to the Galactic plane (top row, $|Z| \leq 0.5$ kpc), to farther above (bottom row, $1 \leq |Z| \leq 2.0$ kpc). The left column shows the median {\feh} for each bin as a function of $\theta$ (point color), compared to the $\pm1\sigma$ expected spread from uncertainty in the measurement of the median (gray shaded region). The right column shows the median metallicity maps as a function of radius $R$ and azimuthal angle $\theta$. The black line is the approximate location of the Galactic Bar, defined as an ellipse with major axis length 10 kpc, a 0.4 axis ratio, and rotated 25 degrees. The black square in the top panel highlights the region shown in more detail in Figure \ref{fig:azimuth_zoom}.}   \label{fig:azimuthal_invariance}
\end{figure*}

To look more closely at the azimuthal symmetry of the disk, the right column of Figure \ref{fig:azimuthal_invariance} is a metallicity map of the disk identical to Figure \ref{fig:metal_map} but displayed in polar coordinates. The spatial bins are sized $\Delta R = 1$ kpc and $\Delta \theta = 10^{\circ}$. As before, the stars are separated into rows based on their height above the Galactic plane. The corresponding panels to the left trace the median metallicity at each radius (y-axis) for different bins in azimuthal angle $\theta$ (point color), restricted to $130 \leq \theta \leq 230$ deg, where there is reasonable coverage with radius. At each radius, the expected spread based on uncertainty of the median measurements is shown as the gray shaded region. The expected spread ($\sigma_{\rm \feh}$) is defined for a given radius as the sum of the uncertainty in the median for each individual $\theta$ bin ($\sigma_{\rm \feh}(R,\theta)$), divided by the number of bins with valid data ($N_{bins,\theta}$), whereas the uncertainty in the median for a given azimuth bin is the standard deviation in {\feh} divided by the square root of the number of stars in that bin:

\begin{equation}
    \label{eq:expected_sigma}
    \sigma_{\rm \feh}(R)^2 = \frac{1}{N_{bins,\theta}}\sum_{\theta = 0}^{360} \sigma_{\rm \feh}(R,\theta)^2
\end{equation}

\begin{equation}
    \label{eq:expected_sigma_theta}
    \sigma_{\rm \feh}(R,\theta)^2 =  \frac{1}{N_{stars}} \sum_{i=0}^{N_{stars}} |({\rm \feh}_i-\overline{\rm \feh})|^2
\end{equation}

Close to the Galactic plane ($|Z| \leq 0.5$ kpc, top row), the Galaxy has little metallicity variation across azimuth near the Solar neighborhood and outward ($R \geq 8$ kpc), although the region of azimuth covered by the observations decreases with increasing radius. Closer to the center of the Galaxy ($R \leq 5$) kpc, the median metallicity varies more with azimuthal angle $\theta$, with the spread possibly slightly exceeding the expected uncertainty. In the middle row higher above the Galactic plane ($0.5 \leq |Z| \leq 1.0$ kpc), a similar trend is seen, where there is more spread with metallicity in azimuth near the center of the Galaxy. There does seem to be an asymmetry in the disk that follows the approximate location of the Galactic bar in these coordinates, with metal-rich stars preferentially residing in the Galactic bar. Higher above the Galactic plane ($1.0 \leq |Z| \leq 2.0$ kpc, bottom row), the variation in azimuth can be attributed entirely to noise from low-number statistics, where the observed spread is all comparable or smaller than the expected spread.

Interactions with satellite galaxies and merger events can also perturb the disk in non-axisymmetric ways, such as warping the disk or introducing kinematic oscillations \citep[e.g.,][]{Gomez_2012,Cheng_2020,Chrobakova_2022}. The restorative force from these perturbations is weaker in the outer Galaxy, so the presence of these features is generally expected to be more obvious at large radii. We find no significant azimuthal asymmetries in metallicity in the outer disk, but as radius increases, our sample covers less range in $\theta$. As such, we are unable to draw any conclusions on the Milky Way's merger history from these metallicity maps alone.

Recent work has detected azimuthal variations in both the gas phase metallicity \citep{Wenger_2019} and stellar metallicity \citep{Inno2019,Poggio2022,Hawkins_2022} possibly corresponding to the Milky Way's spiral arms. Measuring the radial metallicity gradient using Gaia data, \cite{Poggio2022} and \cite{Hawkins_2022} report azimuthal variation in the slope on the order of $0.02-0.1$ dex kpc$^{-1}$. If we measure the slope of the profiles in the left column of Figure \ref{fig:azimuthal_invariance} as a function of azimuthal angle $\theta$, our maximum difference between slopes for the vertical bin closest to the disk is 0.026, which is comparable to the lower end of the variation reported by \cite{Poggio2022}, but may not be significant given the uncertainties in our data.

\begin{figure}
    \centering
\includegraphics[width=0.48\textwidth]{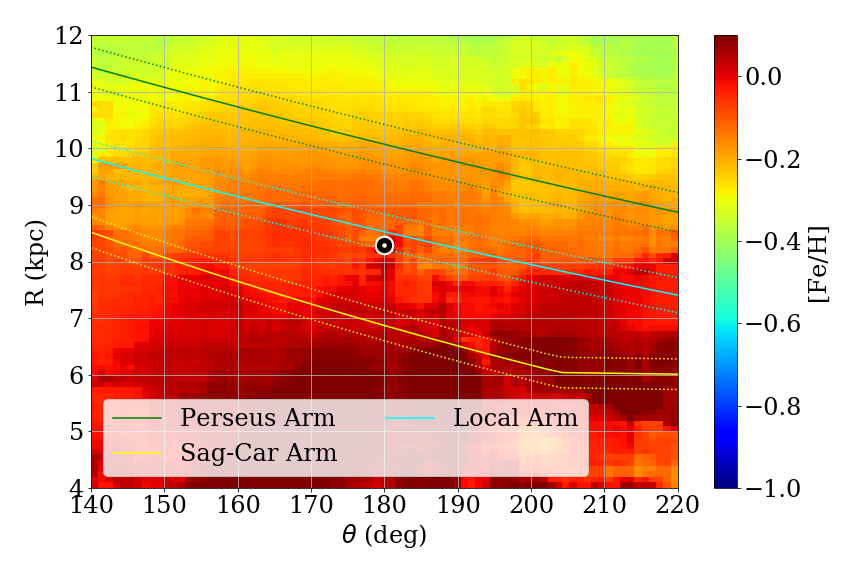}
   \caption{Azimuthal variation in metallicity restricted to a window near the solar neighborhood, outlined by the black rectangle in Figure \ref{fig:azimuthal_invariance}. The metallicity value is calculated as a running median for bin size $\Delta R = 0.5$ kpc and $\Delta \theta = 5^{\circ}$, evaluated every  $\Delta R = 0.1$ kpc and $\Delta \theta = 1^{\circ}$ for smoothing. The approximate location the nearby spiral arms from \cite{Reid2019} are plotted as colored lines.}
   \label{fig:azimuth_zoom}
\end{figure}

In Figure \ref{fig:azimuth_zoom}, we zoom into a region in the solar neighborhood, where the sample has higher number of stars making it possible to study the chemical distribution in more detail. The exact window used is outlined as the black rectangle in the top-right panel of Figure \ref{fig:azimuthal_invariance} for reference. Here, the spatial bins are sized $\Delta R = 0.5$ kpc and $\Delta \theta = 5^{\circ}$ (half the size of the bins in Figure \ref{fig:azimuthal_invariance}), but calculated on a frequency of $\Delta R = 0.1$ kpc and $\Delta \theta = 1^{\circ}$ as a running median for smoothing. The approximate location of the nearby spiral arms from \cite{Reid2019} are plotted as colored lines. There does seem to be a bit of coherent structure signifying that the median metallicity is not symmetric in azimuth, but it does not obviously follow the spiral arms. 



\subsection{Radial and Vertical Metallicity Gradients}
\label{sec:results:metal_gradients}

\begin{figure}
    \centering
    \includegraphics[width=0.45\textwidth]{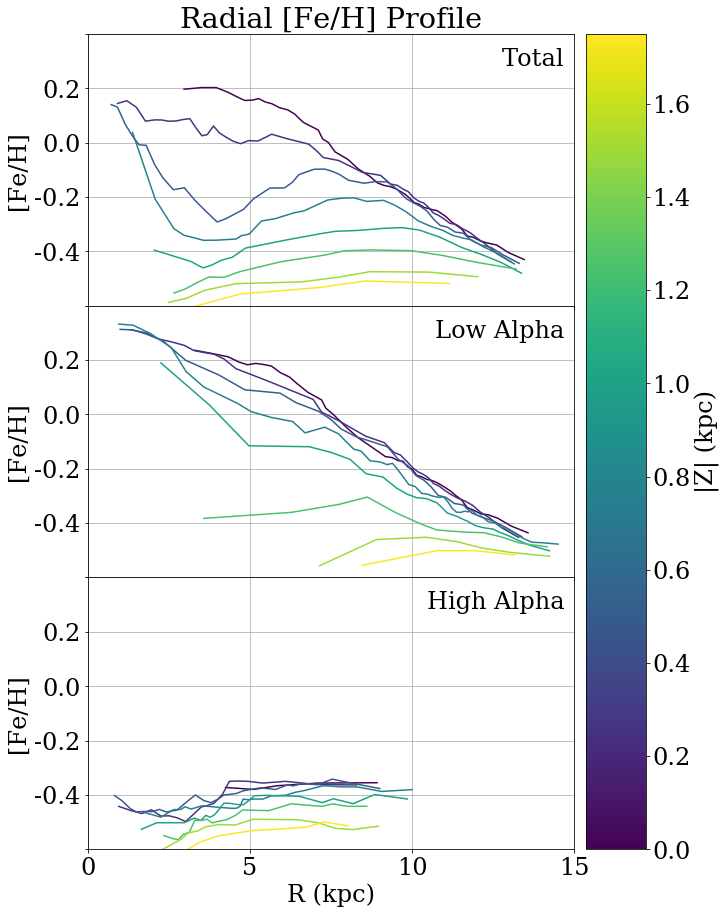}
    \caption{Radial median metallicity profile as a function of height out of the plane (line color), for the total stellar population (top panel), the low-$\alpha$ disk (middle panel), and the high-$\alpha$ disk (bottom panel).}
    \label{fig:radial_metal_gradients}
\end{figure}

\begin{figure}
    \centering
    \includegraphics[width=0.45\textwidth]{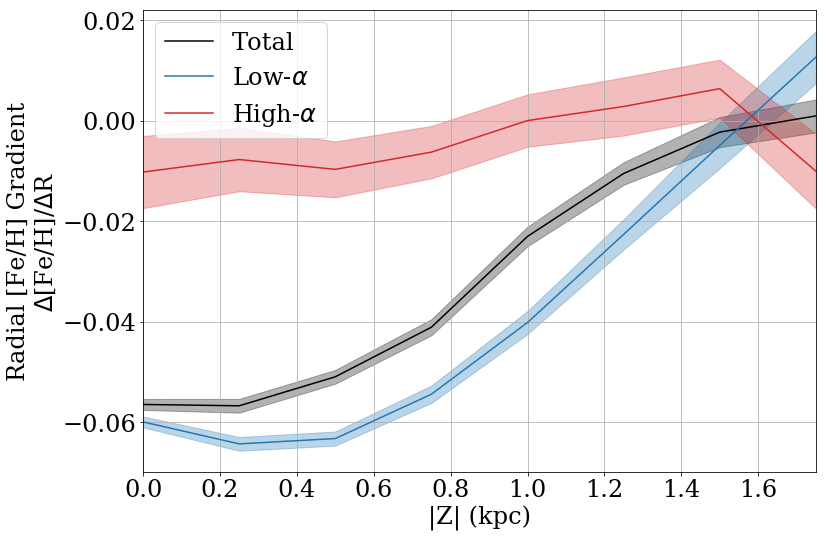}
    \caption{The best-fit slope for each radial metallicity profile in Figure \ref{fig:radial_metal_gradients} in units of dex kpc$^{-1}$, fit with a single line for stars beyond $R>7$ kpc for the total sample (black line), and the low-$\alpha$ (blue line) and high-$\alpha$ (red line) samples independently. The shaded region indicates the $\pm1\sigma$ uncertainty in the slope measurement.}
    \label{fig:radial_metal_gradients_slopes}
\end{figure}

The radial and vertical metallicity gradients in the Milky Way disk have been well-documented observationally, with stars near the center of the Galaxy exhibiting higher metallicity than those at large radii and higher $Z$ \citep[e.g.,][]{Hartkopf1982,Cheng2012,Carrell_2012,Anders2014,Schlesinger2014,Hayden2014, Frankel2019, Katz_2021,gaiamaps2022}. Such a trend is predicted by "inside-out" disk formation models, where stars in the central regions of the Galaxy form earlier on in the Galaxy's history, and the disk subsequently grows outward over time, the global star formation rate consistently decreasing with radius \citep[e.g.,][]{Eggen1962,Larson1976,Matteucci1989,Kobayashi2011,Minchev2015}. However, radial migration could complicate this interpretation because it flattens gradients over time as stars move away from their birth location \citep[e.g.,][]{Sellwood2002,Roskar2008,Wang_2013,Hayden2015, Mackereth2017,Frankel_2018,Frankel_2020, Vickers_2021, Lian2022_migration}. Without radial migration, gradients are predicted to steepen with lookback time, with the oldest stars having the steepest slope and being more centrally concentrated as a result of the inside-out growth of the Galaxy \citep[e.g.,][]{Matteucci1989, Bird_2013,Pilkington_2012, Gibson_2013, Molla_2018}. Radially dependent outflow efficiencies can also have strong impact on the radial gradients \citep{Johnson_2021}, as can radial gas flows within the disk \citep[e.g.,][]{Bilitewski_2012}.
 

We measure the median metallicity profile for the disk by first separating stars into bins of $Z$, and then calculating a running median {\feh} for each bin of $N=200$ data points with an overlap of 50\%, sorted by Galactocentric radii. We do this for the total stellar population and repeat the analysis for the high-$\alpha$ and low-$\alpha$ samples described in Section \ref{sec:data:selection} separately. The resulting radial median metallicity profiles are shown in Figure \ref{fig:radial_metal_gradients} and quantified in Table {\ref{tab:radial_feh_gradients}}. 

The total sample (Fig. \ref{fig:radial_metal_gradients} top row, Table {\ref{tab:radial_feh_gradients}}) shows a negative metallicity gradient close to the Galactic plane, which flattens out at small radii. Moving above the plane, the slope of the gradient flattens until it becomes slightly positive at $|Z| > 1.6$ kpc.

The low-$\alpha$ disk (Fig. \ref{fig:radial_metal_gradients} middle row) shows a steep metallicity gradient everywhere, notably missing the flattening in the inner Galaxy seen in the total population. The low-$\alpha$ disk's metallicity gradient flattens with height $Z$ much like the total population.

The high-$\alpha$ disk (Fig. \ref{fig:radial_metal_gradients} bottom row) exhibits a much flatter or slightly positive metallicity profile whose slope does not change significantly with $Z$. 
The high-$\alpha$ sequence effectively ends at $R \gtrsim 10$ kpc (shown previously in Figure \ref{fig:apogee_mhplots}, meaning there are not enough high-$\alpha$ stars in the outer Galaxy to constrain the metallicity profile past $R \gtrsim 10$ kpc.

The total disk looks like the high-$\alpha$ profile near the center of the Galaxy, and matches the low-$\alpha$ profile in the outer Galaxy. This is due to the relative weights between these two populations at different locations: as shown in Section \ref{sec:results:AlphaDF}, the inner region of the Galaxy is dominated by the high-$\alpha$ sequence, and the outer region is mostly low-$\alpha$ stars.

For each median metallicity profile, we quantify the gradient by fitting a straight line to stars with Galactocentric radius $R \geq 7$ kpc, where the profile reasonably approximates a single line. The best-fit slope for each profile is shown in Figure \ref{fig:radial_metal_gradients_slopes} against height above the plane $Z$, and tabulated in Table \ref{tab:radial_feh_gradients}. The total population and the low-$\alpha$ population have steep negative profiles in the outer Galaxy, which approach zero as it moves above the plane. The high-$\alpha$ slope is close to zero everywhere. Note that if we change the definition of our measured gradient and instead fit without the radial limit of ($R \geq 7$ kpc), the high-$\alpha$ population show a slight positive gradient, consistent with other studies \citep[e.g.,][]{Vickers_2021}. 

These results are generally consistent with previous results from a variety of methodologies. We measure a slope of \textcolor{black}{$-0.056 \pm 0.001$ dex kpc$^{-1}$} for the total population close to the Galactic plane ($|Z|\leq 0.25$ kpc). Using previous data releases of APOGEE, \cite{Feuillet2019} measured the slope of the low-$\alpha$ metallicity gradient to be $-0.059 \pm 0.010$ dex kpc$^{-1}$. Using open clusters as tracers, \cite{Donor2020} measured a radial gradient of $-0.068 \pm 0.001$ dex kpc$^{-1}$ with APOGEE DR16 data, and \cite{Myers_2022} measured $-0.073 \pm 0.002$ dex kpc$^{-1}$ with DR17. Using Gaia DR3 data, \cite{gaiamaps2022} measured a slope of $-0.056 \pm 0.007$ dex kpc$^{-1}$ for their bin closest to the Galactic plane. Additional studies use Cepheid stars as tracers and find similar results, with \cite{Genovali2014} reporting $-0.060 \pm 0.002$  dex kpc$^{-1}$, and \cite{Lemasle2018} reporting $-0.045 \pm 0.007$ dex kpc$^{-1}$. The large sample size, distance range, and high precision of the APOGEE DR17 sample enable us to map the radial, vertical, and $\alpha$-dependence of the metallicity gradient in unprecedented detail. Differences between tracer populations used may explain the slight differences among previous results --- e.g., Cepheids tend to be young stars that are more concentrated to the mid-plane, which alone results in a steeper vertical gradient.

These findings are also generally consistent with \cite{Hayden2014}, who measured the metallicity gradients as a function of position ($R$ and $Z$) in the disk and documented a relatively flat gradient near the center of the Galaxy, a steeper gradient farther out in the disk, and generally flat gradients for the high-$\alpha$ population \citep[][Table 2]{Hayden2014}. Our data set is significantly larger than that of \cite{Hayden2014}, with the inclusion of Southern Hemisphere observations, which leads to better spatial coverage in both the inner and outer regions of the Galaxy and less sensitivity to potential systematics. This may be why our radial gradient \textcolor{black}{$-0.056 \pm 0.001$ dex kpc$^{-1}$} is slightly shallower than the $-0.073\pm 0.003$ dex kpc$^{-1}$ from \cite{Hayden2014} for a comparable spatial zone. 

Numerical simulations from \cite{Rahimi_2014} reproduce similar gradient trends where the radial metallicity gradients flatten with increasing $Z$. Notably, they attribute the slight positive gradient of the Milky Way's thick disk to the flaring of younger populations at large radii. The vertical flaring of the mono-age, mono-abundance disk has been documented extensively in other studies as well \citep[e.g.,][]{Minchev_2015,Bovy2016,Mackereth2017,Lian2022_maps,gaiamaps2022,Robin_2022}).


\begin{table}[]
    \centering    
    \begin{tabular}{|c|c|c|c|}
    \hline 
    $|Z|$ (kpc) & Total & Low-$\alpha$ & High-$\alpha$ \\
    \hline \hline
    0.0 & -0.056 $\pm$ 0.001 & -0.06 $\pm$ 0.001 & -0.01 $\pm$ 0.007 \\
    0.25 & -0.057 $\pm$ 0.001 & -0.064 $\pm$ 0.001 & -0.008 $\pm$ 0.006 \\
    0.5 & -0.051 $\pm$ 0.001 & -0.063 $\pm$ 0.001 & -0.01 $\pm$ 0.006 \\
    0.75 & -0.041 $\pm$ 0.002 & -0.054 $\pm$ 0.002 & -0.006 $\pm$ 0.005 \\
    1.0 & -0.023 $\pm$ 0.002 & -0.04 $\pm$ 0.002 & -0.0 $\pm$ 0.005 \\
    1.25 & -0.01 $\pm$ 0.002 & -0.023 $\pm$ 0.003 & 0.003 $\pm$ 0.006 \\
    1.5 & -0.002 $\pm$ 0.003 & -0.005 $\pm$ 0.005 & 0.006 $\pm$ 0.006 \\
    1.75 & 0.001 $\pm$ 0.003 & 0.013 $\pm$ 0.005 & -0.01 $\pm$ 0.007 \\
\hline
    \end{tabular}
    \caption{Radial metallicity gradients in dex kpc$^{-1}$ as a function of height above the plane $|Z|$  from Figure \ref{fig:radial_metal_gradients_slopes}.
}
    \label{tab:radial_feh_gradients}
\end{table}

The models of \citet{Johnson_2021}, which incorporate radial and vertical redistribution of stars based on a hydrodynamical cosmological simulation of a Milky Way-like galaxy, also show a radial metallicity gradient that flattens with increasing $|Z|$.

\begin{figure}
    \centering
    \includegraphics[width=0.45\textwidth]{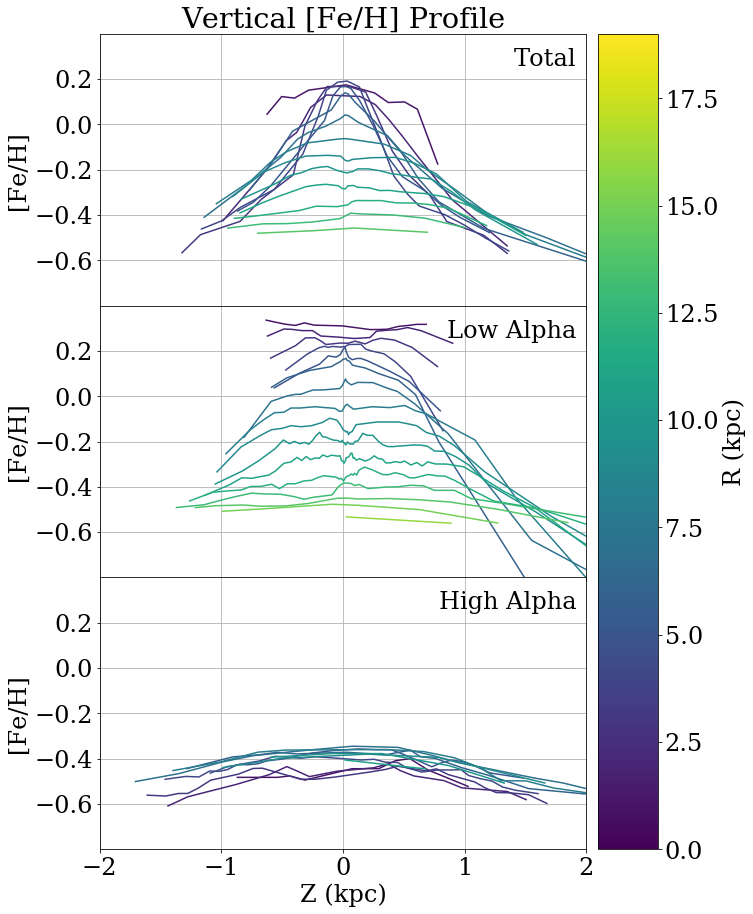}
    \caption{Vertical median metallicity profile as a function of Galactocentric radius (line color) for the total stellar population (top panel), the low-$\alpha$ disk (middle panel), and the high-$\alpha$ disk (bottom panel).}
    \label{fig:vertical_metal_gradients}
\end{figure}

\begin{figure}
    \centering
    \includegraphics[width=0.45\textwidth]{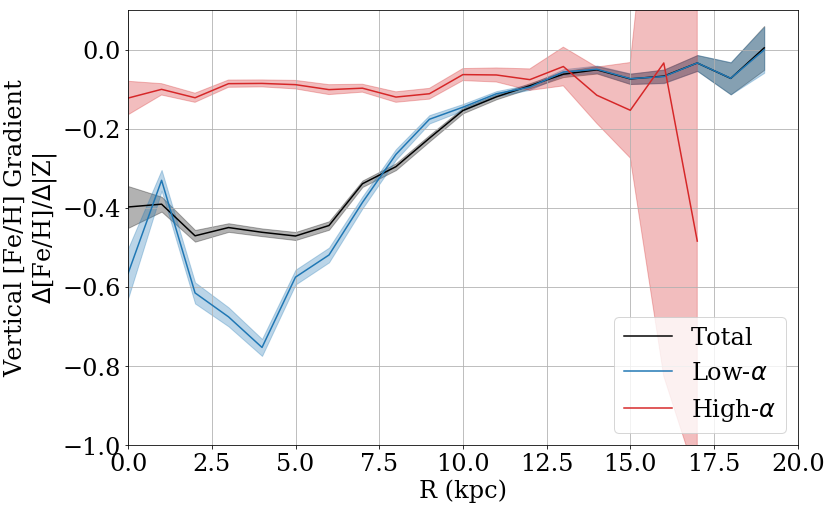}
    \caption{The best-fit slope for each vertical metallicity profile in Figure \ref{fig:vertical_metal_gradients}, fit with a single line for stars beyond $R>7$ kpc for the total sample (black line), and the low-$\alpha$ (blue line) and high-$\alpha$ (red line) samples independently. The shaded region indicates the $\pm1\sigma$ uncertainty in the slope measurement.}
    \label{fig:vertical_metal_gradients_slopes}
\end{figure}

The vertical median metallicity profile of the disk is shown in Figure \ref{fig:vertical_metal_gradients} and Table {\ref{tab:vertical_feh_gradients}}, calculated in the same way as the radial gradients. The best slopes for all stars $|Z| < 2$ kpc are shown in Figure \ref{fig:vertical_metal_gradients_slopes} and Table \ref{tab:vertical_feh_gradients}. Consistent with \cite{gaiamaps2022}, we find the gradients in the disk to be vertically symmetric. The slope measured above the disk ($Z>0$) does not differ significantly from the slope measured below the disk ($Z<0$). The total population has a steep negative vertical gradient close to the center of the Galaxy ($R < 5$ kpc), which flattens moving out in radius. This is generally true for the low-$\alpha$ population as well, although the innermost parts of the Galaxy ($R < 2$ kpc), show a very flat profile, possibly due to the bulge or bar. The high-$\alpha$ population has a shallow negative gradient everywhere, which does not significantly change with radius. Beyond $R\gtrsim 10$ kpc, the vertical gradient for the total, high-$\alpha$ and low-$\alpha$ populations are all close to zero. 

This is also consistent with previous studies \citep[e.g.,][Table 1]{Hayden2014}, where the vertical metallicity gradient approaches 0 as radius increases. Near the solar neighborhood, we report a vertical median metallicity gradient of $-0.315 \pm 0.009$ dex kpc$^{-1}$. \cite{Hayden2014} report a gradient of $-0.31 \pm 0.01$ dex kpc$^{-1}$ in a comparable spatial zone. For the thick disk, \cite{Carrell_2012} measured a vertical metallicity gradient of $-0.113 \pm 0.010$, using stars close to the solar neighborhood with heights $1 \leq |Z| \leq 3$, consistent with our measurement of $-0.112 \pm 0.014$ near the solar neighborhood for the high-$\alpha$ population.



\begin{table}[]
    \centering    
    \begin{tabular}{|c|c|c|c|}
    \hline 
    $R$ (kpc) & Total & Low-$\alpha$ & High-$\alpha$ \\
    \hline \hline 
    0.0 & -0.4 $\pm$ 0.053 & -0.565 $\pm$ 0.063 & -0.126 $\pm$ 0.042 \\
    2.0 & -0.471 $\pm$ 0.015 & -0.616 $\pm$ 0.027 & -0.121 $\pm$ 0.012 \\
    4.0 & -0.462 $\pm$ 0.01 & -0.753 $\pm$ 0.021 & -0.084 $\pm$ 0.009 \\
    6.0 & -0.444 $\pm$ 0.011 & -0.519 $\pm$ 0.019 & -0.1 $\pm$ 0.013 \\
    8.0 & -0.296 $\pm$ 0.009 & -0.265 $\pm$ 0.013 & -0.119 $\pm$ 0.013 \\
    10.0 & -0.153 $\pm$ 0.007 & -0.145 $\pm$ 0.008 & -0.062 $\pm$ 0.015 \\
    12.0 & -0.09 $\pm$ 0.007 & -0.094 $\pm$ 0.007 & -0.075 $\pm$ 0.027 \\
    14.0 & -0.05 $\pm$ 0.009 & -0.049 $\pm$ 0.009 & -0.115 $\pm$ 0.071 \\
    16.0 & -0.066 $\pm$ 0.017 & -0.067 $\pm$ 0.017 & -0.033 $\pm$ 0.792 \\
    18.0 & -0.072 $\pm$ 0.04 & -0.072 $\pm$ 0.041 & \\
    \hline
    \end{tabular}
    \caption{Vertical metallicity gradients in dex kpc$^{-1}$ as a function of Galactocentric Radius $R$  from Figure \ref{fig:vertical_metal_gradients_slopes}.
}
    \label{tab:vertical_feh_gradients}
\end{table}

\begin{figure}
    \centering
    \includegraphics[width=0.45\textwidth]{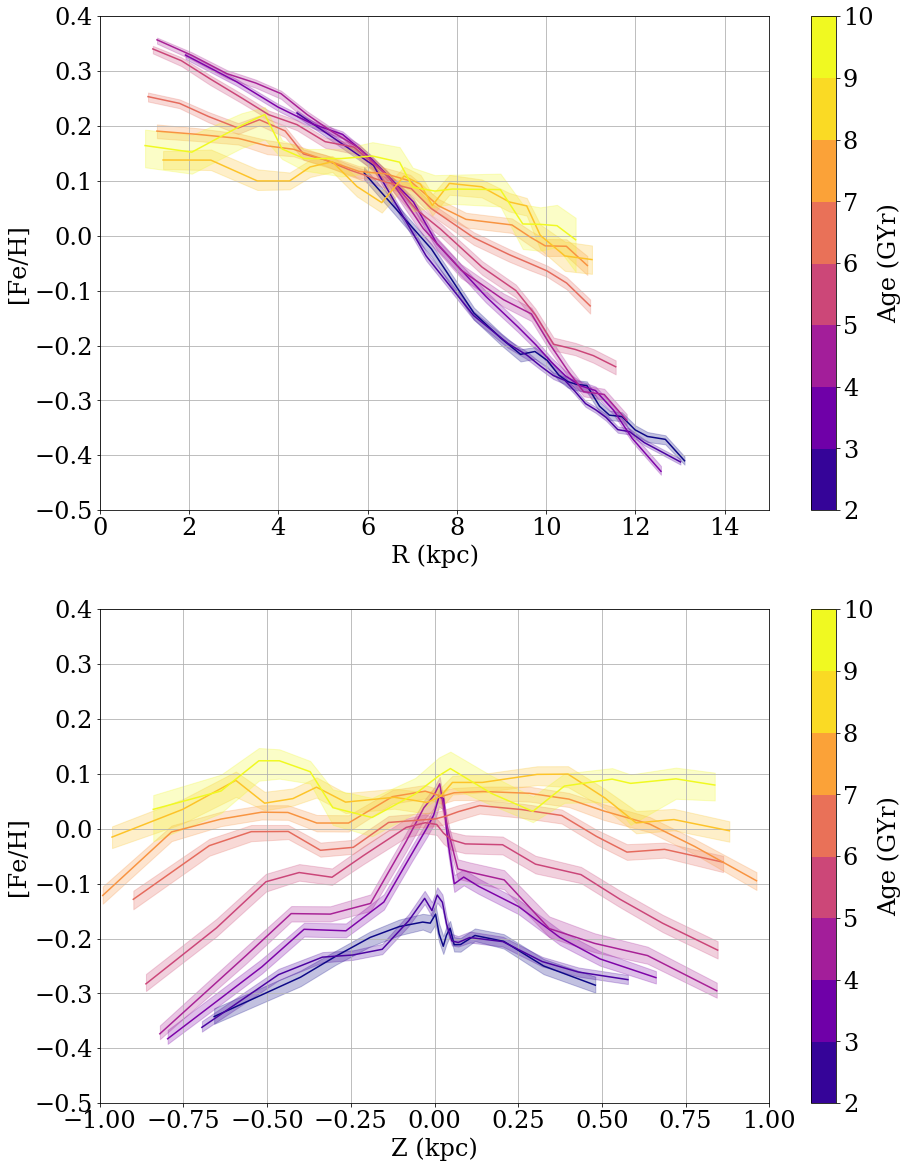}
    \caption{Top panel: the radial $R$ median metallicity profile of the low-$\alpha$ disk, split into different samples of stellar age (line color). Bottom panel: the vertical ($Z$) median metallicity profile of the low-$\alpha$ disk, split into different samples of stellar age (line color).}
    \label{fig:age_metal_gradients}
\end{figure}

Examining the Galaxy's metallicity profile as a function of age provides a direct link to the evolution history of the disk. Figure \ref{fig:age_metal_gradients} shows the radial (top panel) and vertical (bottom panel) metallicity profile for the low-$\alpha$ disk only, separated into bins of stellar age. In both cases, the slope of the profile flattens with increasing age. This trend is also seen in \citet{Anders_2023}, using a similar APOGEE sample with different age and distance estimates. The flattening of the metallicity gradients with age is the opposite of what is predicted by a pure inside-out growth of the Galaxy \citep[e.g.,][]{Matteucci1989,Bird_2013}, where the gradient in the interstellar medium is expected to flatten out over time \citep[e.g.,][]{Pilkington_2012, Gibson_2013, Molla_2018}. The opposite trend, seen here, is commonly attributed to be a signature of radial migration \citep[e.g.,][]{Wang_2013, Magrini_2016, Minchev_2018, Vickers_2021, Anders_2023}. However, an alternate explanation is also presented in \citet{Chiappini_2001}, where a disk formed from pre-enriched gas starts with an initially flat metallicity gradient that steepens over time.

\begin{figure}
    \centering
    \includegraphics[width=0.5\textwidth]{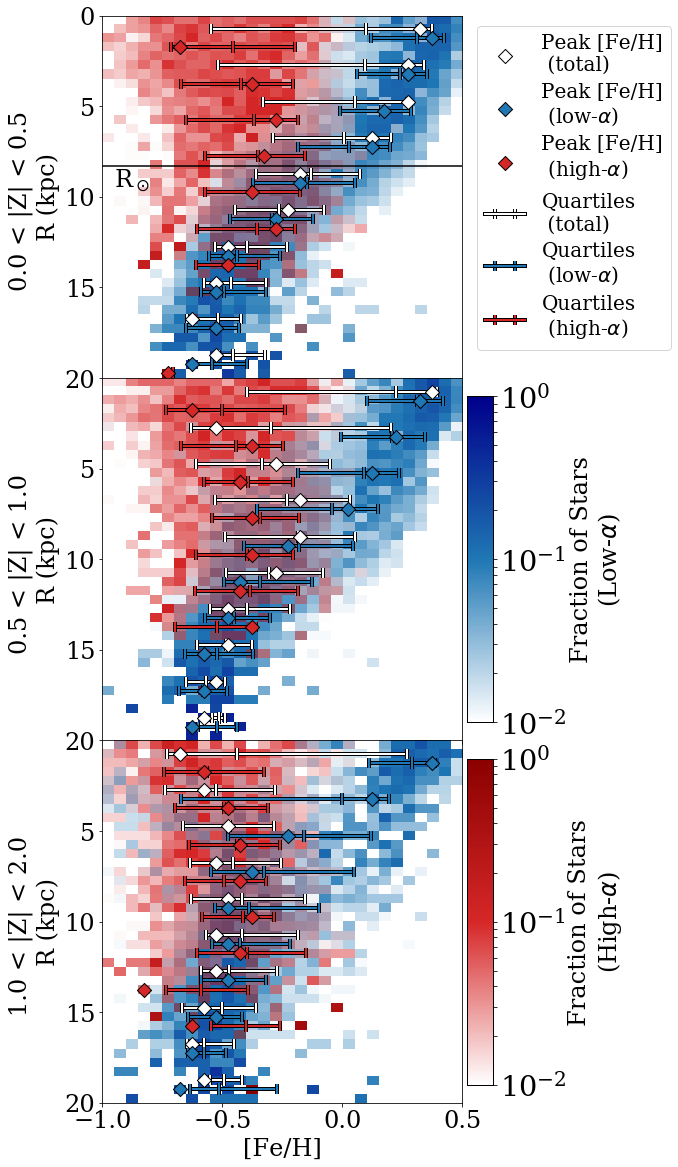}
    \caption{The metallicity distribution function of the Milky Way disk, split into different height bins (panels), from closest to the Galactic plane (top) to farthest beyond (bottom). Each panel shows the fraction of stars at each metallicity {\feh} as a function of Galactocentric radius, further split by color between high-$\alpha$ (red) and low-$\alpha$ (blue) samples. Every third row is annotated with markings for the peak (or mode) of the distribution (white diamond), as well as the 25th, 50th (median), and 75th percentiles (white tick marks) to highlight the shape of the distribution.}
    \label{fig:apogee_MDF_alpha}
\end{figure}




\subsection{Metallicity Distribution Function} \label{sec:results:MDF}

\begin{figure*}
    \centering
    \includegraphics[width=\textwidth]{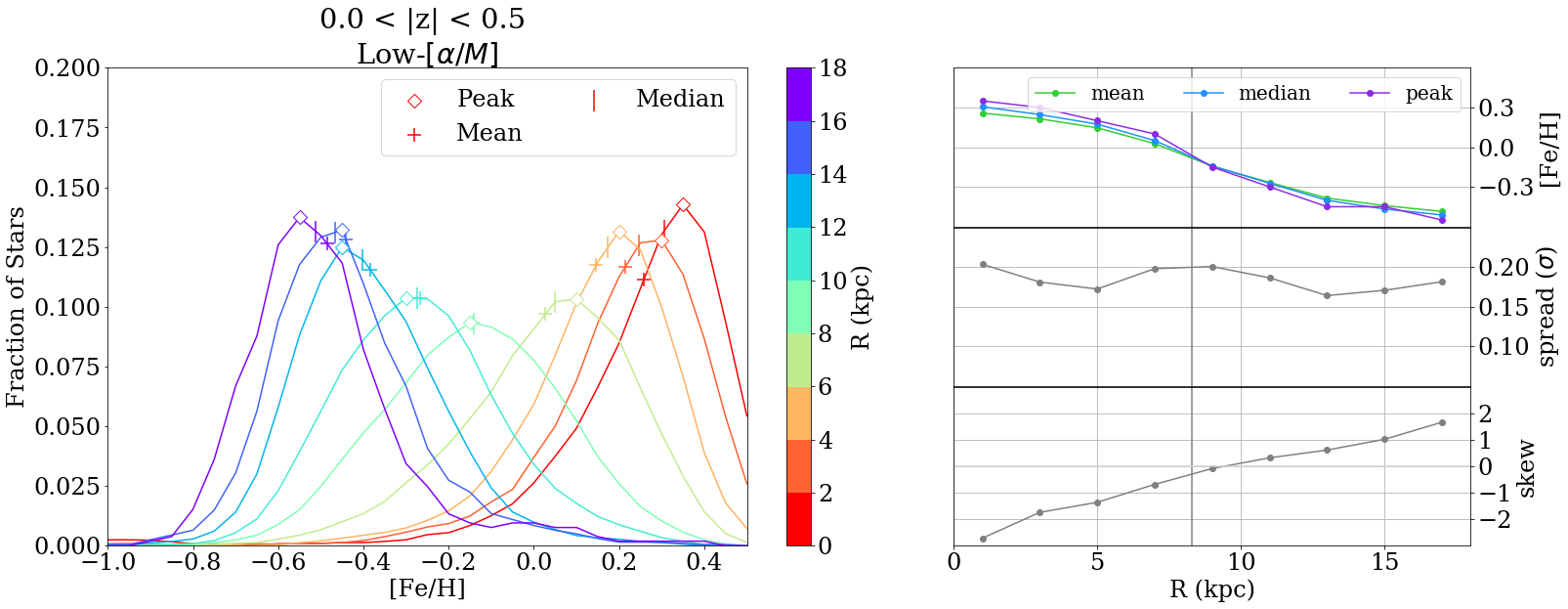}
    \caption{The metallicity distribution function (MDF) and its first three moments, limited to the low-$\alpha$ sample and close to the Galactic plane ($|Z|<0.5$ kpc; equivalent to the left panel in Figure \ref{fig:apogee_MDF_alpha}). \textbf{Left Panel:} The MDF at different radii in the Galaxy (colored lines). The right panels show the first three statistical moments for quantifying this distribution as a function of radius. \textbf{Top Right:} A characteristic {\feh} for each distribution function, measured as a median (blue line), a mean (green line) and the peak (or mode; purple line). \textbf{Middle Right:} The width, $\sigma$, of the distribution. \textbf{Bottom Right:} The skewness of the distribution, where a negative number indicates a left-leaning distribution as shown in the left panel and a positive number corresponds to a right-leaning distribution. The solar position ($R = 8.3$ kpc) is marked by a vertical gray line in all right panels.}
    \label{fig:mdf_stats}
\end{figure*}

While the radial and vertical median metallicity gradients in the disk reveal interesting general trends, more insights can be gleaned from the full metallicity distribution function (MDF) at different locations in the disk.
Specifically, the spread and shape of the underlying MDF can be crucial for characterizing the complex history of the disk more accurately.

Figure \ref{fig:apogee_MDF_alpha} demonstrates how the metallicity distribution function varies with radius for samples at different vertical slices, from closest to the Galactic plane (top panel) to furthest beyond (bottom panel). Every third row is annotated with tick marks denoting the 25th, 50th (median), and 75th percentile of that row's distribution, for the total sample (white), low-$\alpha$ (blue), and high-$\alpha$ (red) samples separately. The diamond point is the peak (or mode) of the distribution. At all heights, the characteristic metallicity (whether median or mode) decreases with radius for the total and low-$\alpha$ populations, but stays roughly constant for the high-$\alpha$ population. 

At smaller radii, there is little overlap between the low-$\alpha$ and high-$\alpha$ MDF. In fact, the high-$\alpha$ population alone is what creates the metal-poor tail of the total MDF. Just outside the solar neighborhood ($8<R<12$ kpc), the MDF of the high-$\alpha$ and low-$\alpha$ populations overlap chemically. Moving above the plane, this overlap starts closer to the center of the Galaxy, around $R = 5$ kpc at $|Z| > 1.0$ kpc.

The shape of the MDF changes as a function of radius; close to the center of the Galaxy, the total and low-$\alpha$ distribution is heavily skewed towards lower metallicities (the peak trends right of the median), and in the outer Galaxy the distribution is skewed towards higher metallicities (the peak is left of the median). This is consistent with the trend seen in previous studies \citep[e.g.,][]{Anders2014,Kordopatis_2015,Hayden2015, Loebman_2016, Katz_2021}.



The high-$\alpha$ MDF is consistently broader than the low-$\alpha$, has a shallower characteristic gradient, and shows less of a skewness trend with radius; the peaks (diamond points) are closer to the median (center tick mark) in general. Stars in the low-$\alpha$ sample transition from being negatively skewed in the inner Galaxy to positively skewed in the outer Galaxy (as shown in Figure \ref{fig:mdf_stats}). However, this trend is not seen as strongly in the high-$\alpha$ disk, even at similar Galactic heights as the thin disk. Because the high-$\alpha$ population is generally older, this could imply that the high-$\alpha$ sample is more "mixed" vertically, meaning the birth location of stars tend to be farther away from their present-day location, largely because they have had more time to migrate radially.

\citet{Kordopatis_2015} also observed an overabundance of metal-rich stars in the solar neighborhood using data from the RAVE survey, and link it to radial migration. \citet{Hayden2015} observed this same trend in APOGEE and present a simple model to show that radial migration could explain the change of the MDF shape with radius, because more stars migrate outward from the inner disk than vice versa. \citet{Loebman_2016} and \citet{Johnson_2021} show that this explanation succeeds quantitatively in models with realistic radial migration from cosmological simulations. These studies also show that the excess metal-rich tail in the MDF disappears at high $|Z|$, in agreement with the results of Figure \ref{fig:apogee_MDF_alpha} and \citet{Hayden2015}.


Some simple statistics can be measured to more fully characterize the MDF as a function of radius, as shown in Figure \ref{fig:mdf_stats} for the case of the low-$\alpha$ disk close to the Galactic plane (equivalent to the top panel of Figure \ref{fig:apogee_MDF_alpha}). The top right plot shows three definitions for a characteristic value for {\feh}; the mean, median, and peak of each distribution as a function of radius. While these values are similar, they are not identical, meaning the measured metallicity gradient of the disk will depend on the parameterization of {\feh} chosen. In the inner disk, the peak {\feh} is up to 0.1 dex higher in metallicity than the mean and median, due to the distributions being skewed metal-rich. In the outer disk, the peak is preferentially more metal-poor. Therefore, a metallicity gradient measured using the peak metallicity as a tracer will have a steeper slope than a gradient measured with the mean metallicity for the same group of stars. 

The middle right panel in Figure \ref{fig:mdf_stats} quantifies the spread of each distribution with radius, as total standard deviation $\sigma$. The outer regions of the disk ($R>10$ kpc) are characterized by narrower distributions with less overall spread, whereas the inner disk MDFs span a larger range of metallicities. 

The bottom right panel quantifies the skewness of each distribution with radius. The MDF in the inner regions of the disk it is negatively skewed, and in the outer regions it is positively skewed, quantifying the trend seen earlier Figure \ref{fig:apogee_MDF_alpha}. 

\subsection{Age Gradients}
\label{sec:results:age_gradients}

The present-day distribution of stellar ages can act as an interesting snapshot as to what the Milky Way might have looked like at different points in time, while also documenting how stars might move and migrate away from their radius of birth.

\begin{figure}
    \centering
    \includegraphics[width=0.45\textwidth]{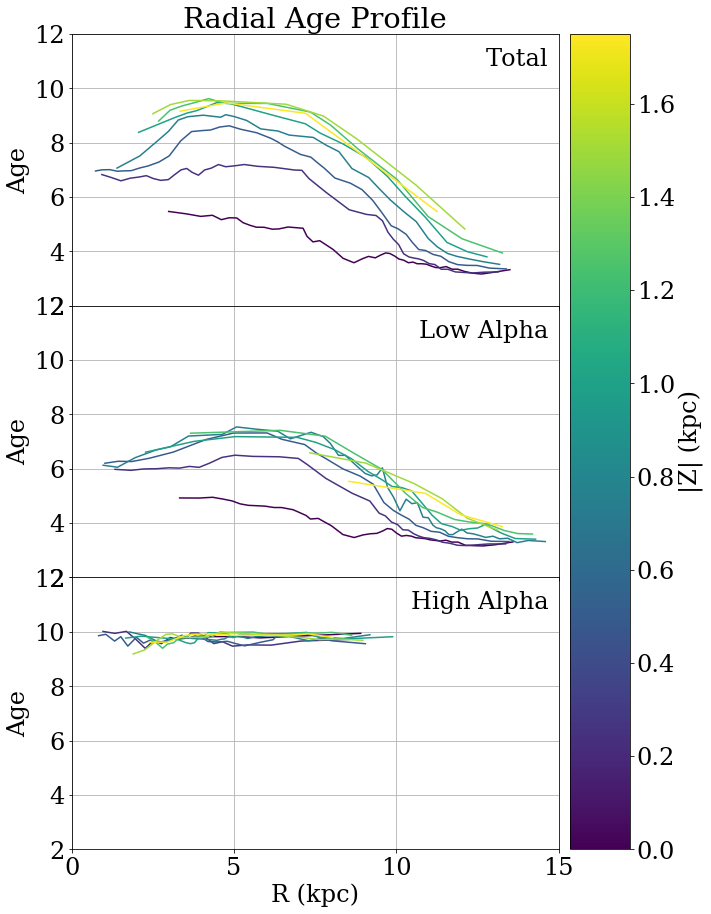}
    \caption{Radial median age profile as a function of height above the plane (line color), for the total stellar population (top panel), the low-$\alpha$ disk (middle panel), and the high-$\alpha$ disk (bottom panel).}
    \label{fig:radial_age_gradients}
\end{figure}

\begin{figure}
    \centering
    \includegraphics[width=0.45\textwidth]{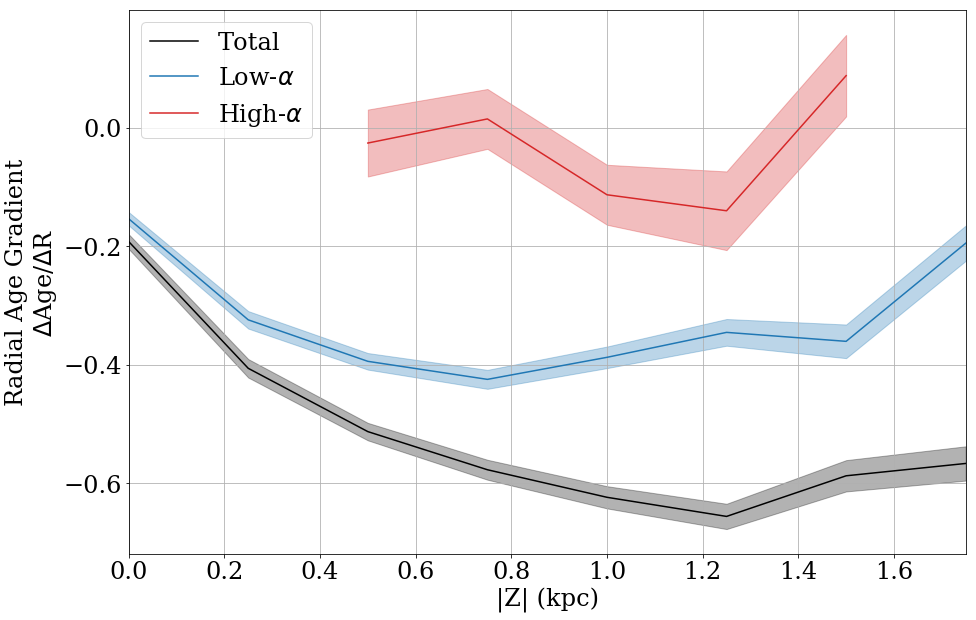}
    \caption{The best-fit slope for each radial age profile in Figure \ref{fig:radial_age_gradients} in units of Gyr kpc$^{-1}$, fit with a single line for stars beyond $R > 7$ kpc for the total sample (black line), and the low-$\alpha$ (blue line) and high-$\alpha$ (red line) samples independently. The shaded region indicates the $\pm 1 \sigma$ uncertainty in the slope measurement.}
    \label{fig:radial_age_gradients_slope}
\end{figure}

The radial median age profile of the Milky Way is presented in Figure \ref{fig:radial_age_gradients}, identical to the way the metallicity gradients in Section \ref{sec:results:metal_gradients} were calculated. The best-fit slope for each profile is shown in Figure \ref{fig:radial_age_gradients_slope}, once again calculated only using stars with $R\geq 7$ kpc where the profile reasonably approximates a straight line.

The median age of the high-$\alpha$ population is generally the same everywhere, with a slope close to 0 at any height above the plane. We note that the median age observed here, $\sim 8.5$ Gyr, is likely too young due to selection effects and the limitations of our age sample, including the metallicity cut of [Fe/H] $\geq -0.7$ dex discussed in Section {\ref{sec:data:ages}}.

The total and low-$\alpha$ stellar populations have a negative radial median age gradient in the outer Galaxy, while in the inner Galaxy the profile flattens out. The measured slope is flattest close to the disk ($z=0$), and becomes negative moving above the plane. The total and low-$\alpha$ stellar populations have a vertical gradient that varies with radius as well. The vertical gradient is steeper closer to the center of the Galaxy, and flatter at large radii, similar to the vertical metallicity gradient. Unlike the high-$\alpha$ sample, the majority of stars in the low-$\alpha$ disk have metallicity [Fe/H] $\geq -0.7$ and are therefore more robust against potential biases caused by the sample selection (see Figure {\ref{fig:distmass_numbers}}).





\subsection{Age Distribution Function} \label{sec:results:AgeDF}

\begin{figure}
    \centering
    \includegraphics[width=0.5\textwidth]{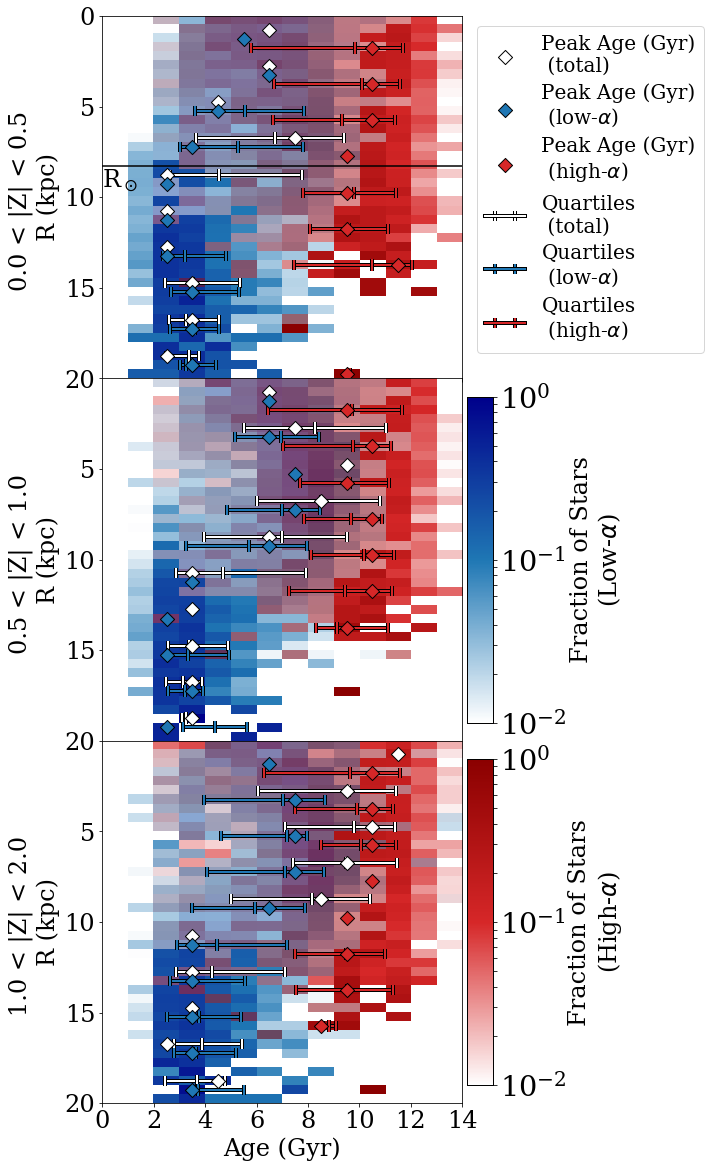}
    \caption{The age distribution function of the Milky Way disk, split into different height bins (panels), from closest to the Galactic plane (top) to farthest beyond (bottom). Each panel shows the fraction of stars at each age as a function of Galactocentric radius, further split by color between high-$\alpha$ (red) and low-$\alpha$ (blue) samples. Every third row is annotated with markings for the peak (or mode) of the distribution (white diamond), as well as the 25th, 50th (median), and 75th percentiles (white tick marks) to highlight the shape of the distribution.}
    \label{fig:apogee_ADF}
\end{figure}




As before with metallicities, more information lies in the shape of the age distribution function at different locations in the Galaxy rather than the gradient alone. Figure \ref{fig:apogee_ADF} depicts the age distribution function (ADF) as a function of Galactocentric radius for different heights in the disk. Close to the Galactic plane (left panel), the peak age of the total population gradually declines, peaking around \textcolor{black}{7.5 Gyr} near the Galactic center and \textcolor{black}{3 Gyr} near the solar radius and outward. The spread of the ADF is fairly broad, spanning up to 5 Gyr at all radii. In the outer Galaxy, the ADF is preferentially skewed toward older ages. \cite{Katz_2021} found the ADF skewed towards younger ages in the inner Galaxy, and skewed towards older ages in the outer Galaxy, but they limited their investigation to just the low-$\alpha$, thin disk sample.

Farther above the Galactic plane (middle and bottom panels of Fig. \ref{fig:apogee_ADF}), the profile does not show a single gradient, but rather has a slight positive age gradient until $R \sim 5$ kpc which transitions into a negative gradient at larger $R$. After $R \sim 12$ kpc, the gradient flattens out. We caution that this flattening of the gradient at large $R$ may be artificially induced by the lack of [Fe/H]$ < -0.7$ stars in our sample when dealing with ages imposed by the sample cuts described in Section \ref{sec:data:ages}. 

Separating further into the low-$\alpha$ and high-$\alpha$ samples reveals slightly different trends. Close to the Galactic plane (top panel), and in the inner Galaxy, there is some minimal overlap between the low-$\alpha$ and high-$\alpha$ samples, but near the solar neighborhood and outward, the ADF is more bimodal and there is more separation between the high-$\alpha$ and low-$\alpha$ ADF. At all radii, the low-$\alpha$ sample has a narrower distribution, and transitions from being skewed toward younger ages in the inner Galaxy to being skewed toward older ages in the outer Galaxy. This is similar behavior to the MDF in Figure \ref{fig:apogee_MDF_alpha}, and also seen by \cite{Katz_2021}. Farther above the plane, there is more overlap between the low-$\alpha$ and high-$\alpha$ samples in the inner Galaxy, but there is still little overlap at larger radii. 

Despite the similarities in these representations of the ADF to the corresponding ones for MDFs, we caution that the age uncertainties (typically $\sim0.15$ dex) are significant relative to the total spread, which is not the case for the [Fe/H] measurements.



\subsection{Age-Metallicity Relation} \label{sec:results:age_metallicity_relation}

\begin{figure*}
    \centering
    \includegraphics[width=\textwidth]{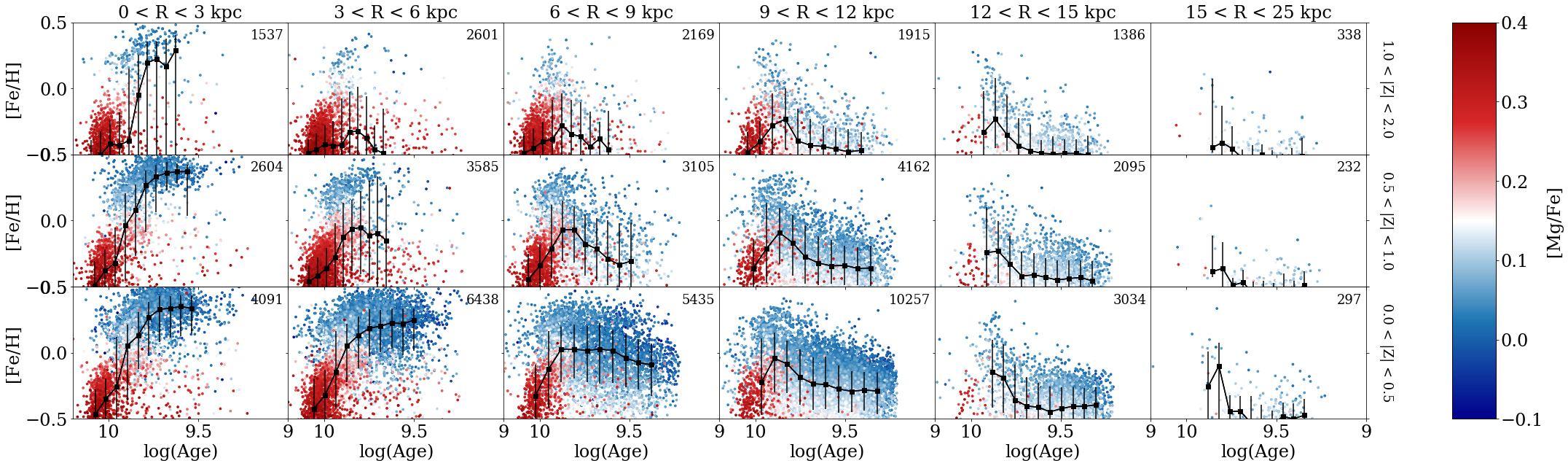}
    \caption{The age-metallicity relation across the Milky Way disk. Panels represent different spatial zones, laid out in the same way as Figure \ref{fig:apogee_mhplots}, with rows corresponding to $Z$ and columns increasing in $R$. The number in the top-right corner of each panel is the number of stars in our sample in that spatial bin. The age and metallicity for individual stars is plotted, colored by {\alpham} abundance. The running median trend is plotted in black square points to guide the eye, with the vertical bars indicating the standard deviation in {\feh} for bins in log(age). The typical (median) uncertainty for any given point is shown in the top right corner of each panel.}
    \label{fig:age_metallicity_relation}
\end{figure*}

The relation between stellar age and metallicity has long been sought after to help constrain chemical evolution models \citep[e.g.,][]{Twarog1980,Edvardsson1993}. In a simple "closed box" system, the metallicity of stars should increase over time, as each generation of stars enriches the interstellar gas from which subsequent generations are born. The actual scenario is much more complex, depending on gas inflow and outflow rates, supernovae yields, stellar migration, and the positionally variable star formation history of the Galaxy. Observations of the age-metallicity relation in the Milky Way include the effects of all of these processes, and they therefore provide a powerful constraint for chemical evolution models to reproduce across different locations in the Galaxy.

Previous studies have found significant scatter in the age-metallicity relation (AMR) near the solar neighborhood, where stars with a single age span a wide range of metallicities, which cannot be attributed to observational errors alone \citep[e.g.,][]{Casagrande2011, Bergemann2014, Aguirre_2018, Lin2018, Grieves_2018, Sahlholdt_2021, Xiang2022, Anders_2023}. The AMR also varies with Galactic location, making it difficult to constrain a single relation that fits the whole disk \citep[e.g.,][]{Hasselquist_2019, Feuillet2019, Casamiquela_2021,Lian2022_migration,Anders_2023}.

The AMR for our sample of stars in the Milky Way disk is shown in Figure \ref{fig:age_metallicity_relation}. The data are split into bins of radius (columns), and height (rows), to demonstrate how the age-metallicity relation varies across Galactic location. Points are colored by $\alpha$-element abundances, which are known to be correlated with age, although the exact trend depends on Galactic position \citep[e.g.,][see also Section \ref{sec:results:age_alpha_relation}]{Haywood_2013,Nissen2015,Bedell2018,Feuillet2018, Hasselquist_2019,Lian2022_migration}. The median trend across different bins in age is tracked as black square points to help guide the eye. Note that the x-axis (stellar age) has been reversed so that older stars are on the left, better expressing the forward flow of time. The x-axis is also presented in log space, which is more representative of the uncertainties in our age estimates (Section \ref{sec:data:ages}). One can also read each panel rotated 90$^{\circ}$ to see the distribution of log(age) at fixed {\feh}.

Consistent with previous studies, there is significant scatter around the age-metallicity relation near the solar neighborhood. There is a slight gradient in $\alpha$-element abundances within that spread; nearly everywhere in the Galaxy, higher-metallicity stars are relatively more $\alpha$-poor.

The age-metallicity relation has less scatter moving towards the inner Galaxy (left column), and above the plane (top row). If the spread in this relation is due to the radial migration of stars, this implies that the in-situ age-metallicity relation of the inner galaxy has been more preserved, and less contaminated by migrated stars; or in other words, more stars migrate outwards than inwards in the Galaxy. This is perhaps not surprising; dynamically, stars in the inner Galaxy migrate outwards, and stars in the outer Galaxy migrate inwards \citep[e.g.,][]{Sellwood2002}. The inner Galaxy is denser than the outer disk, so if one assumes the same rate of migration across all radii, more stars would migrate outwards simply because more stars start in the inner Galaxy. The age-metallicity relation shown in \citet{Sahlholdt_2021} also shows a lower dispersion for the inner regions of the disk (their "Pop C" sample) using GALAH data. 

The AMR is steepest in the inner Galaxy, and flattens moving out in radius. In all panels, the age-metallicity relation flattens out at young stellar ages, perhaps indicating that chemical equilibrium has been reached, where the inflowing gas dilutes the ISM at the same rate as it is being enriched \citep[e.g.,][]{Dalcanton2007, Finlator_2008, Weinberg2017}. In the inner disk, the equilibrium metallicity is higher than the equilibrium metallicity reached in the outer disk and moving away from the midplane. Equilibrium seems to have been reached sooner (at older stellar age) in the outer disk than the inner disk.

\begin{figure*}
    \centering
    \includegraphics[trim={8cm 0 6cm 0},width=\textwidth]{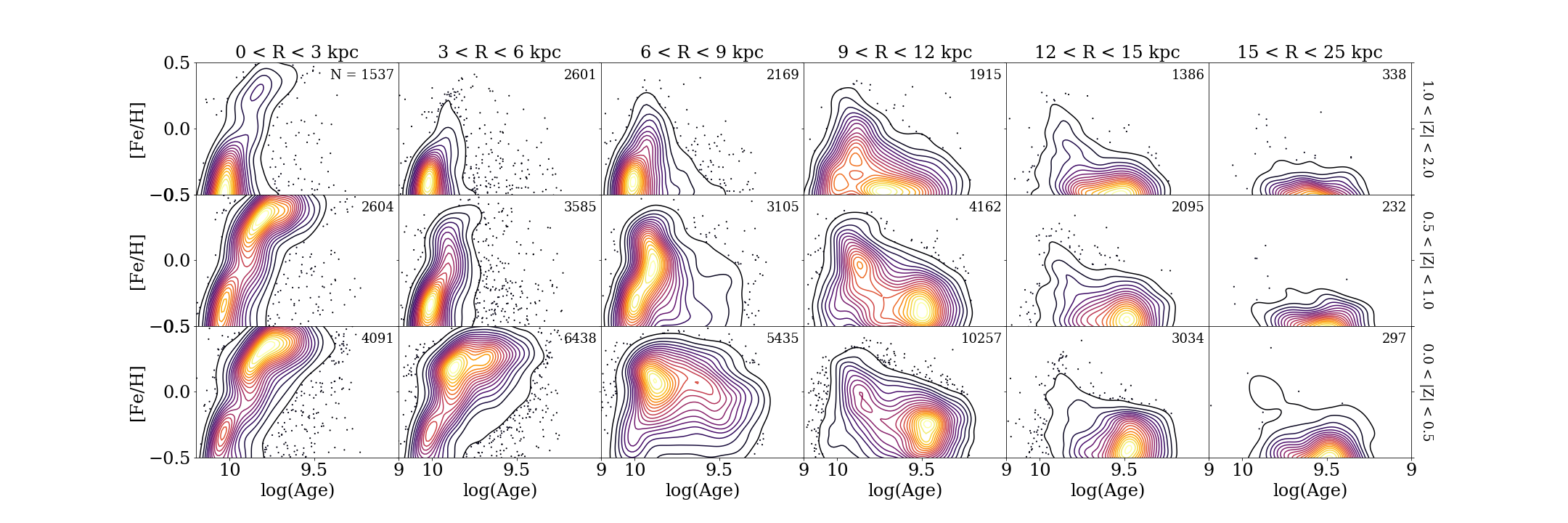}
    \caption{Same as Figure \ref{fig:age_metallicity_relation}, but showing the contours of point density in each panel.}
    \label{fig:age_metallicity_relation_contours}
\end{figure*}

There is a notable inversion of the AMR at large radii in the Galaxy, where older stars trend more metal-rich than the younger stars in the sample. This has been reported before in previous studies \citep[e.g.,][]{Anders2014,Feuillet2018,Hasselquist_2019,Lian2022_migration}. The cosmological simulations of \citet{Lu_2022} explore the possible origins of such an inversion, suggesting that it could be the signature of interactions with a satellite galaxy like the Sagittarius dwarf, and radial migration notably widens the apex.

A recent study from \cite{Xiang2022} documented a disjointed age-metallicity relation for the sum of the total disk, and suggested that the two-infall scenario or a major merger is responsible. In our results, we see some evidence of bimodality in Figure \ref{fig:age_metallicity_relation_contours}, which is a contour map of the point density distribution in Figure \ref{fig:age_metallicity_relation}. In the inner Galaxy ($R < 3$ kpc), the age-metallicity relation has two separate peaks.

\subsection{Age-Alpha Relation and Chemical Clocks} \label{sec:results:age_alpha_relation}

While no clear correlation between age and metallicity relation exists near the solar neighborhood, better correlation has historically been found between age and $\alpha$-element abundances \citep[e.g.,][]{daSilva2012,Haywood_2013,Bensby_2014}, leading some to use $\alpha$-elements as a chemical clock, substituting their abundances when stellar ages are not readily available. Even so, the {\alpham}-age correlation has been found to vary across the Milky Way's disk \citep[e.g.,][]{Aguirre_2018,Feuillet2018,Katz_2021,Vazquez_2022}, extending the metaphor to imply that chemical clocks run in chemical "time zones" throughout the Galaxy. 
Reproducing this variation has been considered a strong constraint on chemical evolution models \citep[e.g.,][]{Haywood_2013,Spitoni_2019,Johnson_2021}.

\begin{figure*}
    \centering
    \includegraphics[width=\textwidth]{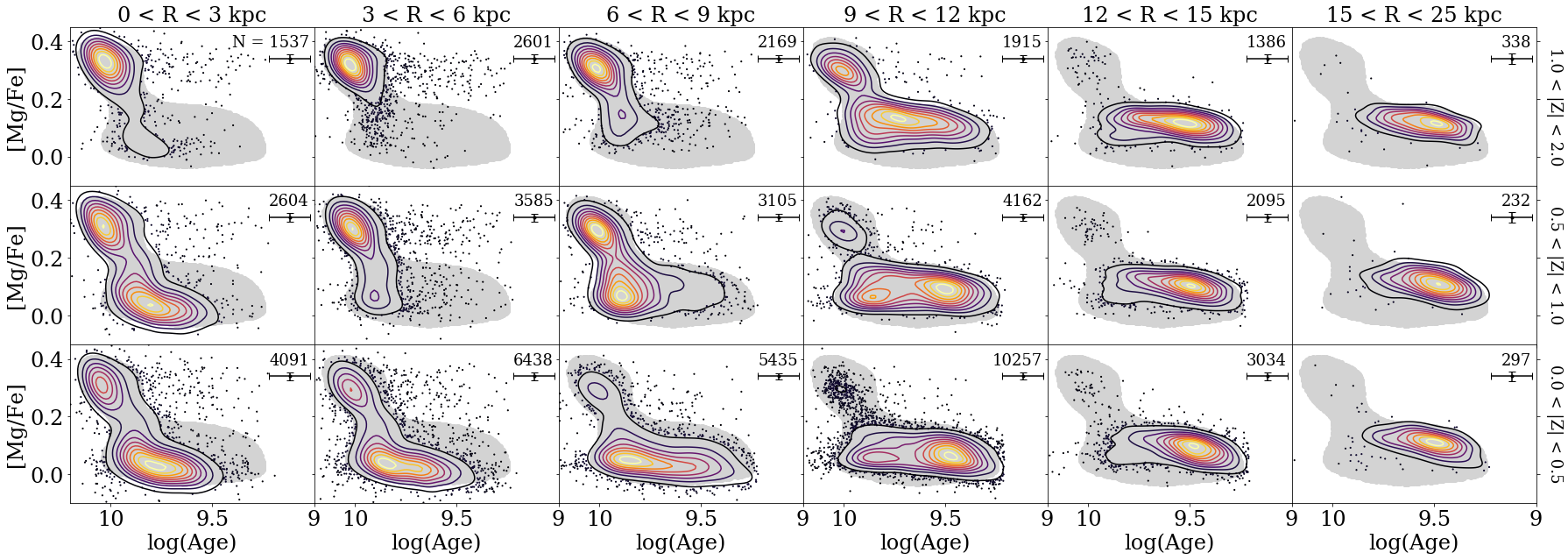}
    \caption{The age-alpha relation across the Milky Way disk. Panels represent different spatial zones, laid out in the same way as Figure \ref{fig:apogee_mhplots}. The contours represent the density of points on the diagram. The gray background shape outlines the 90\% contour for the entire sample, and is the same in all panels for reference. The typical (median) uncertainty for any given point is shown in the top right corner of each panel.}
    \label{fig:age_alpha_relation}
\end{figure*}

Figure {\ref{fig:age_alpha_relation}} shows the relation between age and {\alpham} as a contour plot at various locations throughout the disk. The distribution is double-peaked in nearly all panels, with the older, low-$\alpha$ population most prevalent above the disk and in the inner Galaxy. Near the solar neighborhood and outward, the relation of the low-$\alpha$ sequence is relatively flat, with a large spread in ages corresponding to a small range of {\alpham} abundances. This is consistent with the findings of \cite{Haywood_2013} and \cite{Feuillet2018}. In the inner Galaxy, a small range in {\alpham} abundances corresponds more tightly with a smaller range in age, a phenomenon that applies, though differently, to each the high-$\alpha$ and low-$\alpha$ groups. As in \cite{Haywood_2013}, we observe some age overlap between the two sequences, implying that the low-$\alpha$ sequence in the outer disk began forming stars while the high-$\alpha$ disk was concurrently still forming stars in the center of the Galaxy. 

For the low-$\alpha$ sequence, there is some evolution with Galactic position. The low-$\alpha$ stars are older and more $\alpha$-poor near the center of the Galaxy. In the outer Galaxy, the sequence is more $\alpha$-enhanced and generally younger, although covering a larger spread in ages. Above the plane ($Z > 1$ kpc), the low-$\alpha$ sequence is more $\alpha$-enhanced.

The high-$\alpha$ sequence stays generally in the same location on this diagram regardless of position in the Galaxy. This is similar to the trend seen in Figure \ref{fig:apogee_mhplots} and \ref{fig:alpha_contours}, where the locus of the low-$\alpha$ sequence changes significantly with Galactic position while the high-$\alpha$ sequence stays largely in the same location.

In the $R=9-12$ kpc and low-$|Z|$ zones, the log(age) distribution is bimodal even within the low-$\alpha$ sequence. This could be evidence for a three-phase star formation history. \citet{Sahlholdt_2021} report a similar distribution in the age-metallicity relation of their "Pop A" sample, which probes a similar location in the disk. The younger peak in log(age) appears similar to the recent starburst 2-3 Gyr ago detected independently in \citet{Isern_2019} and \citet{Mor_2019}, although we note that this is the first time to our knowledge that it has been detected in APOGEE data. This recent starburst is thought to be linked with the Sagittarius dwarf spheroidal galaxy's most recent perigalactic passage through the Milky Way's disk \citep{RuizLara_2020,Laporte_2019,Antoja_2020}. The uncertainties in our age estimates are not negligible, but are likely not responsible for this bimodality. Larger uncertainties would blur out the distribution and decrease the observed bimodality.

We remind the reader that using different estimates for stellar ages can change our results. Alternate versions of this figure using different age catalogs are shown in Appendix \ref{sec:app:ages} to emphasize this point. 

\subsection{Chemical Evolution via Chemical Tagging}

\begin{figure}
    \centering
    \includegraphics[width=0.5\textwidth]{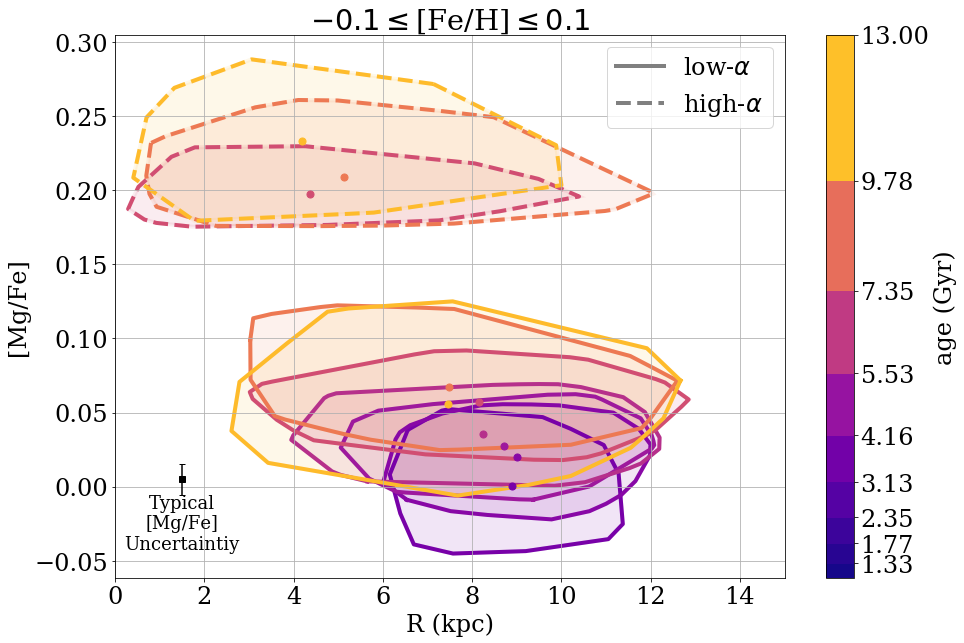}
    \caption{The radial distribution of the low-$\alpha$ (solid lines) and high-$\alpha$ (dashed lines) populations shown as the contour containing 75\% of all points for different bins in stellar age (line color) all at fixed $-0.1 \leq$ [Fe/H] $\leq 0.1$.  The center of each contour is marked as a point. The typical (median) uncertainty {\alpham} is plotted in the bottom left corner.
    }
    \label{fig:chemical_tagging}
\end{figure}

If a group of stars was born together in the same location and at the same time, they should have identical chemical abundances ({\feh} and {\alpham} in our case). This is the basic assumption behind "chemical tagging", used to identify stellar siblings that have been redistributed throughout the Galaxy despite being born together \citep[e.g.,][]{Freeman_2002,Hawkins_2015,Ting_2015,Price_Jones_2020,Buder_2021}. 
Under this assumption, we can potentially track radial migration and the spatial redistribution of a stellar population over time, as well as look at the enrichment history of an area of the Galaxy for fixed metallicity.

Figure \ref{fig:chemical_tagging} shows this evolution for stars close to solar metallicity ($-0.1 \leq$ [Fe/H] $\leq 0.1$ dex). The present-day radial distribution (x-axis) and {\alpham} abundance (y-axis) is shown for different bins in stellar age (line color) for both the low-$\alpha$ (solid lines) and high-$\alpha$ populations as the contour containing 75\% of all points in that bin. Stars at the same age and metallicity should all have same {\alpham} abundance. If stars did not move from where they were born, we would expect to see a tight clump in this space. If stars migrate significantly over time, the shape of the clump should spread out to a broader range of radii, but keep the same {\alpham} abundances (or a "flat slope" in {\alpham}). If the slopes were not flat, it may be indicative of different enrichment histories for different parts of the Galaxy; suggesting a violation of the assumption that stars with the same metallicity and age were born roughly in the same place.

The young, $\alpha$-poor stars in Figure \ref{fig:chemical_tagging} currently reside in a relatively confined clump in radius and {\alpham} as expected ($6 \lesssim R \lesssim 11$ kpc). As stellar age increases, the shape of the clump widens to cover a broader range in radius ($3 \lesssim R \lesssim 12$ kpc) while the {\alpham} abundance remains confined. The high-$\alpha$ sequence does not show this evolution in radial width with time, but notably does not include enough young stars to properly trace this. As shown in Figure \ref{fig:apogee_mhplots_age}, high-$\alpha$ stars tend to be old.

As stellar age decreases, the median {\alpham} value of each subpopulation decreases for both the low-$\alpha$ (solid lines) and high-$\alpha$ (dashed lines) populations. In the low-$\alpha$ population, the median value of {\alpham} decreases from {\alpham} = 0.05 in the oldest age bin ($10 \leq \log(\textrm{age}) \leq 10.1$) to {\alpham}$ = 0.00$ in the youngest age bin ($9 \leq \log(\textrm{age}) \leq 9.1$). This tracks the chemical evolution of a location in the Galaxy, as Type Ia supernovae begin "diluting" the interstellar medium with iron, thereby decreasing the overall {\alpham} ratio over time.

\section{Discussion} 
\label{sec:discussion}

Using large samples of stars to map the Milky Way in different parameter spaces using metallicity, $\alpha$-element abundances, and age, as demonstrated here, has the potential to place strong constraints on chemical evolution models and reveal the major processes which shaped our Galaxy. Directly comparing our results with specific chemical evolution models is beyond the scope of this paper, but in this discussion section we qualitatively compare our results with predictions from the leading classes of chemical evolution models discussed in the Introduction; the "two-infall", "superposition", and "clumpy formation" scenarios. 

The underlying assumption necessary to interpret these results is that in a well-mixed interstellar medium, stars formed at the same time and the same place in the Galaxy will have the same chemical abundances (both metallicity and $\alpha$-elements). Under this assumption, a spread in abundance at present day for stars at a given age at the same location, whether bimodal or not, can only be produced if stars have moved away from their birth location.

\subsection{Superposition and Radial Migration}\label{sec:discussion:superposition}

The "superposition" class of evolution models \citep[e.g.,][]{Schonrich_2009a,Schonrich_2009b,Minchev_2013,Minchev_2014,Minchev2017,Johnson_2021} explain the observed chemical bimodality in the solar vicinity as the superposition of evolutionary tracks for stars born at different Galactocentric radii, with the stars having reached their present-day location in the solar neighborhood through radial migration. Several predictions made by these superposition chemical evolution models are seen in our results.

The metallicity gradient flattening with age (Figure \ref{fig:age_metal_gradients}) is predicted by radial migration \citep[e.g.,][]{Sellwood2002,Roskar2008,Wang_2013, Hayden2015, Mackereth2017,Frankel_2018,Frankel_2020, Vickers_2021, Lian2022_migration}. If stars are formed \textit{in-situ} with a steep metallicity gradient, that gradient will flatten over time as metal-rich inner Galaxy stars migrate outwards and metal-poor outer Galaxy stars migrate inwards, skewing the metallicity distribution at either end of the Galaxy. The older stars in our sample show a flatter gradient than the younger stars; in agreement with this scenario.

The shape of the metallicity distribution function at different locations in the Galaxy (Figures \ref{fig:apogee_MDF_alpha} and \ref{fig:mdf_stats}), specifically the skewness or asymmetry of the MDF, can be a sign of radial migration if a population of stars has a metal-rich tail \citep[e.g.,][]{Hayden2015,Kordopatis_2015,Loebman_2016}). Our data show a skewed MDF in Figure \ref{fig:mdf_stats}, where the inner region of the disk is negatively skewed, and the outer region of the disk is positively skewed. This trend is commonly attributed to radial migration, whereby migrating stars become the metal-rich tails in the MDF at different locations \citep[e.g.,][]{Schonrich_2009a,Schonrich_2009b,Roskar2008,Kordopatis_2015,Loebman_2016,Johnson_2021}. The metal-poor tail in the inner Galaxy's MDF is likely attributed to a spread in ages between the stars, as is predicted by even closed-box chemical evolution tracks when any given location in the Galaxy becomes enriched over time \citep[e.g.,][]{Romano_2013,Vincenzo_2014,Weinberg2017,Toyouchi_2018}. The metal-rich tail of the outer Galaxy's MDF is more difficult to explain with a traditional chemical enrichment track, which leaves radial migration as the most likely culprit.


The inversion in skewness in the MDF occurs around \textcolor{black}{$R = 9.4$} kpc in our data. This could be linked to the Outer Lindblad Resonance (OLR) of the Milky Way disk; a resonance with the Galactic bar driving different dynamical effects throughout the disk, one of which is radial migration \citep{Halle_2015,Michtchenko_2016, Dias_2019,Khoperskov_2020b}. Using Gaia data, \cite{Khoperskov_2020} estimates the OLR to be located at a Galactocentric radius of around 9 kpc. \cite{Khoperskov_2020b} uses a high-resolution N-body simulation to investigate the relationship between the OLR and radial migration, and find that stars from the inner Galaxy migrating outwards become "trapped" in the OLR. When the rotation period of the bar slows down, those stars can escape and migrate farther out. The trapping effect of the OLR could also explain the build-up of metal rich stars at $R\sim 9$ kpc in Figures \ref{fig:global_maps} and \ref{fig:metal_map}.

We also see signs of radial migration in the age-metallicity relation shown in Figure \ref{fig:age_metallicity_relation}. Around the solar neighborhood, there is significant scatter about the age-metallicity trend. If the interstellar medium is always well-mixed, the large spread in metallicity for stars of a given age must mean that some of these stars were not born at their present-day location. This is another consequence of radial migration predicted by the superposition class of models, which explain the spread by emphasizing the difference between the present-day locations of stars and their birth radii \citep[e.g.,][]{Schonrich_2009a,Wang_2013,Minchev_2013,Lian2022_migration}. While any location in the Galaxy should start with a tight age-metallicity relation, migration will blur the present-day relation as metal-rich stars from the inner disk move outward and contaminate the more metal-poor outer disk. Radial migration is most efficient in the plane of the disk, therefore these models predict less spread in the age-metallicity relation at larger vertical heights, an effect also seen in our data.


The turnover in the age-metallicity relation, seen in Figure \ref{fig:age_metallicity_relation} and in \cite{Hasselquist_2019}, can also be explained through radial migration. The older, metal-rich stars were likely formed in the inner Galaxy, and migrated outwards to where they are found today, contaminating the age-metallicity relation. An alternate explanation could be the dilution of the ISM from pristine gas infall, lowering the metallicity of a previously-enriched area of the Galaxy \citep[e.g.,][]{Spitoni_2019,Lu_2022}. However, in this scenario, it is predicted that both the high-$\alpha$ and low-$\alpha$ tracks in metallicity would still decrease with stellar age, with the post-dilution low-$\alpha$ track beginning at a lower metallicity than the high-$\alpha$ track at the same time \citep[e.g.,][]{Spitoni_2019}. That predicted trend is not obvious in our data, and the large spread in metallicity at a given age favors a radial migration explanation. \citet{Lu_2022} further explore the origin of such a turnover using cosmological simulations, and find that even when the turnover in the age-metallicity relation can be directly linked to the infall of a satellite galaxy, radial migration can widen the shape of the peak.

Vertical motions should also be considered. In the age-$\alpha$ relation of Figure \ref{fig:age_alpha_relation}, a population of stars with the same age and in the same present-day location can have a large spread in {\alpham}, most dramatically seen in the inner Galaxy ($R<3$ kpc). This violates the underlying assumption mentioned earlier, meaning stars must have moved around to create that spread. However, in this case, \textit{radial} migration is an unsatisfactory explanation, as stars from the outer Galaxy migrating inward is expected to be a less frequent occurrence than the other way around, simply a consequence from the density profile of the Galaxy \citep[e.g.,][]{Sellwood2002}. Instead, the \textit{vertical} motion of stars could result in the observed spread if stars born above the mid plane are currently found near $Z=0$. Even stars formed above the plane will inevitably have vertical motions that cause their orbits to cross the plane, meaning this could be a natural consequence of a star's vertical orbit. Related vertical motions could be linked with dynamical heating \citep[e.g.,][]{Spitzer1951,Barbanis1967,Lacey1984,Mackereth2019} or the "upside-down" formation of the disk \citep[e.g.,][]{Toth1992,Quinn1993,Hanninen_2002,Brook_2004,Freudenburg_2017,Bird_2013,Bird_2021}. Using a quantity like guiding radius ($R_{\rm guide}$) and maximum height $|Z_{\rm max}|$ calculated from parameterized stellar orbits, instead of the present-day $R$ and $|Z|$ we use here, may remove contamination by thick disk stars currently "passing through" the thin disk from these figures. Some studies, including \citet[][]{Boeche_2013},
\citet[][]{Katz_2021}, and \citet[][]{Spitoni_2022b}, have looked at these quantities, and found similar overall trends.

The evolution of radial distribution with stellar age seen in Figure \ref{fig:chemical_tagging} is yet more evidence for radial migration. The youngest, low-$\alpha$ stars indicate that for a population of fixed age, {\feh}, and {\alpham}, they are expected to be born at a similar radius in the Galaxy. As stellar age increases, the population redistributes into a larger range of $R$, showing that stars migrate radially over time. The high-$\alpha$ stars, which cover a similar range in Galactic radius despite a wide window in age ($5 \leq$ age $ \leq 13$ Gyr), may suggest an "upper limit" on the efficiency of radial migration and the timescales over which stars can migrate on average; see e.g., \citet{Frankel_2020} and \citet{Lian2022_migration} for such an analysis. Similarly, the relative distribution of stars within the low-$\alpha$ contours may hint at the importance of direction in radial migration (i.e., what fraction of stars migrate outwards instead of inwards), although such a discussion is beyond the scope of this work.

This wide variety of results suggests that significant stellar migration occurs in the Milky Way disk, most influencing the trends seen close to the Galactic plane and at larger radii. However, there are open questions remaining about the nature of the inner disk, where an apparent bimodality exists that is not easily explained by migration models.


\subsection{Two-Infall or Major Merger}\label{sec:discussion:twoinfall}

The "two-infall" class of evolution models \citep[e.g.,][]{Chiappini1997, Chiappini_2001,Lian2020a,Lian2020b,Lian_2020c,Spitoni_2019,Spitoni_2020,Spitoni_2021,Spitoni_2022b} suggest that the two sequences in {\alpham}-{\feh} space formed sequentially in time. Under this scenario, the thick disk was formed first during the initial collapse of the Galaxy, and after some time delay a second infall of gas fed the creation of the thin disk. These models predict several of the observed trends in our results, most notably in areas where radial migration is not as efficient, including the inner Galaxy and at greater heights above the Galactic plane.

Figure {\ref{fig:apogee_mhplots}} shows that the $\alpha$-bimodality persists throughout the majority of the disk. This is significant because while the "superposition" class of chemical evolution models can produce the low-$\alpha$ "sequence" and broad distribution of {\alpham} in the solar neighborhood using only radial migration \citep[e.g.,][]{Schonrich_2009a,Schonrich_2009b,Minchev_2013,Minchev_2014,Nidever2014,Sharma_2021b,Johnson_2021}, radial migration is known to be most efficient close to the Galactic plane and in the outer disk. Therefore, the bimodality in other parts of the Galaxy is more difficult to explain with radial migration alone. Previous studies by \cite{Freudenburg_2017} and \cite{Zasowski_2019} report that the shape of the MDF and the {\alpham}-{\feh} trends seen in the inner disk ($3 < R < 5$ kpc) could be modeled well using a single evolutionary track in an "upside-down" disk formation model. 

Our results are consistent with this in the {\alpham}-{\feh} realm, but when expanded to ages, we show that two distinct tracks of {\alpham} are observed even in the inner Galaxy. This is also reported in \cite{Queiroz_2021}. 



Near the solar neighborhood, radial migration models predict bimodality by explaining the high-$\alpha$ sequence as contaminants from the inner Galaxy; due to the intrinsic density profile of the Galaxy, more stars are expected to migrate outwards than inwards, so the inner Galaxy should display less contamination from the low-$\alpha$ sequence. This is in contradiction with our findings, where the $\alpha$-bimodality persists throughout the majority of the disk. \cite{Johnson_2021} also report that the bimodality reproduced by their superposition model is weaker than the observed bimodality in the Milky Way; in particular, the model overproduces intermediate-$\alpha$ stars compared to observations. \cite{Chen_2022} find greater success in producing bimodality with radial migration. However, a two-infall (or multi-infall) model may be needed to explain the bimodality in the inner Galaxy.


The age-metallicity relation in Figure \ref{fig:age_metallicity_relation} produces some trends that are better explained by the two-infall model than by radial migration. \cite{Minchev_2013,Minchev_2014} report that while the scatter around the age-metallicity relation can be attributed to radial migration, the overall slope is only weakly affected. In our data, consistent with other recent studies \citep[][]{Feuillet2019,Hasselquist_2019}, the slope of the age-metallicity relation varies significantly with Galactic radius. This is reproduced with a two-infall model, where the low-$\alpha$, post-infall disk has a shallower slope in the age-metallicity relation compared to the high-$\alpha$ population, due to the continuous inflow of gas diluting the disk that was not present during the formation of the original high-$\alpha$ disk \citep[][]{Spitoni_2019,Spitoni_2020,Spitoni_2021}.

The apparent disk bimodality is not only observed in {\alpham}-{\feh} chemistry, but also in the age-metallicity relation. Recent work by \cite{Xiang2022} show a disjointed age-metallicity relation for the sum of the total disk, which is only possible to reproduce in a two-infall scenario or with a major merger event. In our study, the age-metallicity relation as a function of Galactic position (Fig. \ref{fig:age_metallicity_relation}) also shows a possible bimodality, most apparent in the $0 < R < 3 $ kpc range of the inner disk. 

The age-[Mg/Fe] relation in Figure \ref{fig:age_alpha_relation} is perhaps more convincing evidence for the two-infall model, with clear bimodality in the relation persisting across nearly the entire disk. In the outer disk ($9<R<15$), there appears to be an additional bimodality within the low-$\alpha$ sequence, suggesting a three-phase star history similar to that detected in \cite{Sahlholdt_2021} using a sample of stars from the GALAH survey. The uncertainties in our age estimates are not negligible, but are likely not responsible for this bimodality. Larger data uncertainties would blur out the distribution and decrease the observed bimodality. Due to the age uncertainties, the "true" age bimodality in the Milky Way may be stronger than what is shown in our analysis. The peak of our "third" starburst is around 2-3 Gyr, consistent with the recent burst \citet{Isern_2019} and \citet{Mor_2019} report, possibly linked to the most recent interaction with the Sagittarius dwarf spheroidal galaxy as it passed through the disk \citep[e.g.,][]{RuizLara_2020,Laporte_2019,Antoja_2020}

In the age-[Mg/Fe] relation, the age overlap between the high-$\alpha$ and low-$\alpha$ sequences are impossible to explain with a single evolutionary track. In this interpretation, we do caution that if the transition from high-$\alpha$ to low-$\alpha$ is fast, uncertainties in age determination could produce an artificial impression of age overlap. \cite{Haywood_2013} also detects an overlap in age, and claim that the dichotomy in the solar neighborhood can be reproduced by the two-infall scenario, and that little to no radial migration is needed. 

For the inner disk, the spread in {\alpham} for a given age is likely due to the vertical motions of stars as discussed previously in Section \ref{sec:discussion:superposition}. However, if the spread was purely from vertical blurring, a continuous spread would be expected, and not the bimodal distribution seen here.

As discussed in Appendix \ref{sec:app:ages}, the choice of stellar age catalog does significantly change the appearance of the age-[Mg/Fe] relation across the disk, suggesting that stellar age estimates are not yet robust enough to draw strong conclusions from this relation. Nevertheless, several of the key features observed here (the age overlap and the bimodality within the low-$\alpha$ sequence) do persist across different age catalogs. 

In summary, the bimodality in {\alpham} and stellar ages persisting across the inner disk is not easily explained through radial migration, which is most efficient at larger radii. A multi-phase star formation history, such as those presented in the two-infall model, better predicts the trends observed in the inner Galaxy.


\subsection{Clumpy Star Formation Models}

The "clumpy star formation" models \citep[e.g.,][]{Clarke_2019,Silva_2020,Amarante_2020} predict that the two sequences in $\alpha$-element abundances formed simultaneously but in different modes: the high-$\alpha$ sequence formed in rapidly-enriched gaseous clumps and the low-$\alpha$ formed in a less efficient smooth disk. This clumpy phase of early disk formation is predicted by numerical simulations \mbox{\citep[e.g.,][]{Bournaud_2007}} and often seen in observations of high redshift galaxies \mbox{\citep[e.g.,][]{Elmegreen2005}}.

One major result that the clump star formation models predict is the temporal overlap between the two $\alpha$ sequences: The low-$\alpha$ disk starts forming at the same time as the high-$\alpha$ sequence, meaning there should be some overlap in stellar ages between the two sequences. We see this overlap in Figure \ref{fig:age_metallicity_relation} and \ref{fig:age_alpha_relation}, where stars at around $\log$(age) $\sim 9.7$ Gyr span a significant range in both metallicity and $\alpha$-element abundances. This age overlap has also been observed in previous studies \citep[e.g.,][]{Haywood_2013,Hayden_2017,Aguirre_2018,Gent2022}. Unfortunately, high uncertainties in age estimates could artificially produce this overlap.

The age distribution in the {\alpham}-{\feh} plane shown in Figure \ref{fig:apogee_mhplots_age} also possibly points to this formation scenario. As discussed in Section \ref{sec:results:AlphaDF}, the distribution of stellar ages within the low-$\alpha$ sequence suggests that it is not a single evolutionary sequence. Stellar age is more closely correlated with {\alpham} instead of {\feh}. A strikingly similar trend is notably predicted under the clumpy formation scenario \citep[][Fig. 9]{Clarke_2019}.

One potential avenue for further investigating the difference between the two-infall class of models and the clumpy star formation models lies not within the Milky Way but in other galaxies. If the two-infall model is true, a chemical bimodality would only be present in galaxies that experienced significant gas infall both at early and late times, meaning it would be a rare phenomenon only affecting approximately 5\% of galaxies with comparable mass to the Milky Way \citep[e.g.,][]{Mackereth_2018,Gebek_2022}. In contrast, the clumpy star formation models predict that chemical bimodality would be more common in galaxies with comparable mass to the Milky Way, because star formation clumps are observed in more than 60\% of high-redshift galaxies \citep[e.g.,][]{Guo_2015}. Such an analysis could be done using spatially-resolved stellar population deconstruction of an edge-on disk, the likes of which are only recently becoming achievable observationally \citep[e.g.,][]{Martig_2021}.

\section{Conclusions} \label{sec:conclusions}

The large sample size, extensive spatial coverage, more complete sampling of the inner Galaxy and mid plane inaccessible to optical spectroscopic surveys, and precise abundance measurements for stars in the final data release of APOGEE can help provide strong constraints on Galactic formation and evolution models, in particular for its disk populations. We present results from the final data release of the combined SDSS APOGEE and APOGEE-2 surveys that explore the chemical and age structure of the Milky Way's disk, measure gradients and distribution functions, and link these new observational constraints to predictions from different chemical evolution models.

Our main conclusions are as follows:

\begin{itemize}
    \item \textbf{Cartography:}
    Overall maps of the Milky Way disk exhibit negative radial age and metallicity gradients. The bar/bulge stands out as more metal-rich and $\alpha$-poor in the inner Galaxy compared to stars at similar radii but different azimuthal angles \citep[e.g.,][]{Wegg_2019,Zasowski_2019,Hasselquist_2020,Eilers2021}.
    \item \textbf{{\alpham} Distribution:} The distribution of $\alpha$-element abundances reveals the chemically bimodal disk structure in the Milky Way \citep[e.g.,][]{Furhmann1998, Bensby2005, Reddy2006, Lee2011, Hayden2015,Kordopatis_2015_GaiaESO}. The low-$\alpha$ disk is thinner (in $Z$) and more radially extended than the high-$\alpha$ disk \citep[e.g.,][]{Yoshii1982, Gilmore1983, Bensby_2011, Bovy2016}. The locus of the low-$\alpha$ sequence varies with radius. Stellar ages within the low-$\alpha$ sequence do not seem to track a true single sequence, implying a superposition of evolutionary tracks.
    \item \textbf{Azimuthal Variance of Metallicity:} We find no significant evidence of large-scale azimuthal asymmetry in most of the disk, although the Galactic bar stands out as metal-rich in the mid-height plane. In the solar neighborhood, we see some coherent, non-axisymmetric structure in metallicity, although it does not obviously correlate with the spiral arms as it does in some studies \citep[e.g.,][]{Inno2019,Poggio2022,Hawkins_2022}\ . 
    \item \textbf{Metallicity Gradients:} The Milky Way's full radial metallicity gradient is flat near the center of the Galaxy, and steepens farther out in radius. The high-$\alpha$ disk displays a nearly flat metallicity profile everywhere in the Galaxy, and the low-$\alpha$ disk has a negative gradient that is shallower at high $Z$ than it is close to the plane. We measure the overall radial metallicity gradient of the disk $R \geq 7$ kpc to be \textcolor{black}{$-0.056 \pm 0.001$} dex kpc$^{-1}$. The overall vertical metallicity gradient of the disk at the solar neighborhood is \textcolor{black}{$-0.296 \pm 0.01$} dex kpc$^{-1}$. Both the radial and vertical metallicity gradients flatten with increasing stellar age. These values are consistent with previous studies \citep[e.g.,][]{Hartkopf1982,Anders2014,Hayden2014,Frankel2019,Katz_2021,Vickers_2021,gaiamaps2022}.
    \item \textbf{Metallicity Distribution Function:} The MDF of the inner Galaxy has the widest spread, but this narrows with radius. The shape of the MDF skews strongly for the low-$\alpha$ disk, transitioning around $R \sim 9.4$ kpc from having a metal-poor tail in the inner Galaxy to having a metal-rich tail in the outer Galaxy \citep[e.g.,][]{Anders2014,Kordopatis_2015,Hayden2015,Loebman_2016,Katz_2021}.
    \item \textbf{Age Gradients:} Like the metallicity gradient, the age profile of the disk is flat in the inner Galaxy but transitions to a negative gradient in the outer Galaxy. The outer Galaxy's gradient is steeper at higher $Z$ for the low-$\alpha$ population, and flat everywhere for the high-$\alpha$ stars \citep[e.g.,][]{Bergemann2014,Martig_2016b,Katz_2021,Anders_2023}.
    \item \textbf{Age Distribution Function:} The ADF for the low-$\alpha$ disk changes in skewness similar to the MDF \citep[e.g.,][]{Katz_2021}, with the inner Galaxy skewed towards younger ages, and the outer Galaxy skewed towards older ages. Above the plane $(|Z|>1$ kpc), there is significant overlap between the ADF of the low-$\alpha$ and high-$\alpha$ populations \citep[e.g.,][]{Haywood_2013,Hayden_2017,Aguirre_2018,Gent2022}, which does not hold closer to the plane. 
    \item \textbf{Age-Metallicity Relation:} The AMR exhibits significant spread near the solar neighborhood, but is more tightly constrained in the inner Galaxy and at larger vertical heights \citep[e.g.,][]{Casagrande2011,Bergemann2014,Feuillet2018}. The slope of the age-metallicity relation varies with radius, and there exists a population of older, metal-rich stars around the solar neighborhood that are likely present due to radial migration \citep[e.g.,][]{Hasselquist_2019, Sahlholdt_2021,Lian2022_migration}. The age-metallicity relation suggests that the outer disk began forming low-$\alpha$ stars while the high-$\alpha$ sequence was still forming in the inner disk \citep[e.g.,][]{Haywood_2013,Aguirre_2018,Gent2022}.
    \item \textbf{Age-Alpha Relation and Chemical Clocks:} The age-alpha relation appears bimodal nearly everywhere in the Galaxy. The low-$\alpha$ sequence evolves significantly with Galactic position, while the high-$\alpha$ sequence displays a constant trend independent of Galactic position  \citep[e.g.,][]{Haywood_2013,Feuillet2018,Katz_2021}. There may be evidence of a three-phase star formation history \citep[e.g.,][]{Sahlholdt_2021} just outside the solar neighborhood ($9< R < 12$ kpc).\\
\end{itemize}

Our results suggest that radial migration is an important process in shaping the present-day appearance of the disk, especially at large radii and close to the Galactic plane. However, stellar migration alone cannot explain the bimodal nature of the $\alpha$-element abundances or the distribution of stellar ages in the disk. A non-continuous evolution model, such as the two-infall scenario or clumpy star formation, appears necessary to explain the trends seen in the inner Galaxy. 

\section*{Acknowledgements}

We would like to thank the anonymous reviewer for their useful feedback towards improving this manuscript.

J.I., C.P, J.A.H. and A.S.M gratefully acknowledge support from NSF grant AST-1909897. D.H.W. and J.A.J. gratefully acknowledge support from NSF grant AST-1909841.

T.C.B. acknowledges partial support for this work from grant PHY 14-30152; Physics Frontier Center/JINA Center for the Evolution of the Elements (JINA-CEE), awarded by the US National Science Foundation, and OISE-1927130: The International Research Network for Nuclear Astrophysics (IReNA), awarded by the US National Science Foundation.

J.G.F-T gratefully acknowledges the grant support provided by Proyecto Fondecyt Iniciaci\'on No. 11220340, and also from ANID Concurso de Fomento a la Vinculaci\'on Internacional para Instituciones de Investigaci\'on Regionales (Modalidad corta duraci\'on) Proyecto No. FOVI210020, and from the Joint Committee ESO-Government of Chile 2021 (ORP 023/2021), and from Becas Santander Movilidad Internacional Profesores 2022, Banco Santander Chile.

C.A.P is thankful for funding from the Spanish government through grants AYA2014-56359-P, AYA2017-86389-P and PID2020-117493GB-100.

Funding for the Sloan Digital Sky Survey IV has been provided by the Alfred P. Sloan Foundation, the U.S. Department of Energy Office of Science, and the Participating Institutions. SDSS acknowledges support and resources from the Center for High-Performance Computing at the University of Utah. The SDSS web site is \url{www.sdss.org}.

SDSS is managed by the Astrophysical Research Consortium for the Participating Institutions of the SDSS Collaboration including the Brazilian Participation Group, the Carnegie Institution for Science, Carnegie Mellon University, Center for Astrophysics | Harvard \& Smithsonian (CfA), the Chilean Participation Group, the French Participation Group, Instituto de Astrof{\'i}sica de Canarias, The Johns Hopkins University, Kavli Institute for the Physics and Mathematics of the Universe (IPMU) / University of Tokyo, the Korean Participation Group, Lawrence Berkeley National Laboratory, Leibniz Institut f{\"u}r Astrophysik Potsdam (AIP), Max-Planck-Institut f{\"u}r Astronomie (MPIA Heidelberg), Max-Planck-Institut f{\"u}r Astrophysik (MPA Garching), Max-Planck-Institut f{\"u}r Extraterrestrische Physik (MPE), National Astronomical Observatories of China, New Mexico State University, New York University, University of Notre Dame, Observat{\'o}rio Nacional / MCTI, The Ohio State University, Pennsylvania State University, Shanghai Astronomical Observatory, United Kingdom Participation Group, Universidad Nacional Aut{\'o}noma de M{\'e}xico, University of Arizona, University of Colorado Boulder, University of Oxford, University of Portsmouth, University of Utah, University of Virginia, University of Washington, University of Wisconsin, Vanderbilt University, and Yale University.


\bibliography{sample631}{}

\begin{thebibliography}{}
\expandafter\ifx\csname natexlab\endcsname\relax\def\natexlab#1{#1}\fi
\providecommand{\url}[1]{\href{#1}{#1}}
\providecommand{\dodoi}[1]{doi:~\href{http://doi.org/#1}{\nolinkurl{#1}}}
\providecommand{\doeprint}[1]{\href{http://ascl.net/#1}{\nolinkurl{http://ascl.net/#1}}}
\providecommand{\doarXiv}[1]{\href{https://arxiv.org/abs/#1}{\nolinkurl{https://arxiv.org/abs/#1}}}

\bibitem[{{Abdurro'uf} {et~al.}(2022){Abdurro'uf}, {Accetta}, {Aerts}, {Silva
  Aguirre}, {Ahumada}, {Ajgaonkar}, {Filiz Ak}, {Alam}, {Allende Prieto},
  {Almeida}, {Anders}, {Anderson}, {Andrews}, {Anguiano}, {Aquino-Ort{\'\i}z},
  {Arag{\'o}n-Salamanca}, {Argudo-Fern{\'a}ndez}, {Ata}, {Aubert},
  {Avila-Reese}, {Badenes}, {Barb{\'a}}, {Barger}, {Barrera-Ballesteros},
  {Beaton}, {Beers}, {Belfiore}, {Bender}, {Bernardi}, {Bershady}, {Beutler},
  {Bidin}, {Bird}, {Bizyaev}, {Blanc}, {Blanton}, {Boardman}, {Bolton},
  {Boquien}, {Borissova}, {Bovy}, {Brandt}, {Brown}, {Brownstein}, {Brusa},
  {Buchner}, {Bundy}, {Burchett}, {Bureau}, {Burgasser}, {Cabang}, {Campbell},
  {Cappellari}, {Carlberg}, {Wanderley}, {Carrera}, {Cash}, {Chen}, {Chen},
  {Cherinka}, {Chiappini}, {Choi}, {Chojnowski}, {Chung}, {Clerc}, {Cohen},
  {Comerford}, {Comparat}, {da Costa}, {Covey}, {Crane}, {Cruz-Gonzalez},
  {Culhane}, {Cunha}, {Dai}, {Damke}, {Darling}, {Davidson}, {Davies},
  {Dawson}, {De Lee}, {Diamond-Stanic}, {Cano-D{\'\i}az}, {S{\'a}nchez},
  {Donor}, {Duckworth}, {Dwelly}, {Eisenstein}, {Elsworth}, {Emsellem},
  {Eracleous}, {Escoffier}, {Fan}, {Farr}, {Feng}, {Fern{\'a}ndez-Trincado},
  {Feuillet}, {Filipp}, {Fillingham}, {Frinchaboy}, {Fromenteau}, {Galbany},
  {Garc{\'\i}a}, {Garc{\'\i}a-Hern{\'a}ndez}, {Ge}, {Geisler}, {Gelfand},
  {G{\'e}ron}, {Gibson}, {Goddy}, {Godoy-Rivera}, {Grabowski}, {Green},
  {Greener}, {Grier}, {Griffith}, {Guo}, {Guy}, {Hadjara}, {Harding},
  {Hasselquist}, {Hayes}, {Hearty}, {Hern{\'a}ndez}, {Hill}, {Hogg},
  {Holtzman}, {Horta}, {Hsieh}, {Hsu}, {Hsu}, {Huber}, {Huertas-Company},
  {Hutchinson}, {Hwang}, {Ibarra-Medel}, {Chitham}, {Ilha}, {Imig}, {Jaekle},
  {Jayasinghe}, {Ji}, {Johnson}, {Jones}, {J{\"o}nsson}, {Katkov}, {Khalatyan},
  {Kinemuchi}, {Kisku}, {Knapen}, {Kneib}, {Kollmeier}, {Kong}, {Kounkel},
  {Kreckel}, {Krishnarao}, {Lacerna}, {Lane}, {Langgin}, {Lavender}, {Law},
  {Lazarz}, {Leung}, {Leung}, {Lewis}, {Li}, {Li}, {Lian}, {Liang}, {Lin},
  {Lin}, {Lin}, {Lintott}, {Long}, {Longa-Pe{\~n}a}, {L{\'o}pez-Cob{\'a}},
  {Lu}, {Lundgren}, {Luo}, {Mackereth}, {de la Macorra}, {Mahadevan},
  {Majewski}, {Manchado}, {Mandeville}, {Maraston}, {Margalef-Bentabol},
  {Masseron}, {Masters}, {Mathur}, {McDermid}, {Mckay}, {Merloni},
  {Merrifield}, {Meszaros}, {Miglio}, {Di Mille}, {Minniti}, {Minsley},
  {Monachesi}, {Moon}, {Mosser}, {Mulchaey}, {Muna}, {Mu{\~n}oz}, {Myers},
  {Myers}, {Nadathur}, {Nair}, {Nandra}, {Neumann}, {Newman}, {Nidever},
  {Nikakhtar}, {Nitschelm}, {O'Connell}, {Garma-Oehmichen}, {Luan Souza de
  Oliveira}, {Olney}, {Oravetz}, {Ortigoza-Urdaneta}, {Osorio}, {Otter},
  {Pace}, {Padilla}, {Pan}, {Pan}, {Parikh}, {Parker}, {Peirani}, {Pe{\~n}a
  Ram{\'\i}rez}, {Penny}, {Percival}, {Perez-Fournon}, {Pinsonneault},
  {Poidevin}, {Poovelil}, {Price-Whelan}, {B{\'a}rbara de Andrade Queiroz},
  {Raddick}, {Ray}, {Rembold}, {Riddle}, {Riffel}, {Riffel}, {Rix}, {Robin},
  {Rodr{\'\i}guez-Puebla}, {Roman-Lopes}, {Rom{\'a}n-Z{\'u}{\~n}iga}, {Rose},
  {Ross}, {Rossi}, {Rubin}, {Salvato}, {S{\'a}nchez}, {S{\'a}nchez-Gallego},
  {Sanderson}, {Santana Rojas}, {Sarceno}, {Sarmiento}, {Sayres}, {Sazonova},
  {Schaefer}, {Schiavon}, {Schlegel}, {Schneider}, {Schultheis}, {Schwope},
  {Serenelli}, {Serna}, {Shao}, {Shapiro}, {Sharma}, {Shen}, {Shetrone}, {Shu},
  {Simon}, {Skrutskie}, {Smethurst}, {Smith}, {Sobeck}, {Spoo}, {Sprague},
  {Stark}, {Stassun}, {Steinmetz}, {Stello}, {Stone-Martinez},
  {Storchi-Bergmann}, {Stringfellow}, {Stutz}, {Su}, {Taghizadeh-Popp},
  {Talbot}, {Tayar}, {Telles}, {Teske}, {Thakar}, {Theissen}, {Tkachenko},
  {Thomas}, {Tojeiro}, {Hernandez Toledo}, {Troup}, {Trump}, {Trussler},
  {Turner}, {Tuttle}, {Unda-Sanzana}, {V{\'a}zquez-Mata}, {Valentini},
  {Valenzuela}, {Vargas-Gonz{\'a}lez}, {Vargas-Maga{\~n}a}, {Alfaro},
  {Villanova}, {Vincenzo}, {Wake}, {Warfield}, {Washington}, {Weaver},
  {Weijmans}, {Weinberg}, {Weiss}, {Westfall}, {Wild}, {Wilde}, {Wilson},
  {Wilson}, {Wilson}, {Wolf}, {Wood-Vasey}, {Yan}, {Zamora}, {Zasowski},
  {Zhang}, {Zhao}, {Zheng}, {Zheng}, \& {Zhu}}]{SDSSdr17}
{Abdurro'uf}, {Accetta}, K., {Aerts}, C., {et~al.} 2022, \apjs, 259, 35,
  \dodoi{10.3847/1538-4365/ac4414}

\bibitem[{Aguirre {et~al.}(2018)Aguirre, Bojsen-Hansen, Slumstrup, Casagrande,
  Kawata, Ciuc{\'{a} }, Handberg, Lund, Mosumgaard, Huber, Johnson,
  Pinsonneault, Serenelli, Stello, Tayar, Bird, Cassisi, Hon, Martig, Nissen,
  Rix, Schönrich, Sahlholdt, Trick, \& Yu}]{Aguirre_2018}
Aguirre, V.~S., Bojsen-Hansen, M., Slumstrup, D., {et~al.} 2018, Monthly
  Notices of the Royal Astronomical Society, \dodoi{10.1093/mnras/sty150}

\bibitem[{{Allende Prieto} {et~al.}(2006){Allende Prieto}, {Beers}, {Wilhelm},
  {Newberg}, {Rockosi}, {Yanny}, \& {Lee}}]{AllendePrieto_2006}
{Allende Prieto}, C., {Beers}, T.~C., {Wilhelm}, R., {et~al.} 2006, \apj, 636,
  804, \dodoi{10.1086/498131}

\bibitem[{{Amarante} {et~al.}(2020){Amarante}, {Beraldo e Silva}, {Debattista},
  \& {Smith}}]{Amarante_2020}
{Amarante}, J. A.~S., {Beraldo e Silva}, L., {Debattista}, V.~P., \& {Smith},
  M.~C. 2020, \apjl, 891, L30, \dodoi{10.3847/2041-8213/ab78a4}

\bibitem[{{Anders} {et~al.}(2018){Anders}, {Chiappini}, {Santiago},
  {Matijevi{\v{c}}}, {Queiroz}, {Steinmetz}, \& {Guiglion}}]{Anders_2018}
{Anders}, F., {Chiappini}, C., {Santiago}, B.~X., {et~al.} 2018, \aap, 619,
  A125, \dodoi{10.1051/0004-6361/201833099}

\bibitem[{{Anders} {et~al.}(2014){Anders}, {Chiappini}, {Santiago},
  {Rocha-Pinto}, {Girardi}, {da Costa}, {Maia}, {Steinmetz}, {Minchev},
  {Schultheis}, {Boeche}, {Miglio}, {Montalb{\'a}n}, {Schneider}, {Beers},
  {Cunha}, {Allende Prieto}, {Balbinot}, {Bizyaev}, {Brauer}, {Brinkmann},
  {Frinchaboy}, {Garc{\'\i}a P{\'e}rez}, {Hayden}, {Hearty}, {Holtzman},
  {Johnson}, {Kinemuchi}, {Majewski}, {Malanushenko}, {Malanushenko},
  {Nidever}, {O'Connell}, {Pan}, {Robin}, {Schiavon}, {Shetrone}, {Skrutskie},
  {Smith}, {Stassun}, \& {Zasowski}}]{Anders2014}
---. 2014, \aap, 564, A115, \dodoi{10.1051/0004-6361/201323038}

\bibitem[{Anders {et~al.}(2023)Anders, Gispert, Ratcliffe, Chiappini, Minchev,
  Nepal, Queiroz, Amarante, Antoja, Casali, Casamiquela, Khalatyan, Miglio,
  Perottoni, \& Schultheis}]{Anders_2023}
Anders, F., Gispert, P., Ratcliffe, B., {et~al.} 2023, Spectroscopic age
  estimates for 180 000 APOGEE red-giant stars: Precise spatial and kinematic
  trends with age in the Galactic disc.
\newblock \doarXiv{2304.08276}

\bibitem[{{Antoja} {et~al.}(2020){Antoja}, {Ramos}, {Mateu}, {Helmi}, {Anders},
  {Jordi}, \& {Carballo-Bello}}]{Antoja_2020}
{Antoja}, T., {Ramos}, P., {Mateu}, C., {et~al.} 2020, \aap, 635, L3,
  \dodoi{10.1051/0004-6361/201937145}

\bibitem[{{Barbanis} \& {Woltjer}(1967)}]{Barbanis1967}
{Barbanis}, B., \& {Woltjer}, L. 1967, \apj, 150, 461, \dodoi{10.1086/149349}

\bibitem[{Beaton {et~al.}(2021)Beaton, Oelkers, Hayes, Covey, Chojnowski, Lee,
  Sobeck, Majewski, Cohen, Fern{\'{a}}ndez-Trincado, Longa-Pe{\~{n}}a,
  O'Connell, Santana, Stringfellow, Zasowski, Aerts, Anguiano, Bender,
  Ca{\~{n}}as, Cunha, Donor, Fleming, Frinchaboy, Feuillet, Harding,
  Hasselquist, Holtzman, Johnson, Kollmeier, Kounkel, Mahadevan, Price-Whelan,
  Rojas-Arriagada, Rom{\'{a}}n-Z{\'{u}}{\~{n}}iga, Schlafly, Schultheis,
  Shetrone, Simon, Stassun, Stutz, Tayar, Teske, Tkachenko, Troup, Albareti,
  Bizyaev, Bovy, Burgasser, Comparat, Downes, Geisler, Inno, Manchado, Ness,
  Pinsonneault, Prada, Roman-Lopes, Simonian, Smith, Yan, \&
  Zamora}]{Beaton_2021}
Beaton, R.~L., Oelkers, R.~J., Hayes, C.~R., {et~al.} 2021, The Astronomical
  Journal, 162, 302, \dodoi{10.3847/1538-3881/ac260c}

\bibitem[{{Bedell} {et~al.}(2018){Bedell}, {Bean}, {Mel{\'e}ndez}, {Spina},
  {Ram{\'\i}rez}, {Asplund}, {Alves-Brito}, {dos Santos}, {Dreizler}, {Yong},
  {Monroe}, \& {Casagrande}}]{Bedell2018}
{Bedell}, M., {Bean}, J.~L., {Mel{\'e}ndez}, J., {et~al.} 2018, \apj, 865, 68,
  \dodoi{10.3847/1538-4357/aad908}

\bibitem[{Bensby {et~al.}(2011)Bensby, Alves-Brito, Oey, Yong, \& Mel{\'{e}
  }ndez}]{Bensby_2011}
Bensby, T., Alves-Brito, A., Oey, M.~S., Yong, D., \& Mel{\'{e} }ndez, J. 2011,
  The Astrophysical Journal, 735, L46, \dodoi{10.1088/2041-8205/735/2/l46}

\bibitem[{{Bensby} {et~al.}(2005){Bensby}, {Feltzing}, {Lundstr{\"o}m}, \&
  {Ilyin}}]{Bensby2005}
{Bensby}, T., {Feltzing}, S., {Lundstr{\"o}m}, I., \& {Ilyin}, I. 2005, \aap,
  433, 185, \dodoi{10.1051/0004-6361:20040332}

\bibitem[{{Bensby} {et~al.}(2014){Bensby}, {Feltzing}, \& {Oey}}]{Bensby_2014}
{Bensby}, T., {Feltzing}, S., \& {Oey}, M.~S. 2014, \aap, 562, A71,
  \dodoi{10.1051/0004-6361/201322631}

\bibitem[{Bensby {et~al.}(2007)Bensby, Zenn, Oey, \& Feltzing}]{Bensby_2007}
Bensby, T., Zenn, A.~R., Oey, M.~S., \& Feltzing, S. 2007, The Astrophysical
  Journal, 663, L13, \dodoi{10.1086/519792}

\bibitem[{{Beraldo e Silva} {et~al.}(2020){Beraldo e Silva}, {Debattista},
  {Khachaturyants}, \& {Nidever}}]{Silva_2020}
{Beraldo e Silva}, L., {Debattista}, V.~P., {Khachaturyants}, T., \& {Nidever},
  D. 2020, \mnras, 492, 4716, \dodoi{10.1093/mnras/staa065}

\bibitem[{{Bergemann} {et~al.}(2014){Bergemann}, {Ruchti}, {Serenelli},
  {Feltzing}, {Alves-Brito}, {Asplund}, {Bensby}, {Gruyters}, {Heiter},
  {Hourihane}, {Korn}, {Lind}, {Marino}, {Jofre}, {Nordlander}, {Ryde},
  {Worley}, {Gilmore}, {Randich}, {Ferguson}, {Jeffries}, {Micela},
  {Negueruela}, {Prusti}, {Rix}, {Vallenari}, {Alfaro}, {Allende Prieto},
  {Bragaglia}, {Koposov}, {Lanzafame}, {Pancino}, {Recio-Blanco}, {Smiljanic},
  {Walton}, {Costado}, {Franciosini}, {Hill}, {Lardo}, {de Laverny}, {Magrini},
  {Maiorca}, {Masseron}, {Morbidelli}, {Sacco}, {Kordopatis}, \&
  {Tautvai{\v{s}}ien{\.{e}}}}]{Bergemann2014}
{Bergemann}, M., {Ruchti}, G.~R., {Serenelli}, A., {et~al.} 2014, \aap, 565,
  A89, \dodoi{10.1051/0004-6361/201423456}

\bibitem[{Bilitewski \& Schönrich(2012)}]{Bilitewski_2012}
Bilitewski, T., \& Schönrich, R. 2012, Monthly Notices of the Royal
  Astronomical Society, 426, 2266, \dodoi{10.1111/j.1365-2966.2012.21827.x}

\bibitem[{Bird {et~al.}(2013)Bird, Kazantzidis, Weinberg, Guedes, Callegari,
  Mayer, \& Madau}]{Bird_2013}
Bird, J.~C., Kazantzidis, S., Weinberg, D.~H., {et~al.} 2013, The Astrophysical
  Journal, 773, 43, \dodoi{10.1088/0004-637x/773/1/43}

\bibitem[{Bird {et~al.}(2021)Bird, Loebman, Weinberg, Brooks, Quinn, \&
  Christensen}]{Bird_2021}
Bird, J.~C., Loebman, S.~R., Weinberg, D.~H., {et~al.} 2021, Monthly Notices of
  the Royal Astronomical Society, 503, 1815, \dodoi{10.1093/mnras/stab289}

\bibitem[{{Bland-Hawthorn} \& {Gerhard}(2016)}]{BlandHawthorn2016}
{Bland-Hawthorn}, J., \& {Gerhard}, O. 2016, \araa, 54, 529,
  \dodoi{10.1146/annurev-astro-081915-023441}

\bibitem[{{Blanton} {et~al.}(2017){Blanton}, {Bershady}, {Abolfathi},
  {Albareti}, {Allende Prieto}, {Almeida}, {Alonso-Garc{\'\i}a}, {Anders},
  {Anderson}, {Andrews}, {Aquino-Ort{\'\i}z}, {Arag{\'o}n-Salamanca},
  {Argudo-Fern{\'a}ndez}, {Armengaud}, {Aubourg}, {Avila-Reese}, {Badenes},
  {Bailey}, {Barger}, {Barrera-Ballesteros}, {Bartosz}, {Bates}, {Baumgarten},
  {Bautista}, {Beaton}, {Beers}, {Belfiore}, {Bender}, {Berlind}, {Bernardi},
  {Beutler}, {Bird}, {Bizyaev}, {Blanc}, {Blomqvist}, {Bolton}, {Boquien},
  {Borissova}, {van den Bosch}, {Bovy}, {Brandt}, {Brinkmann}, {Brownstein},
  {Bundy}, {Burgasser}, {Burtin}, {Busca}, {Cappellari}, {Delgado Carigi},
  {Carlberg}, {Carnero Rosell}, {Carrera}, {Chanover}, {Cherinka}, {Cheung},
  {G{\'o}mez Maqueo Chew}, {Chiappini}, {Choi}, {Chojnowski}, {Chuang},
  {Chung}, {Cirolini}, {Clerc}, {Cohen}, {Comparat}, {da Costa}, {Cousinou},
  {Covey}, {Crane}, {Croft}, {Cruz-Gonzalez}, {Garrido Cuadra}, {Cunha},
  {Damke}, {Darling}, {Davies}, {Dawson}, {de la Macorra}, {Dell'Agli}, {De
  Lee}, {Delubac}, {Di Mille}, {Diamond-Stanic}, {Cano-D{\'\i}az}, {Donor},
  {Downes}, {Drory}, {du Mas des Bourboux}, {Duckworth}, {Dwelly}, {Dyer},
  {Ebelke}, {Eigenbrot}, {Eisenstein}, {Emsellem}, {Eracleous}, {Escoffier},
  {Evans}, {Fan}, {Fern{\'a}ndez-Alvar}, {Fernandez-Trincado}, {Feuillet},
  {Finoguenov}, {Fleming}, {Font-Ribera}, {Fredrickson}, {Freischlad},
  {Frinchaboy}, {Fuentes}, {Galbany}, {Garcia-Dias},
  {Garc{\'\i}a-Hern{\'a}ndez}, {Gaulme}, {Geisler}, {Gelfand},
  {Gil-Mar{\'\i}n}, {Gillespie}, {Goddard}, {Gonzalez-Perez}, {Grabowski},
  {Green}, {Grier}, {Gunn}, {Guo}, {Guy}, {Hagen}, {Hahn}, {Hall}, {Harding},
  {Hasselquist}, {Hawley}, {Hearty}, {Gonzalez Hern{\'a}ndez}, {Ho}, {Hogg},
  {Holley-Bockelmann}, {Holtzman}, {Holzer}, {Huehnerhoff}, {Hutchinson},
  {Hwang}, {Ibarra-Medel}, {da Silva Ilha}, {Ivans}, {Ivory}, {Jackson},
  {Jensen}, {Johnson}, {Jones}, {J{\"o}nsson}, {Jullo}, {Kamble}, {Kinemuchi},
  {Kirkby}, {Kitaura}, {Klaene}, {Knapp}, {Kneib}, {Kollmeier}, {Lacerna},
  {Lane}, {Lang}, {Law}, {Lazarz}, {Lee}, {Le Goff}, {Liang}, {Li}, {Li},
  {Lian}, {Lima}, {Lin}, {Lin}, {Bertran de Lis}, {Liu}, {de Icaza Lizaola},
  {Long}, {Lucatello}, {Lundgren}, {MacDonald}, {Deconto Machado}, {MacLeod},
  {Mahadevan}, {Geimba Maia}, {Maiolino}, {Majewski}, {Malanushenko},
  {Malanushenko}, {Manchado}, {Mao}, {Maraston}, {Marques-Chaves}, {Masseron},
  {Masters}, {McBride}, {McDermid}, {McGrath}, {McGreer}, {Medina Pe{\~n}a},
  {Melendez}, {Merloni}, {Merrifield}, {Meszaros}, {Meza}, {Minchev},
  {Minniti}, {Miyaji}, {More}, {Mulchaey}, {M{\"u}ller-S{\'a}nchez}, {Muna},
  {Munoz}, {Myers}, {Nair}, {Nandra}, {Correa do Nascimento}, {Negrete},
  {Ness}, {Newman}, {Nichol}, {Nidever}, {Nitschelm}, {Ntelis}, {O'Connell},
  {Oelkers}, {Oravetz}, {Oravetz}, {Pace}, {Padilla}, {Palanque-Delabrouille},
  {Alonso Palicio}, {Pan}, {Parejko}, {Parikh}, {P{\^a}ris}, {Park}, {Patten},
  {Peirani}, {Pellejero-Ibanez}, {Penny}, {Percival}, {Perez-Fournon},
  {Petitjean}, {Pieri}, {Pinsonneault}, {Pisani}, {Poleski}, {Prada},
  {Prakash}, {Queiroz}, {Raddick}, {Raichoor}, {Barboza Rembold}, {Richstein},
  {Riffel}, {Riffel}, {Rix}, {Robin}, {Rockosi}, {Rodr{\'\i}guez-Torres},
  {Roman-Lopes}, {Rom{\'a}n-Z{\'u}{\~n}iga}, {Rosado}, {Ross}, {Rossi}, {Ruan},
  {Ruggeri}, {Rykoff}, {Salazar-Albornoz}, {Salvato}, {S{\'a}nchez}, {Aguado},
  {S{\'a}nchez-Gallego}, {Santana}, {Santiago}, {Sayres}, {Schiavon}, {da Silva
  Schimoia}, {Schlafly}, {Schlegel}, {Schneider}, {Schultheis}, {Schuster},
  {Schwope}, {Seo}, {Shao}, {Shen}, {Shetrone}, {Shull}, {Simon}, {Skinner},
  {Skrutskie}, {Slosar}, {Smith}, {Sobeck}, {Sobreira}, {Somers}, {Souto},
  {Stark}, {Stassun}, {Stauffer}, {Steinmetz}, {Storchi-Bergmann},
  {Streblyanska}, {Stringfellow}, {Su{\'a}rez}, {Sun}, {Suzuki}, {Szigeti},
  {Taghizadeh-Popp}, {Tang}, {Tao}, {Tayar}, {Tembe}, {Teske}, {Thakar},
  {Thomas}, {Thompson}, {Tinker}, {Tissera}, {Tojeiro}, {Hernandez Toledo}, {de
  la Torre}, {Tremonti}, {Troup}, {Valenzuela}, {Martinez Valpuesta},
  {Vargas-Gonz{\'a}lez}, {Vargas-Maga{\~n}a}, {Vazquez}, {Villanova}, {Vivek},
  {Vogt}, {Wake}, {Walterbos}, {Wang}, {Weaver}, {Weijmans}, {Weinberg},
  {Westfall}, {Whelan}, {Wild}, {Wilson}, {Wood-Vasey}, {Wylezalek}, {Xiao},
  {Yan}, {Yang}, {Ybarra}, {Y{\`e}che}, {Zakamska}, {Zamora}, {Zarrouk},
  {Zasowski}, {Zhang}, {Zhao}, {Zheng}, {Zheng}, {Zhou}, {Zhou}, {Zhu},
  {Zoccali}, \& {Zou}}]{Blanton_2017}
{Blanton}, M.~R., {Bershady}, M.~A., {Abolfathi}, B., {et~al.} 2017, \aj, 154,
  28, \dodoi{10.3847/1538-3881/aa7567}

\bibitem[{Boeche {et~al.}(2013)Boeche, Siebert, Piffl, Just, Steinmetz, Sharma,
  Kordopatis, Gilmore, Chiappini, Williams, Grebel, Bland-Hawthorn, Gibson,
  Munari, Siviero, Bienaym{\'{e} }, Navarro, Parker, Reid, Seabroke, Watson,
  Wyse, \& Zwitter}]{Boeche_2013}
Boeche, C., Siebert, A., Piffl, T., {et~al.} 2013, Astronomy {$\&$}
  Astrophysics, 559, A59, \dodoi{10.1051/0004-6361/201322085}

\bibitem[{{Bournaud} {et~al.}(2007){Bournaud}, {Elmegreen}, \&
  {Elmegreen}}]{Bournaud_2007}
{Bournaud}, F., {Elmegreen}, B.~G., \& {Elmegreen}, D.~M. 2007, \apj, 670, 237,
  \dodoi{10.1086/522077}

\bibitem[{Bovy {et~al.}(2012{\natexlab{a}})Bovy, Rix, \& Hogg}]{Bovy_2012}
Bovy, J., Rix, H.-W., \& Hogg, D.~W. 2012{\natexlab{a}}, The Astrophysical
  Journal, 751, 131, \dodoi{10.1088/0004-637x/751/2/131}

\bibitem[{Bovy {et~al.}(2012{\natexlab{b}})Bovy, Rix, Liu, Hogg, Beers, \&
  Lee}]{Bovy2012}
Bovy, J., Rix, H.-W., Liu, C., {et~al.} 2012{\natexlab{b}}, The Astrophysical
  Journal, 753, 148, \dodoi{10.1088/0004-637x/753/2/148}

\bibitem[{Bovy {et~al.}(2016)Bovy, Rix, Schlafly, Nidever, Holtzman, Shetrone,
  \& Beers}]{Bovy2016}
Bovy, J., Rix, H.-W., Schlafly, E.~F., {et~al.} 2016, The Astrophysical
  Journal, 823, 30, \dodoi{10.3847/0004-637x/823/1/30}

\bibitem[{{Bowen} \& {Vaughan}(1973)}]{lco25m_Bowen1973}
{Bowen}, I.~S., \& {Vaughan}, A.~H., J. 1973, \ao, 12, 1430,
  \dodoi{10.1364/AO.12.001430}

\bibitem[{{Bregman}(1980)}]{Bregman_1980}
{Bregman}, J.~N. 1980, \apj, 236, 577, \dodoi{10.1086/157776}

\bibitem[{{Brook} {et~al.}(2004){Brook}, {Kawata}, {Gibson}, \&
  {Freeman}}]{Brook_2004}
{Brook}, C.~B., {Kawata}, D., {Gibson}, B.~K., \& {Freeman}, K.~C. 2004, \apj,
  612, 894, \dodoi{10.1086/422709}

\bibitem[{Buder {et~al.}(2018)Buder, Asplund, Duong, Kos, Lind, Ness, Sharma,
  Bland-Hawthorn, Casey, De~Silva, \& et~al.}]{GALAH_Buder2018}
Buder, S., Asplund, M., Duong, L., {et~al.} 2018, Monthly Notices of the Royal
  Astronomical Society, 478, 4513–4552, \dodoi{10.1093/mnras/sty1281}

\bibitem[{Buder {et~al.}(2021)Buder, Lind, Ness, Feuillet, Horta, Monty, Buck,
  Nordlander, Bland-Hawthorn, Casey, Silva, D'Orazi, Freeman, Hayden, Kos,
  Martell, Lewis, Lin, Schlesinger, Sharma, Simpson, Stello, Zucker, Zwitter,
  Ciuc{\u{a}}, Horner, Kobayashi, Ting, \& and}]{Buder_2021}
Buder, S., Lind, K., Ness, M.~K., {et~al.} 2021, Monthly Notices of the Royal
  Astronomical Society, 510, 2407, \dodoi{10.1093/mnras/stab3504}

\bibitem[{Carrell {et~al.}(2012)Carrell, Chen, \& Zhao}]{Carrell_2012}
Carrell, K., Chen, Y., \& Zhao, G. 2012, The Astronomical Journal, 144, 185,
  \dodoi{10.1088/0004-6256/144/6/185}

\bibitem[{{Casagrande} {et~al.}(2011){Casagrande}, {Sch{\"o}nrich}, {Asplund},
  {Cassisi}, {Ram{\'\i}rez}, {Mel{\'e}ndez}, {Bensby}, \&
  {Feltzing}}]{Casagrande2011}
{Casagrande}, L., {Sch{\"o}nrich}, R., {Asplund}, M., {et~al.} 2011, \aap, 530,
  A138, \dodoi{10.1051/0004-6361/201016276}

\bibitem[{Casamiquela {et~al.}(2021)Casamiquela, Soubiran, Jofr{\'{e} },
  Chiappini, Lagarde, Tarricq, Carrera, Jordi, Balaguer-N{\'{u}}{\~{n}}ez,
  Carbajo-Hijarrubia, \& Blanco-Cuaresma}]{Casamiquela_2021}
Casamiquela, L., Soubiran, C., Jofr{\'{e} }, P., {et~al.} 2021, Astronomy
  {$\&$} Astrophysics, 652, A25, \dodoi{10.1051/0004-6361/202039951}

\bibitem[{Chen {et~al.}(2022)Chen, Hayden, Sharma, Bland-Hawthorn, Kobayashi,
  \& Karakas}]{Chen_2022}
Chen, B., Hayden, M.~R., Sharma, S., {et~al.} 2022, Chemical Evolution with
  Radial Mixing Redux: Extending beyond the Solar Neighborhood,  arXiv,
  \dodoi{10.48550/ARXIV.2204.11413}

\bibitem[{{Cheng} {et~al.}(2012){Cheng}, {Rockosi}, {Morrison},
  {Sch{\"o}nrich}, {Lee}, {Beers}, {Bizyaev}, {Pan}, \&
  {Schneider}}]{Cheng2012}
{Cheng}, J.~Y., {Rockosi}, C.~M., {Morrison}, H.~L., {et~al.} 2012, \apj, 746,
  149, \dodoi{10.1088/0004-637X/746/2/149}

\bibitem[{Cheng {et~al.}(2020)Cheng, Anguiano, Majewski, Hayes, Arras,
  Chiappini, Hasselquist, de~Andrade~Queiroz, Nitschelm,
  Garc{\'{\i}}a-Hern{\'{a}}ndez, Lane, Roman-Lopes, \& Frinchaboy}]{Cheng_2020}
Cheng, X., Anguiano, B., Majewski, S.~R., {et~al.} 2020, The Astrophysical
  Journal, 905, 49, \dodoi{10.3847/1538-4357/abc3c2}

\bibitem[{{Chiappini} {et~al.}(1997){Chiappini}, {Matteucci}, \&
  {Gratton}}]{Chiappini1997}
{Chiappini}, C., {Matteucci}, F., \& {Gratton}, R. 1997, \apj, 477, 765,
  \dodoi{10.1086/303726}

\bibitem[{Chiappini {et~al.}(2001)Chiappini, Matteucci, \&
  Romano}]{Chiappini_2001}
Chiappini, C., Matteucci, F., \& Romano, D. 2001, The Astrophysical Journal,
  554, 1044, \dodoi{10.1086/321427}

\bibitem[{Choi {et~al.}(2016)Choi, Dotter, Conroy, Cantiello, Paxton, \&
  Johnson}]{Choi_2016}
Choi, J., Dotter, A., Conroy, C., {et~al.} 2016, The Astrophysical Journal,
  823, 102, \dodoi{10.3847/0004-637x/823/2/102}

\bibitem[{Chrob{\'a}kov{\'a} {et~al.}(2022)Chrob{\'a}kov{\'a}, Nagy, \&
  L{\'o}pez-Corredoira}]{Chrobakova_2022}
Chrob{\'a}kov{\'a}, {\v{Z}}., Nagy, R., \& L{\'o}pez-Corredoira, M. 2022, Warp
  and flare of the Galactic disc revealed with supergiants by Gaia EDR3,
  arXiv, \dodoi{10.48550/ARXIV.2206.08230}

\bibitem[{Ciuc{\u{a} } {et~al.}(2021)Ciuc{\u{a} }, Kawata, Miglio, Davies, \&
  Grand}]{Ciuc_2021}
Ciuc{\u{a} }, I., Kawata, D., Miglio, A., Davies, G.~R., \& Grand, R. J.~J.
  2021, Monthly Notices of the Royal Astronomical Society, 503, 2814,
  \dodoi{10.1093/mnras/stab639}

\bibitem[{{Clarke} {et~al.}(2019){Clarke}, {Debattista}, {Nidever}, {Loebman},
  {Simons}, {Kassin}, {Du}, {Ness}, {Fisher}, {Quinn}, {Wadsley}, {Freeman}, \&
  {Popescu}}]{Clarke_2019}
{Clarke}, A.~J., {Debattista}, V.~P., {Nidever}, D.~L., {et~al.} 2019, \mnras,
  484, 3476, \dodoi{10.1093/mnras/stz104}

\bibitem[{{da Silva} {et~al.}(2012){da Silva}, {Porto de Mello}, {Milone}, {da
  Silva}, {Ribeiro}, \& {Rocha-Pinto}}]{daSilva2012}
{da Silva}, R., {Porto de Mello}, G.~F., {Milone}, A.~C., {et~al.} 2012, \aap,
  542, A84, \dodoi{10.1051/0004-6361/201118751}

\bibitem[{{Dalcanton}(2007)}]{Dalcanton2007}
{Dalcanton}, J.~J. 2007, \apj, 658, 941, \dodoi{10.1086/508913}

\bibitem[{Dias {et~al.}(2019)Dias, Monteiro, L{\'{e} }pine, \&
  Barros}]{Dias_2019}
Dias, W.~S., Monteiro, H., L{\'{e} }pine, J. R.~D., \& Barros, D.~A. 2019,
  Monthly Notices of the Royal Astronomical Society, 486, 5726,
  \dodoi{10.1093/mnras/stz1196}

\bibitem[{{Donor} {et~al.}(2020){Donor}, {Frinchaboy}, {Cunha}, {O'Connell},
  {Allende Prieto}, {Almeida}, {Anders}, {Beaton}, {Bizyaev}, {Brownstein},
  {Carrera}, {Chiappini}, {Cohen}, {Garc{\'\i}a-Hern{\'a}ndez}, {Geisler},
  {Hasselquist}, {J{\"o}nsson}, {Lane}, {Majewski}, {Minniti}, {Bidin}, {Pan},
  {Roman-Lopes}, {Sobeck}, \& {Zasowski}}]{Donor2020}
{Donor}, J., {Frinchaboy}, P.~M., {Cunha}, K., {et~al.} 2020, \aj, 159, 199,
  \dodoi{10.3847/1538-3881/ab77bc}

\bibitem[{{Edvardsson} {et~al.}(1993){Edvardsson}, {Andersen}, {Gustafsson},
  {Lambert}, {Nissen}, \& {Tomkin}}]{Edvardsson1993}
{Edvardsson}, B., {Andersen}, J., {Gustafsson}, B., {et~al.} 1993, \aap, 275,
  101

\bibitem[{{Eggen} {et~al.}(1962){Eggen}, {Lynden-Bell}, \&
  {Sandage}}]{Eggen1962}
{Eggen}, O.~J., {Lynden-Bell}, D., \& {Sandage}, A.~R. 1962, \apj, 136, 748,
  \dodoi{10.1086/147433}

\bibitem[{{Eilers} {et~al.}(2021){Eilers}, {Hogg}, {Rix}, {Ness},
  {Price-Whelan}, {Meszaros}, \& {Nitschelm}}]{Eilers2021}
{Eilers}, A.-C., {Hogg}, D.~W., {Rix}, H.-W., {et~al.} 2021, arXiv e-prints,
  arXiv:2112.03295.
\newblock \doarXiv{2112.03295}

\bibitem[{{Elmegreen} {et~al.}(2005){Elmegreen}, {Elmegreen}, {Vollbach},
  {Foster}, \& {Ferguson}}]{Elmegreen2005}
{Elmegreen}, B.~G., {Elmegreen}, D.~M., {Vollbach}, D.~R., {Foster}, E.~R., \&
  {Ferguson}, T.~E. 2005, \apj, 634, 101, \dodoi{10.1086/496952}

\bibitem[{Feuillet {et~al.}(2019)Feuillet, Frankel, Lind, Frinchaboy,
  García-Hernández, Lane, Nitschelm, \& Roman-Lopes}]{Feuillet2019}
Feuillet, D.~K., Frankel, N., Lind, K., {et~al.} 2019, Monthly Notices of the
  Royal Astronomical Society, 489, 1742–1752, \dodoi{10.1093/mnras/stz2221}

\bibitem[{{Feuillet} {et~al.}(2018){Feuillet}, {Bovy}, {Holtzman}, {Weinberg},
  {Garc{\'\i}a-Hern{\'a}ndez}, {Hearty}, {Majewski}, {Roman-Lopes}, {Rybizki},
  \& {Zamora}}]{Feuillet2018}
{Feuillet}, D.~K., {Bovy}, J., {Holtzman}, J., {et~al.} 2018, \mnras, 477,
  2326, \dodoi{10.1093/mnras/sty779}

\bibitem[{Finlator \& Dav{\'{e} }(2008)}]{Finlator_2008}
Finlator, K., \& Dav{\'{e} }, R. 2008, Monthly Notices of the Royal
  Astronomical Society, 385, 2181, \dodoi{10.1111/j.1365-2966.2008.12991.x}

\bibitem[{Frankel {et~al.}(2018)Frankel, Rix, Ting, Ness, \&
  Hogg}]{Frankel_2018}
Frankel, N., Rix, H.-W., Ting, Y.-S., Ness, M., \& Hogg, D.~W. 2018, The
  Astrophysical Journal, 865, 96, \dodoi{10.3847/1538-4357/aadba5}

\bibitem[{{Frankel} {et~al.}(2019){Frankel}, {Sanders}, {Rix}, {Ting}, \&
  {Ness}}]{Frankel2019}
{Frankel}, N., {Sanders}, J., {Rix}, H.-W., {Ting}, Y.-S., \& {Ness}, M. 2019,
  \apj, 884, 99, \dodoi{10.3847/1538-4357/ab4254}

\bibitem[{Frankel {et~al.}(2020)Frankel, Sanders, Ting, \& Rix}]{Frankel_2020}
Frankel, N., Sanders, J., Ting, Y.-S., \& Rix, H.-W. 2020, The Astrophysical
  Journal, 896, 15, \dodoi{10.3847/1538-4357/ab910c}

\bibitem[{{Fraternali}(2017)}]{Fraternali_2017}
{Fraternali}, F. 2017, in Astrophysics and Space Science Library, Vol. 430, Gas
  Accretion onto Galaxies, ed. A.~{Fox} \& R.~{Dav{\'e}}, 323,
  \dodoi{10.1007/978-3-319-52512-9_14}

\bibitem[{Freeman \& Bland-Hawthorn(2002)}]{Freeman_2002}
Freeman, K., \& Bland-Hawthorn, J. 2002, Annual Review of Astronomy and
  Astrophysics, 40, 487, \dodoi{10.1146/annurev.astro.40.060401.093840}

\bibitem[{Freudenburg {et~al.}(2017)Freudenburg, Weinberg, Hayden, \&
  Holtzman}]{Freudenburg_2017}
Freudenburg, J. K.~C., Weinberg, D.~H., Hayden, M.~R., \& Holtzman, J.~A. 2017,
  The Astrophysical Journal, 849, 17, \dodoi{10.3847/1538-4357/aa8c03}

\bibitem[{{Fuhrmann}(1998)}]{Furhmann1998}
{Fuhrmann}, K. 1998, \aap, 338, 161

\bibitem[{{Gaia Collaboration} {et~al.}(2018){Gaia Collaboration}, {Brown},
  {Vallenari}, {Prusti}, {de Bruijne}, {Babusiaux}, {Bailer-Jones}, {Biermann},
  {Evans}, {Eyer}, {Jansen}, {Jordi}, {Klioner}, {Lammers}, {Lindegren},
  {Luri}, {Mignard}, {Panem}, {Pourbaix}, {Randich}, {Sartoretti}, {Siddiqui},
  {Soubiran}, {van Leeuwen}, {Walton}, {Arenou}, {Bastian}, {Cropper},
  {Drimmel}, {Katz}, {Lattanzi}, {Bakker}, {Cacciari}, {Casta{\~n}eda},
  {Chaoul}, {Cheek}, {De Angeli}, {Fabricius}, {Guerra}, {Holl}, {Masana},
  {Messineo}, {Mowlavi}, {Nienartowicz}, {Panuzzo}, {Portell}, {Riello},
  {Seabroke}, {Tanga}, {Th{\'e}venin}, {Gracia-Abril}, {Comoretto},
  {Garcia-Reinaldos}, {Teyssier}, {Altmann}, {Andrae}, {Audard},
  {Bellas-Velidis}, {Benson}, {Berthier}, {Blomme}, {Burgess}, {Busso},
  {Carry}, {Cellino}, {Clementini}, {Clotet}, {Creevey}, {Davidson}, {De
  Ridder}, {Delchambre}, {Dell'Oro}, {Ducourant},
  {Fern{\'a}ndez-Hern{\'a}ndez}, {Fouesneau}, {Fr{\'e}mat}, {Galluccio},
  {Garc{\'\i}a-Torres}, {Gonz{\'a}lez-N{\'u}{\~n}ez}, {Gonz{\'a}lez-Vidal},
  {Gosset}, {Guy}, {Halbwachs}, {Hambly}, {Harrison}, {Hern{\'a}ndez},
  {Hestroffer}, {Hodgkin}, {Hutton}, {Jasniewicz}, {Jean-Antoine-Piccolo},
  {Jordan}, {Korn}, {Krone-Martins}, {Lanzafame}, {Lebzelter}, {L{\"o}ffler},
  {Manteiga}, {Marrese}, {Mart{\'\i}n-Fleitas}, {Moitinho}, {Mora}, {Muinonen},
  {Osinde}, {Pancino}, {Pauwels}, {Petit}, {Recio-Blanco}, {Richards},
  {Rimoldini}, {Robin}, {Sarro}, {Siopis}, {Smith}, {Sozzetti}, {S{\"u}veges},
  {Torra}, {van Reeven}, {Abbas}, {Abreu Aramburu}, {Accart}, {Aerts},
  {Altavilla}, {{\'A}lvarez}, {Alvarez}, {Alves}, {Anderson}, {Andrei},
  {Anglada Varela}, {Antiche}, {Antoja}, {Arcay}, {Astraatmadja}, {Bach},
  {Baker}, {Balaguer-N{\'u}{\~n}ez}, {Balm}, {Barache}, {Barata}, {Barbato},
  {Barblan}, {Barklem}, {Barrado}, {Barros}, {Barstow}, {Bartholom{\'e}
  Mu{\~n}oz}, {Bassilana}, {Becciani}, {Bellazzini}, {Berihuete}, {Bertone},
  {Bianchi}, {Bienaym{\'e}}, {Blanco-Cuaresma}, {Boch}, {Boeche}, {Bombrun},
  {Borrachero}, {Bossini}, {Bouquillon}, {Bourda}, {Bragaglia}, {Bramante},
  {Breddels}, {Bressan}, {Brouillet}, {Br{\"u}semeister}, {Brugaletta},
  {Bucciarelli}, {Burlacu}, {Busonero}, {Butkevich}, {Buzzi}, {Caffau},
  {Cancelliere}, {Cannizzaro}, {Cantat-Gaudin}, {Carballo}, {Carlucci},
  {Carrasco}, {Casamiquela}, {Castellani}, {Castro-Ginard}, {Charlot},
  {Chemin}, {Chiavassa}, {Cocozza}, {Costigan}, {Cowell}, {Crifo}, {Crosta},
  {Crowley}, {Cuypers}, {Dafonte}, {Damerdji}, {Dapergolas}, {David}, {David},
  {de Laverny}, {De Luise}, {De March}, {de Martino}, {de Souza}, {de Torres},
  {Debosscher}, {del Pozo}, {Delbo}, {Delgado}, {Delgado}, {Di Matteo},
  {Diakite}, {Diener}, {Distefano}, {Dolding}, {Drazinos}, {Dur{\'a}n},
  {Edvardsson}, {Enke}, {Eriksson}, {Esquej}, {Eynard Bontemps}, {Fabre},
  {Fabrizio}, {Faigler}, {Falc{\~a}o}, {Farr{\`a}s Casas}, {Federici},
  {Fedorets}, {Fernique}, {Figueras}, {Filippi}, {Findeisen}, {Fonti},
  {Fraile}, {Fraser}, {Fr{\'e}zouls}, {Gai}, {Galleti}, {Garabato},
  {Garc{\'\i}a-Sedano}, {Garofalo}, {Garralda}, {Gavel}, {Gavras}, {Gerssen},
  {Geyer}, {Giacobbe}, {Gilmore}, {Girona}, {Giuffrida}, {Glass}, {Gomes},
  {Granvik}, {Gueguen}, {Guerrier}, {Guiraud}, {Guti{\'e}rrez-S{\'a}nchez},
  {Haigron}, {Hatzidimitriou}, {Hauser}, {Haywood}, {Heiter}, {Helmi}, {Heu},
  {Hilger}, {Hobbs}, {Hofmann}, {Holland}, {Huckle}, {Hypki}, {Icardi},
  {Jan{\ss}en}, {Jevardat de Fombelle}, {Jonker}, {Juh{\'a}sz}, {Julbe},
  {Karampelas}, {Kewley}, {Klar}, {Kochoska}, {Kohley}, {Kolenberg},
  {Kontizas}, {Kontizas}, {Koposov}, {Kordopatis}, {Kostrzewa-Rutkowska},
  {Koubsky}, {Lambert}, {Lanza}, {Lasne}, {Lavigne}, {Le Fustec}, {Le
  Poncin-Lafitte}, {Lebreton}, {Leccia}, {Leclerc}, {Lecoeur-Taibi},
  {Lenhardt}, {Leroux}, {Liao}, {Licata}, {Lindstr{\o}m}, {Lister}, {Livanou},
  {Lobel}, {L{\'o}pez}, {Managau}, {Mann}, {Mantelet}, {Marchal}, {Marchant},
  {Marconi}, {Marinoni}, {Marschalk{\'o}}, {Marshall}, {Martino}, {Marton},
  {Mary}, {Massari}, {Matijevi{\v{c}}}, {Mazeh}, {McMillan}, {Messina},
  {Michalik}, {Millar}, {Molina}, {Molinaro}, {Moln{\'a}r}, {Montegriffo},
  {Mor}, {Morbidelli}, {Morel}, {Morris}, {Mulone}, {Muraveva}, {Musella},
  {Nelemans}, {Nicastro}, {Noval}, {O'Mullane}, {Ord{\'e}novic},
  {Ord{\'o}{\~n}ez-Blanco}, {Osborne}, {Pagani}, {Pagano}, {Pailler},
  {Palacin}, {Palaversa}, {Panahi}, {Pawlak}, {Piersimoni}, {Pineau}, {Plachy},
  {Plum}, {Poggio}, {Poujoulet}, {Pr{\v{s}}a}, {Pulone}, {Racero}, {Ragaini},
  {Rambaux}, {Ramos-Lerate}, {Regibo}, {Reyl{\'e}}, {Riclet}, {Ripepi}, {Riva},
  {Rivard}, {Rixon}, {Roegiers}, {Roelens}, {Romero-G{\'o}mez}, {Rowell},
  {Royer}, {Ruiz-Dern}, {Sadowski}, {Sagrist{\`a} Sell{\'e}s}, {Sahlmann},
  {Salgado}, {Salguero}, {Sanna}, {Santana-Ros}, {Sarasso}, {Savietto},
  {Schultheis}, {Sciacca}, {Segol}, {Segovia}, {S{\'e}gransan}, {Shih},
  {Siltala}, {Silva}, {Smart}, {Smith}, {Solano}, {Solitro}, {Sordo}, {Soria
  Nieto}, {Souchay}, {Spagna}, {Spoto}, {Stampa}, {Steele},
  {Steidelm{\"u}ller}, {Stephenson}, {Stoev}, {Suess}, {Surdej}, {Szabados},
  {Szegedi-Elek}, {Tapiador}, {Taris}, {Tauran}, {Taylor}, {Teixeira},
  {Terrett}, {Teyssandier}, {Thuillot}, {Titarenko}, {Torra Clotet}, {Turon},
  {Ulla}, {Utrilla}, {Uzzi}, {Vaillant}, {Valentini}, {Valette}, {van Elteren},
  {Van Hemelryck}, {van Leeuwen}, {Vaschetto}, {Vecchiato}, {Veljanoski},
  {Viala}, {Vicente}, {Vogt}, {von Essen}, {Voss}, {Votruba}, {Voutsinas},
  {Walmsley}, {Weiler}, {Wertz}, {Wevers}, {Wyrzykowski}, {Yoldas},
  {{\v{Z}}erjal}, {Ziaeepour}, {Zorec}, {Zschocke}, {Zucker}, {Zurbach}, \&
  {Zwitter}}]{GaiaDR2_Brown2018}
{Gaia Collaboration}, {Brown}, A.~G.~A., {Vallenari}, A., {et~al.} 2018, \aap,
  616, A1, \dodoi{10.1051/0004-6361/201833051}

\bibitem[{{Gaia Collaboration} {et~al.}(2021){Gaia Collaboration}, {Brown},
  {Vallenari}, {Prusti}, {de Bruijne}, {Babusiaux}, {Biermann}, {Creevey},
  {Evans}, {Eyer}, {Hutton}, {Jansen}, {Jordi}, {Klioner}, {Lammers},
  {Lindegren}, {Luri}, {Mignard}, {Panem}, {Pourbaix}, {Randich}, {Sartoretti},
  {Soubiran}, {Walton}, {Arenou}, {Bailer-Jones}, {Bastian}, {Cropper},
  {Drimmel}, {Katz}, {Lattanzi}, {van Leeuwen}, {Bakker}, {Cacciari},
  {Casta{\~n}eda}, {De Angeli}, {Ducourant}, {Fabricius}, {Fouesneau},
  {Fr{\'e}mat}, {Guerra}, {Guerrier}, {Guiraud}, {Jean-Antoine Piccolo},
  {Masana}, {Messineo}, {Mowlavi}, {Nicolas}, {Nienartowicz}, {Pailler},
  {Panuzzo}, {Riclet}, {Roux}, {Seabroke}, {Sordo}, {Tanga}, {Th{\'e}venin},
  {Gracia-Abril}, {Portell}, {Teyssier}, {Altmann}, {Andrae}, {Bellas-Velidis},
  {Benson}, {Berthier}, {Blomme}, {Brugaletta}, {Burgess}, {Busso}, {Carry},
  {Cellino}, {Cheek}, {Clementini}, {Damerdji}, {Davidson}, {Delchambre},
  {Dell'Oro}, {Fern{\'a}ndez-Hern{\'a}ndez}, {Galluccio}, {Garc{\'\i}a-Lario},
  {Garcia-Reinaldos}, {Gonz{\'a}lez-N{\'u}{\~n}ez}, {Gosset}, {Haigron},
  {Halbwachs}, {Hambly}, {Harrison}, {Hatzidimitriou}, {Heiter},
  {Hern{\'a}ndez}, {Hestroffer}, {Hodgkin}, {Holl}, {Jan{\ss}en}, {Jevardat de
  Fombelle}, {Jordan}, {Krone-Martins}, {Lanzafame}, {L{\"o}ffler}, {Lorca},
  {Manteiga}, {Marchal}, {Marrese}, {Moitinho}, {Mora}, {Muinonen}, {Osborne},
  {Pancino}, {Pauwels}, {Petit}, {Recio-Blanco}, {Richards}, {Riello},
  {Rimoldini}, {Robin}, {Roegiers}, {Rybizki}, {Sarro}, {Siopis}, {Smith},
  {Sozzetti}, {Ulla}, {Utrilla}, {van Leeuwen}, {van Reeven}, {Abbas}, {Abreu
  Aramburu}, {Accart}, {Aerts}, {Aguado}, {Ajaj}, {Altavilla}, {{\'A}lvarez},
  {{\'A}lvarez Cid-Fuentes}, {Alves}, {Anderson}, {Anglada Varela}, {Antoja},
  {Audard}, {Baines}, {Baker}, {Balaguer-N{\'u}{\~n}ez}, {Balbinot}, {Balog},
  {Barache}, {Barbato}, {Barros}, {Barstow}, {Bartolom{\'e}}, {Bassilana},
  {Bauchet}, {Baudesson-Stella}, {Becciani}, {Bellazzini}, {Bernet}, {Bertone},
  {Bianchi}, {Blanco-Cuaresma}, {Boch}, {Bombrun}, {Bossini}, {Bouquillon},
  {Bragaglia}, {Bramante}, {Breedt}, {Bressan}, {Brouillet}, {Bucciarelli},
  {Burlacu}, {Busonero}, {Butkevich}, {Buzzi}, {Caffau}, {Cancelliere},
  {C{\'a}novas}, {Cantat-Gaudin}, {Carballo}, {Carlucci}, {Carnerero},
  {Carrasco}, {Casamiquela}, {Castellani}, {Castro-Ginard}, {Castro Sampol},
  {Chaoul}, {Charlot}, {Chemin}, {Chiavassa}, {Cioni}, {Comoretto}, {Cooper},
  {Cornez}, {Cowell}, {Crifo}, {Crosta}, {Crowley}, {Dafonte}, {Dapergolas},
  {David}, {David}, {de Laverny}, {De Luise}, {De March}, {De Ridder}, {de
  Souza}, {de Teodoro}, {de Torres}, {del Peloso}, {del Pozo}, {Delbo},
  {Delgado}, {Delgado}, {Delisle}, {Di Matteo}, {Diakite}, {Diener},
  {Distefano}, {Dolding}, {Eappachen}, {Edvardsson}, {Enke}, {Esquej}, {Fabre},
  {Fabrizio}, {Faigler}, {Fedorets}, {Fernique}, {Fienga}, {Figueras},
  {Fouron}, {Fragkoudi}, {Fraile}, {Franke}, {Gai}, {Garabato},
  {Garcia-Gutierrez}, {Garc{\'\i}a-Torres}, {Garofalo}, {Gavras}, {Gerlach},
  {Geyer}, {Giacobbe}, {Gilmore}, {Girona}, {Giuffrida}, {Gomel}, {Gomez},
  {Gonzalez-Santamaria}, {Gonz{\'a}lez-Vidal}, {Granvik},
  {Guti{\'e}rrez-S{\'a}nchez}, {Guy}, {Hauser}, {Haywood}, {Helmi}, {Hidalgo},
  {Hilger}, {H{\l}adczuk}, {Hobbs}, {Holland}, {Huckle}, {Jasniewicz},
  {Jonker}, {Juaristi Campillo}, {Julbe}, {Karbevska}, {Kervella}, {Khanna},
  {Kochoska}, {Kontizas}, {Kordopatis}, {Korn}, {Kostrzewa-Rutkowska},
  {Kruszy{\'n}ska}, {Lambert}, {Lanza}, {Lasne}, {Le Campion}, {Le Fustec},
  {Lebreton}, {Lebzelter}, {Leccia}, {Leclerc}, {Lecoeur-Taibi}, {Liao},
  {Licata}, {Lindstr{\o}m}, {Lister}, {Livanou}, {Lobel}, {Madrero Pardo},
  {Managau}, {Mann}, {Marchant}, {Marconi}, {Marcos Santos}, {Marinoni},
  {Marocco}, {Marshall}, {Martin Polo}, {Mart{\'\i}n-Fleitas}, {Masip},
  {Massari}, {Mastrobuono-Battisti}, {Mazeh}, {McMillan}, {Messina},
  {Michalik}, {Millar}, {Mints}, {Molina}, {Molinaro}, {Moln{\'a}r},
  {Montegriffo}, {Mor}, {Morbidelli}, {Morel}, {Morris}, {Mulone}, {Munoz},
  {Muraveva}, {Murphy}, {Musella}, {Noval}, {Ord{\'e}novic}, {Orr{\`u}},
  {Osinde}, {Pagani}, {Pagano}, {Palaversa}, {Palicio}, {Panahi}, {Pawlak},
  {Pe{\~n}alosa Esteller}, {Penttil{\"a}}, {Piersimoni}, {Pineau}, {Plachy},
  {Plum}, {Poggio}, {Poretti}, {Poujoulet}, {Pr{\v{s}}a}, {Pulone}, {Racero},
  {Ragaini}, {Rainer}, {Raiteri}, {Rambaux}, {Ramos}, {Ramos-Lerate}, {Re
  Fiorentin}, {Regibo}, {Reyl{\'e}}, {Ripepi}, {Riva}, {Rixon}, {Robichon},
  {Robin}, {Roelens}, {Rohrbasser}, {Romero-G{\'o}mez}, {Rowell}, {Royer},
  {Rybicki}, {Sadowski}, {Sagrist{\`a} Sell{\'e}s}, {Sahlmann}, {Salgado},
  {Salguero}, {Samaras}, {Sanchez Gimenez}, {Sanna}, {Santove{\~n}a},
  {Sarasso}, {Schultheis}, {Sciacca}, {Segol}, {Segovia}, {S{\'e}gransan},
  {Semeux}, {Shahaf}, {Siddiqui}, {Siebert}, {Siltala}, {Slezak}, {Smart},
  {Solano}, {Solitro}, {Souami}, {Souchay}, {Spagna}, {Spoto}, {Steele},
  {Steidelm{\"u}ller}, {Stephenson}, {S{\"u}veges}, {Szabados}, {Szegedi-Elek},
  {Taris}, {Tauran}, {Taylor}, {Teixeira}, {Thuillot}, {Tonello}, {Torra},
  {Torra}, {Turon}, {Unger}, {Vaillant}, {van Dillen}, {Vanel}, {Vecchiato},
  {Viala}, {Vicente}, {Voutsinas}, {Weiler}, {Wevers}, {Wyrzykowski}, {Yoldas},
  {Yvard}, {Zhao}, {Zorec}, {Zucker}, {Zurbach}, \&
  {Zwitter}}]{GaiaDR3_Brown2021}
---. 2021, \aap, 650, C3, \dodoi{10.1051/0004-6361/202039657e}

\bibitem[{{Gaia Collaboration} {et~al.}(2022){Gaia Collaboration},
  Recio-Blanco, Kordopatis, de~Laverny, Palicio, Spagna, Spina, Katz,
  Fiorentin, Poggio, McMillan, Vallenari, Lattanzi, Seabroke, Casamiquela,
  Bragaglia, Antoja, Bailer-Jones, Andrae, Fouesneau, Cropper, Cantat-Gaudin,
  Heiter, Bijaoui, Brown, Prusti, de~Bruijne, Arenou, Babusiaux, Biermann,
  Creevey, Ducourant, Evans, Eyer, Guerra, Hutton, Jordi, Klioner, Lammers,
  Lindegren, Luri, Mignard, Panem, Pourbaix, Randich, Sartoretti, Soubiran,
  Tanga, Walton, Bastian, Drimmel, Jansen, van Leeuwen, Bakker, Cacciari,
  Castañeda, Angeli, Fabricius, Frémat, Galluccio, Guerrier, Masana,
  Messineo, Mowlavi, Nicolas, Nienartowicz, Pailler, Panuzzo, Riclet, Roux,
  Sordo, Thévenin, Gracia-Abril, Portell, Teyssier, Altmann, Audard,
  Bellas-Velidis, Benson, Berthier, Blomme, Burgess, Busonero, Busso, Cánovas,
  Carry, Cellino, Cheek, Clementini, Damerdji, Davidson, de~Teodoro, Campos,
  Delchambre, Dell'Oro, Esquej, Fernández-Hernández, Fraile, Garabato,
  García-Lario, Gosset, Haigron, Halbwachs, Hambly, Harrison, Hernández,
  Hestroffer, Hodgkin, Holl, Janßen, de~Fombelle, Jordan, Krone-Martins,
  Lanzafame, Löffler, Marchal, Marrese, Moitinho, Muinonen, Osborne, Pancino,
  Pauwels, Reylé, Riello, Rimoldini, Roegiers, Rybizki, Sarro, Siopis, Smith,
  Sozzetti, Utrilla, van Leeuwen, Abbas, Ábrahám, Aramburu, Aerts, Aguado,
  Ajaj, Aldea-Montero, Altavilla, Álvarez, Alves, Anders, Anderson, Varela,
  Baines, Baker, Balaguer-Núñez, Balbinot, Balog, Barache, Barbato, Barros,
  Barstow, Bartolomé, Bassilana, Bauchet, Becciani, Bellazzini, Berihuete,
  Bernet, Bertone, Bianchi, Binnenfeld, Blanco-Cuaresma, Boch, Bombrun,
  Bossini, Bouquillon, Bramante, Breedt, Bressan, Brouillet, Brugaletta,
  Bucciarelli, Burlacu, Butkevich, Buzzi, Caffau, Cancelliere, Carballo,
  Carlucci, Carnerero, Carrasco, Castellani, Castro-Ginard, Chaoul, Charlot,
  Chemin, Chiaramida, Chiavassa, Chornay, Comoretto, Contursi, Cooper, Cornez,
  Cowell, Crifo, Crosta, Crowley, Dafonte, Dapergolas, David, Luise, March,
  Ridder, de~Souza, de~Torres, del Peloso, del Pozo, Delbo, Delgado, Delisle,
  Demouchy, Dharmawardena, Matteo, Diakite, Diener, Distefano, Dolding,
  Edvardsson, Enke, Fabre, Fabrizio, Faigler, Fedorets, Fernique, Figueras,
  Fournier, Fouron, Fragkoudi, Gai, Garcia-Gutierrez, Garcia-Reinaldos,
  García-Torres, Garofalo, Gavel, Gavras, Gerlach, Geyer, Giacobbe, Gilmore,
  Girona, Giuffrida, Gomel, Gomez, González-Núñez, González-Santamaría,
  González-Vidal, Granvik, Guillout, Guiraud, Gutiérrez-Sánchez, Guy,
  Hatzidimitriou, Hauser, Haywood, Helmer, Helmi, Sarmiento, Hidalgo,
  Hładczuk, Hobbs, Holland, Huckle, Jardine, Jasniewicz, Piccolo,
  Jiménez-Arranz, Campillo, Julbe, Karbevska, Kervella, Khanna, Korn,
  Kóspál, Kostrzewa-Rutkowska, Kruszyńska, Kun, Laizeau, Lambert, Lanza,
  Lasne, Campion, Lebreton, Lebzelter, Leccia, Leclerc, Lecoeur-Taibi, Liao,
  Licata, Lindstrøm, Lister, Livanou, Lobel, Lorca, Loup, Pardo, Romeo,
  Managau, Mann, Manteiga, Marchant, Marconi, Marcos, Santos, Pina, Marinoni,
  Marocco, Marshall, Polo, Martín-Fleitas, Marton, Mary, Masip, Massari,
  Mastrobuono-Battisti, Mazeh, Messina, Michalik, Millar, Mints, Molina,
  Molinaro, Molnár, Monari, Monguió, Montegriffo, Montero, Mor, Mora,
  Morbidelli, Morel, Morris, Muraveva, Murphy, Musella, Nagy, Noval, Ocaña,
  Ogden, Ordenovic, Osinde, Pagani, Pagano, Palaversa, Pallas-Quintela, Panahi,
  Payne-Wardenaar, Esteller, Penttilä, Pichon, Piersimoni, Pineau, Plachy,
  Plum, Prša, Pulone, Racero, Ragaini, Rainer, Raiteri, Ramos, Ramos-Lerate,
  Regibo, Richards, Diaz, Ripepi, Riva, Rix, Rixon, Robichon, Robin, Robin,
  Roelens, Rogues, Rohrbasser, Romero-Gómez, Rowell, Royer, Mieres, Rybicki,
  Sadowski, Núñez, Sellés, Sahlmann, Salguero, Samaras, Gimenez, Sanna,
  Santoveña, Sarasso, Schultheis, Sciacca, Segol, Segovia, Ségransan, Semeux,
  Shahaf, Siddiqui, Siebert, Siltala, Silvelo, Slezak, Slezak, Smart, Snaith,
  Solano, Solitro, Souami, Souchay, Spoto, Steele, Steidelmüller, Stephenson,
  Süveges, Surdej, Szabados, Szegedi-Elek, Taris, Taylor, Teixeira, Tolomei,
  Tonello, Torra, Torra, Elipe, Trabucchi, Tsounis, Turon, Ulla, Unger,
  Vaillant, van Dillen, van Reeven, Vanel, Vecchiato, Viala, Vicente,
  Voutsinas, Weiler, Wevers, Wyrzykowski, Yoldas, Yvard, Zhao, Zorec, Zucker,
  \& Zwitter}]{gaiamaps2022}
{Gaia Collaboration}, Recio-Blanco, A., Kordopatis, G., {et~al.} 2022, Gaia
  Data Release 3: Chemical cartography of the Milky Way.
\newblock \doarXiv{2206.05534}

\bibitem[{{Garc{\'\i}a P{\'e}rez} {et~al.}(2016){Garc{\'\i}a P{\'e}rez},
  {Allende Prieto}, {Holtzman}, {Shetrone}, {M{\'e}sz{\'a}ros}, {Bizyaev},
  {Carrera}, {Cunha}, {Garc{\'\i}a-Hern{\'a}ndez}, {Johnson}, {Majewski},
  {Nidever}, {Schiavon}, {Shane}, {Smith}, {Sobeck}, {Troup}, {Zamora},
  {Weinberg}, {Bovy}, {Eisenstein}, {Feuillet}, {Frinchaboy}, {Hayden},
  {Hearty}, {Nguyen}, {O'Connell}, {Pinsonneault}, {Wilson}, \&
  {Zasowski}}]{aspcap}
{Garc{\'\i}a P{\'e}rez}, A.~E., {Allende Prieto}, C., {Holtzman}, J.~A.,
  {et~al.} 2016, \aj, 151, 144, \dodoi{10.3847/0004-6256/151/6/144}

\bibitem[{Gebek \& Matthee(2022)}]{Gebek_2022}
Gebek, A., \& Matthee, J. 2022, The Astrophysical Journal, 924, 73,
  \dodoi{10.3847/1538-4357/ac350b}

\bibitem[{{Genovali} {et~al.}(2014){Genovali}, {Lemasle}, {Bono}, {Romaniello},
  {Fabrizio}, {Ferraro}, {Iannicola}, {Laney}, {Nonino}, {Bergemann},
  {Buonanno}, {Fran{\c{c}}ois}, {Inno}, {Kudritzki}, {Matsunaga}, {Pedicelli},
  {Primas}, \& {Th{\'e}venin}}]{Genovali2014}
{Genovali}, K., {Lemasle}, B., {Bono}, G., {et~al.} 2014, \aap, 566, A37,
  \dodoi{10.1051/0004-6361/201323198}

\bibitem[{Gent {et~al.}(2022)Gent, Eitner, Laporte, Serenelli, Koposov, \&
  Bergemann}]{Gent2022}
Gent, M.~R., Eitner, P., Laporte, C. F.~P., {et~al.} 2022, The Prince and The
  Pauper: Co-evolution of the thin and thick disc in the Milky Way,  arXiv,
  \dodoi{10.48550/ARXIV.2206.10949}

\bibitem[{Gibson {et~al.}(2013)Gibson, Pilkington, Brook, Stinson, \&
  Bailin}]{Gibson_2013}
Gibson, B.~K., Pilkington, K., Brook, C.~B., Stinson, G.~S., \& Bailin, J.
  2013, Astronomy {$\&$} Astrophysics, 554, A47,
  \dodoi{10.1051/0004-6361/201321239}

\bibitem[{{Gilmore} \& {Reid}(1983)}]{Gilmore1983}
{Gilmore}, G., \& {Reid}, N. 1983, \mnras, 202, 1025,
  \dodoi{10.1093/mnras/202.4.1025}

\bibitem[{Gilmore {et~al.}(2022)Gilmore, Randich, Worley, Hourihane, Gonneau,
  Sacco, Lewis, Magrini, Francois, Jeffries, Koposov, Bragaglia, Alfaro,
  Prieto, Blomme, Korn, Lanzafame, Pancino, Recio-Blanco, Smiljanic, Van~Eck,
  Zwitter, Bensby, Flaccomio, Irwin, Franciosini, Morbidelli, Damiani, Bonito,
  Friel, Vink, Prisinzano, Abbas, Hatzidimitriou, Held, Jordi, Paunzen, Spagna,
  Jackson, Apellaniz, Asplund, Bonifacio, Feltzing, Binney, Drew, Ferguson,
  Micela, Negueruela, Prusti, Rix, Vallenari, Bergemann, Casey, de~Laverny,
  Frasca, Hill, Lind, Sbordone, Sousa, Adibekyan, Caffau, Daflon, Feuillet,
  Gebran, Hernandez, Guiglion, Herrero, Lobel, Merle, Mikolaitis, Montes,
  Morel, Ruchti, Soubiran, Tabernero, Tautvaisiene, Traven, Valentini, Van~der
  Swaelmen, Villanova, Vazquez, Bayo, Biazzo, Carraro, Edvardsson, Heiter,
  Jofre, Marconi, Martayan, Masseron, Monaco, Walton, Zaggia, Borsen-Koch,
  Alves, Balaguer-Nunez, Barklem, Barrado, Bellazzini, Berlanas, Binks,
  Bressan, Capuzzo-Dolcetta, Casagrande, Casamiquela, Collins, D'Orazi, Dantas,
  Debattista, Delgado-Mena, Di~Marcantonio, Drazdauskas, Evans, Famaey,
  Franchini, Fremat, Fu, Geisler, Gerhard, Solares, Grebel, Albarran,
  Jimenez-Esteban, Jonsson, Khachaturyants, Kordopatis, Kos, Lagarde, Ludwig,
  Mahy, Mapelli, Marfil, Martell, Messina, Miglio, Minchev, Moitinho,
  Montalban, Monteiro, Morossi, Mowlavi, Mucciarelli, Murphy, Nardetto,
  Ortolani, Paletou, Palous, Pickering, Quirrenbach, Fiorentin, Read, Romano,
  Ryde, Sanna, Santos, Seabroke, Spina, Steinmetz, Stonkute, Sutorius,
  Thevenin, Tosi, Tsantaki, Wright, Wyse, Zoccali, Zorec, \&
  Zucker}]{Gilmore_GaiaESO}
Gilmore, G., Randich, S., Worley, C.~C., {et~al.} 2022, The Gaia-ESO Public
  Spectroscopic Survey: Motivation, implementation, GIRAFFE data processing,
  analysis, and final data products,  arXiv, \dodoi{10.48550/ARXIV.2208.05432}

\bibitem[{{G{\'o}mez} {et~al.}(2012){G{\'o}mez}, {Minchev}, {O'Shea}, {Lee},
  {Beers}, {An}, {Bullock}, {Purcell}, \& {Villalobos}}]{Gomez_2012}
{G{\'o}mez}, F.~A., {Minchev}, I., {O'Shea}, B.~W., {et~al.} 2012, \mnras, 423,
  3727, \dodoi{10.1111/j.1365-2966.2012.21176.x}

\bibitem[{Grieves {et~al.}(2018)Grieves, Ge, Thomas, Willis, Ma,
  Lorenzo-Oliveira, Queiroz, Ghezzi, Chiappini, Anders, Dutra-Ferreira,
  de~Mello, Santiago, da~Costa, Ogando, del Peloso, Tan, Schneider, Pepper,
  Stassun, Zhao, Bizyaev, \& Pan}]{Grieves_2018}
Grieves, N., Ge, J., Thomas, N., {et~al.} 2018, Monthly Notices of the Royal
  Astronomical Society, 481, 3244, \dodoi{10.1093/mnras/sty2431}

\bibitem[{{Gunn} {et~al.}(2006){Gunn}, {Siegmund}, {Mannery}, {Owen}, {Hull},
  {Leger}, {Carey}, {Knapp}, {York}, {Boroski}, {Kent}, {Lupton}, {Rockosi},
  {Evans}, {Waddell}, {Anderson}, {Annis}, {Barentine}, {Bartoszek}, {Bastian},
  {Bracker}, {Brewington}, {Briegel}, {Brinkmann}, {Brown}, {Carr},
  {Czarapata}, {Drennan}, {Dombeck}, {Federwitz}, {Gillespie}, {Gonzales},
  {Hansen}, {Harvanek}, {Hayes}, {Jordan}, {Kinney}, {Klaene}, {Kleinman},
  {Kron}, {Kresinski}, {Lee}, {Limmongkol}, {Lindenmeyer}, {Long}, {Loomis},
  {McGehee}, {Mantsch}, {Neilsen}, {Neswold}, {Newman}, {Nitta}, {Peoples},
  {Pier}, {Prieto}, {Prosapio}, {Rivetta}, {Schneider}, {Snedden}, \&
  {Wang}}]{apo25m_Gunn2006}
{Gunn}, J.~E., {Siegmund}, W.~A., {Mannery}, E.~J., {et~al.} 2006, \aj, 131,
  2332, \dodoi{10.1086/500975}

\bibitem[{Guo {et~al.}(2015)Guo, Ferguson, Bell, Koo, Conselice, Giavalisco,
  Kassin, Lu, Lucas, Mandelker, McIntosh, Primack, Ravindranath, Barro,
  Ceverino, Dekel, Faber, Fang, Koekemoer, Noeske, Rafelski, \&
  Straughn}]{Guo_2015}
Guo, Y., Ferguson, H.~C., Bell, E.~F., {et~al.} 2015, The Astrophysical
  Journal, 800, 39, \dodoi{10.1088/0004-637x/800/1/39}

\bibitem[{Halle {et~al.}(2015)Halle, Matteo, Haywood, \& Combes}]{Halle_2015}
Halle, A., Matteo, P.~D., Haywood, M., \& Combes, F. 2015, Astronomy {$\&$}
  Astrophysics, 578, A58, \dodoi{10.1051/0004-6361/201525612}

\bibitem[{{Hartkopf} \& {Yoss}(1982)}]{Hartkopf1982}
{Hartkopf}, W.~I., \& {Yoss}, K.~M. 1982, \aj, 87, 1679, \dodoi{10.1086/113261}

\bibitem[{Hasselquist {et~al.}(2019)Hasselquist, Holtzman, Shetrone, Tayar,
  Weinberg, Feuillet, Cunha, Pinsonneault, Johnson, Bird, Beers, Schiavon,
  Minchev, Fern{\'{a} }ndez-Trincado, Garc{\'{\i}}a-Hern{\'{a}}ndez, Nitschelm,
  \& Zamora}]{Hasselquist_2019}
Hasselquist, S., Holtzman, J.~A., Shetrone, M., {et~al.} 2019, The
  Astrophysical Journal, 871, 181, \dodoi{10.3847/1538-4357/aaf859}

\bibitem[{Hasselquist {et~al.}(2020)Hasselquist, Zasowski, Feuillet,
  Schultheis, Nataf, Anguiano, Beaton, Beers, Cohen, Cunha,
  Fern{\'{a}}ndez-Trincado, Garc{\'{\i}}a-Hern{\'{a}}ndez, Geisler, Holtzman,
  Johnson, Lane, Majewski, Bidin, Nitschelm, Roman-Lopes, Schiavon, Smith, \&
  Sobeck}]{Hasselquist_2020}
Hasselquist, S., Zasowski, G., Feuillet, D.~K., {et~al.} 2020, The
  Astrophysical Journal, 901, 109, \dodoi{10.3847/1538-4357/abaeee}

\bibitem[{Hawkins(2022)}]{Hawkins_2022}
Hawkins, K. 2022, Chemical Cartography with LAMOST and Gaia Reveal Azimuthal
  and Spiral Structure in the Galactic Disk,  arXiv,
  \dodoi{10.48550/ARXIV.2207.04542}

\bibitem[{Hawkins {et~al.}(2015)Hawkins, Jofr{\'{e} }, Masseron, \&
  Gilmore}]{Hawkins_2015}
Hawkins, K., Jofr{\'{e} }, P., Masseron, T., \& Gilmore, G. 2015, Monthly
  Notices of the Royal Astronomical Society, 453, 758,
  \dodoi{10.1093/mnras/stv1586}

\bibitem[{Hayden {et~al.}(2017)Hayden, Recio-Blanco, de~Laverny, Mikolaitis, \&
  Worley}]{Hayden_2017}
Hayden, M.~R., Recio-Blanco, A., de~Laverny, P., Mikolaitis, S., \& Worley,
  C.~C. 2017, Astronomy {$\&$} Astrophysics, 608, L1,
  \dodoi{10.1051/0004-6361/201731494}

\bibitem[{{Hayden} {et~al.}(2014){Hayden}, {Holtzman}, {Bovy}, {Majewski},
  {Johnson}, {Allende Prieto}, {Beers}, {Cunha}, {Frinchaboy}, {Garc{\'\i}a
  P{\'e}rez}, {Girardi}, {Hearty}, {Lee}, {Nidever}, {Schiavon}, {Schlesinger},
  {Schneider}, {Schultheis}, {Shetrone}, {Smith}, {Zasowski}, {Bizyaev},
  {Feuillet}, {Hasselquist}, {Kinemuchi}, {Malanushenko}, {Malanushenko},
  {O'Connell}, {Pan}, \& {Stassun}}]{Hayden2014}
{Hayden}, M.~R., {Holtzman}, J.~A., {Bovy}, J., {et~al.} 2014, \aj, 147, 116,
  \dodoi{10.1088/0004-6256/147/5/116}

\bibitem[{Hayden {et~al.}(2015)Hayden, Bovy, Holtzman, Nidever, Bird, Weinberg,
  Andrews, Majewski, Prieto, Anders, Beers, Bizyaev, Chiappini, Cunha,
  Frinchaboy, García-Herńandez, García~Pérez, Girardi, Harding, Hearty,
  Johnson, Mészáros, Minchev, O’Connell, Pan, Robin, Schiavon, Schneider,
  Schultheis, Shetrone, Skrutskie, Steinmetz, Smith, Wilson, Zamora, \&
  Zasowski}]{Hayden2015}
Hayden, M.~R., Bovy, J., Holtzman, J.~A., {et~al.} 2015, The Astrophysical
  Journal, 808, 132, \dodoi{10.1088/0004-637x/808/2/132}

\bibitem[{Hayden {et~al.}(2020)Hayden, Sharma, Bland-Hawthorn, Spina, Buder,
  Asplund, Casey, De~Silva, D'Orazi, Freeman, Kos, Lewis, Lin, Lind, Martell,
  Schlesinger, Simpson, Zucker, Zwitter, Chen, Cotar, Feuillet, Horner, Joyce,
  Nordlander, Stello, Tepper-Garcia, Ting, Wang, \& Wittenmyer}]{Hayden_2020}
Hayden, M.~R., Sharma, S., Bland-Hawthorn, J., {et~al.} 2020, The GALAH Survey:
  Chemical Clocks,  arXiv, \dodoi{10.48550/ARXIV.2011.13745}

\bibitem[{Haywood {et~al.}(2016)Haywood, Lehnert, Matteo, Snaith, Schultheis,
  Katz, \& G{\'{o} }mez}]{Haywood_2016}
Haywood, M., Lehnert, M.~D., Matteo, P.~D., {et~al.} 2016, Astronomy {$\&$}
  Astrophysics, 589, A66, \dodoi{10.1051/0004-6361/201527567}

\bibitem[{Haywood {et~al.}(2018)Haywood, Matteo, Lehnert, Snaith, Fragkoudi, \&
  Khoperskov}]{Haywood_2018}
Haywood, M., Matteo, P.~D., Lehnert, M., {et~al.} 2018, Astronomy {$\&$}
  Astrophysics, 618, A78, \dodoi{10.1051/0004-6361/201731363}

\bibitem[{Haywood {et~al.}(2013)Haywood, Matteo, Lehnert, Katz, \& G{\'{o}
  }mez}]{Haywood_2013}
Haywood, M., Matteo, P.~D., Lehnert, M.~D., Katz, D., \& G{\'{o} }mez, A. 2013,
  Astronomy {$\&$} Astrophysics, 560, A109, \dodoi{10.1051/0004-6361/201321397}

\bibitem[{Haywood {et~al.}(2019)Haywood, Snaith, Lehnert, Matteo, \&
  Khoperskov}]{Haywood_2019}
Haywood, M., Snaith, O., Lehnert, M.~D., Matteo, P.~D., \& Khoperskov, S. 2019,
  Astronomy {$\&$} Astrophysics, 625, A105, \dodoi{10.1051/0004-6361/201834155}

\bibitem[{Helmi {et~al.}(2018)Helmi, Babusiaux, Koppelman, Massari, Veljanoski,
  \& Brown}]{Helmi_2018}
Helmi, A., Babusiaux, C., Koppelman, H.~H., {et~al.} 2018, Nature, 563, 85,
  \dodoi{10.1038/s41586-018-0625-x}

\bibitem[{{Holtzman} {et~al.}(2015){Holtzman}, {Shetrone}, {Johnson}, {Allende
  Prieto}, {Anders}, {Andrews}, {Beers}, {Bizyaev}, {Blanton}, {Bovy},
  {Carrera}, {Chojnowski}, {Cunha}, {Eisenstein}, {Feuillet}, {Frinchaboy},
  {Galbraith-Frew}, {Garc{\'\i}a P{\'e}rez}, {Garc{\'\i}a-Hern{\'a}ndez},
  {Hasselquist}, {Hayden}, {Hearty}, {Ivans}, {Majewski}, {Martell},
  {Meszaros}, {Muna}, {Nidever}, {Nguyen}, {O'Connell}, {Pan}, {Pinsonneault},
  {Robin}, {Schiavon}, {Shane}, {Sobeck}, {Smith}, {Troup}, {Weinberg},
  {Wilson}, {Wood-Vasey}, {Zamora}, \& {Zasowski}}]{Holtzman_2015}
{Holtzman}, J.~A., {Shetrone}, M., {Johnson}, J.~A., {et~al.} 2015, \aj, 150,
  148, \dodoi{10.1088/0004-6256/150/5/148}

\bibitem[{Holtzman {et~al.}(2018)Holtzman, Hasselquist, Shetrone, Cunha,
  Prieto, Anguiano, Bizyaev, Bovy, Casey, Edvardsson, Johnson, Jönsson,
  Meszaros, Smith, Sobeck, Zamora, Chojnowski, Fernandez-Trincado, Hernandez,
  Majewski, Pinsonneault, Souto, Stringfellow, Tayar, Troup, \&
  Zasowski}]{Holtzman_2018}
Holtzman, J.~A., Hasselquist, S., Shetrone, M., {et~al.} 2018, The Astronomical
  Journal, 156, 125, \dodoi{10.3847/1538-3881/aad4f9}

\bibitem[{Huang {et~al.}(2020)Huang, Schönrich, Zhang, Wu, Chen, Wang, Xiang,
  Wang, Yuan, Li, Sun, Li, \& Liu}]{Huang_2020}
Huang, Y., Schönrich, R., Zhang, H., {et~al.} 2020, The Astrophysical Journal
  Supplement Series, 249, 29, \dodoi{10.3847/1538-4365/ab994f}

\bibitem[{Hänninen \& Flynn(2002)}]{Hanninen_2002}
Hänninen, J., \& Flynn, C. 2002, Monthly Notices of the Royal Astronomical
  Society, 337, 731, \dodoi{10.1046/j.1365-8711.2002.05956.x}

\bibitem[{{Iben}(1965)}]{Iben_1965}
{Iben}, Icko, J. 1965, \apj, 142, 1447, \dodoi{10.1086/148429}

\bibitem[{{Inno} {et~al.}(2019){Inno}, {Urbaneja}, {Matsunaga}, {Bono},
  {Nonino}, {Debattista}, {Sormani}, {Bergemann}, {da Silva}, {Lemasle},
  {Romaniello}, \& {Rix}}]{Inno2019}
{Inno}, L., {Urbaneja}, M.~A., {Matsunaga}, N., {et~al.} 2019, \mnras, 482, 83,
  \dodoi{10.1093/mnras/sty2661}

\bibitem[{Isern(2019)}]{Isern_2019}
Isern, J. 2019, The Astrophysical Journal, 878, L11,
  \dodoi{10.3847/2041-8213/ab238e}

\bibitem[{Johnson {et~al.}(2021)Johnson, Weinberg, Vincenzo, Bird, Loebman,
  Brooks, Quinn, Christensen, \& Griffith}]{Johnson_2021}
Johnson, J.~W., Weinberg, D.~H., Vincenzo, F., {et~al.} 2021, Monthly Notices
  of the Royal Astronomical Society, 508, 4484, \dodoi{10.1093/mnras/stab2718}

\bibitem[{{Juri{\'c}} {et~al.}(2008){Juri{\'c}}, {Ivezi{\'c}}, {Brooks},
  {Lupton}, {Schlegel}, {Finkbeiner}, {Padmanabhan}, {Bond}, {Sesar},
  {Rockosi}, {Knapp}, {Gunn}, {Sumi}, {Schneider}, {Barentine}, {Brewington},
  {Brinkmann}, {Fukugita}, {Harvanek}, {Kleinman}, {Krzesinski}, {Long},
  {Neilsen}, {Nitta}, {Snedden}, \& {York}}]{Juric2008}
{Juri{\'c}}, M., {Ivezi{\'c}}, {\v{Z}}., {Brooks}, A., {et~al.} 2008, \apj,
  673, 864, \dodoi{10.1086/523619}

\bibitem[{Jönsson {et~al.}(2018)Jönsson, Prieto, Holtzman, Feuillet, Hawkins,
  Cunha, Mészáros, Hasselquist, Fernández-Trincado, García-Hernández, \&
  et~al.}]{aspcap2018}
Jönsson, H., Prieto, C.~A., Holtzman, J.~A., {et~al.} 2018, The Astronomical
  Journal, 156, 126, \dodoi{10.3847/1538-3881/aad4f5}

\bibitem[{Jönsson {et~al.}(2020)Jönsson, Holtzman, Prieto, Cunha, Garc{\'{\i}
  }a-Hern{\'{a}}ndez, Hasselquist, Masseron, Osorio, Shetrone, Smith,
  Stringfellow, Bizyaev, Edvardsson, Majewski, M{\'{e}}sz{\'{a}}ros, Souto,
  Zamora, Beaton, Bovy, Donor, Pinsonneault, Poovelil, \&
  Sobeck}]{Jonsson_2020}
Jönsson, H., Holtzman, J.~A., Prieto, C.~A., {et~al.} 2020, The Astronomical
  Journal, 160, 120, \dodoi{10.3847/1538-3881/aba592}

\bibitem[{Katz {et~al.}(2021)Katz, G{\'{o} }mez, Haywood, Snaith, \&
  Matteo}]{Katz_2021}
Katz, D., G{\'{o} }mez, A., Haywood, M., Snaith, O., \& Matteo, P.~D. 2021,
  Astronomy {$\&$} Astrophysics, 655, A111, \dodoi{10.1051/0004-6361/202140453}

\bibitem[{Kawata \& Chiappini(2016)}]{Kawata_2016}
Kawata, D., \& Chiappini, C. 2016, Astronomische Nachrichten, 337, 976,
  \dodoi{10.1002/asna.201612421}

\bibitem[{Khoperskov {et~al.}(2020{\natexlab{a}})Khoperskov, Gerhard, Matteo,
  Haywood, Katz, Khrapov, Khoperskov, \& Arnaboldi}]{Khoperskov_2020}
Khoperskov, S., Gerhard, O., Matteo, P.~D., {et~al.} 2020{\natexlab{a}},
  Astronomy {$\&$} Astrophysics, 634, L8, \dodoi{10.1051/0004-6361/201936645}

\bibitem[{{Khoperskov} {et~al.}(2021){Khoperskov}, {Haywood}, {Snaith}, {Di
  Matteo}, {Lehnert}, {Vasiliev}, {Naroenkov}, \& {Berczik}}]{Khoperskov_2021}
{Khoperskov}, S., {Haywood}, M., {Snaith}, O., {et~al.} 2021, \mnras, 501,
  5176, \dodoi{10.1093/mnras/staa3996}

\bibitem[{Khoperskov {et~al.}(2020{\natexlab{b}})Khoperskov, Matteo, Haywood,
  G{\'{o} }mez, \& Snaith}]{Khoperskov_2020b}
Khoperskov, S., Matteo, P.~D., Haywood, M., G{\'{o} }mez, A., \& Snaith, O.~N.
  2020{\natexlab{b}}, Astronomy {$\&$} Astrophysics, 638, A144,
  \dodoi{10.1051/0004-6361/201937188}

\bibitem[{{Kobayashi} \& {Nakasato}(2011)}]{Kobayashi2011}
{Kobayashi}, C., \& {Nakasato}, N. 2011, \apj, 729, 16,
  \dodoi{10.1088/0004-637X/729/1/16}

\bibitem[{{Kordopatis} {et~al.}(2013){Kordopatis}, {Gilmore}, {Wyse},
  {Steinmetz}, {Siebert}, {Bienaym{\'e}}, {McMillan}, {Minchev}, {Zwitter},
  {Gibson}, {Seabroke}, {Grebel}, {Bland-Hawthorn}, {Boeche}, {Freeman},
  {Munari}, {Navarro}, {Parker}, {Reid}, \& {Siviero}}]{Kordopatis2013}
{Kordopatis}, G., {Gilmore}, G., {Wyse}, R.~F.~G., {et~al.} 2013, \mnras, 436,
  3231, \dodoi{10.1093/mnras/stt1804}

\bibitem[{Kordopatis {et~al.}(2015{\natexlab{a}})Kordopatis, Wyse, Gilmore,
  Recio-Blanco, de~Laverny, Hill, Adibekyan, Heiter, Minchev, Famaey, Bensby,
  Feltzing, Guiglion, Korn, Mikolaitis, Schultheis, Vallenari, Bayo, Carraro,
  Flaccomio, Franciosini, Hourihane, Jofr{\'{e}}, Koposov, Lardo, Lewis, Lind,
  Magrini, Morbidelli, Pancino, Randich, Sacco, Worley, \&
  Zaggia}]{Kordopatis_2015_GaiaESO}
Kordopatis, G., Wyse, R. F.~G., Gilmore, G., {et~al.} 2015{\natexlab{a}},
  Astronomy {\&} Astrophysics, 582, A122, \dodoi{10.1051/0004-6361/201526258}

\bibitem[{Kordopatis {et~al.}(2015{\natexlab{b}})Kordopatis, Binney, Gilmore,
  Wyse, Belokurov, McMillan, Hatfield, Grebel, Steinmetz, Navarro, Seabroke,
  Minchev, Chiappini, Bienaym{\'{e} }, Bland-Hawthorn, Freeman, Gibson, Helmi,
  Munari, Parker, Reid, Siebert, Siviero, \& Zwitter}]{Kordopatis_2015}
Kordopatis, G., Binney, J., Gilmore, G., {et~al.} 2015{\natexlab{b}}, Monthly
  Notices of the Royal Astronomical Society, 447, 3526,
  \dodoi{10.1093/mnras/stu2726}

\bibitem[{Kubryk {et~al.}(2015)Kubryk, Prantzos, \& Athanassoula}]{Kubryk_2015}
Kubryk, M., Prantzos, N., \& Athanassoula, E. 2015, Astronomy {$\&$}
  Astrophysics, 580, A126, \dodoi{10.1051/0004-6361/201424171}

\bibitem[{{Lacey}(1984)}]{Lacey1984}
{Lacey}, C.~G. 1984, \mnras, 208, 687, \dodoi{10.1093/mnras/208.4.687}

\bibitem[{{Laporte} {et~al.}(2019){Laporte}, {Minchev}, {Johnston}, \&
  {G{\'o}mez}}]{Laporte_2019}
{Laporte}, C. F.~P., {Minchev}, I., {Johnston}, K.~V., \& {G{\'o}mez}, F.~A.
  2019, \mnras, 485, 3134, \dodoi{10.1093/mnras/stz583}

\bibitem[{{Larson}(1976)}]{Larson1976}
{Larson}, R.~B. 1976, \mnras, 176, 31, \dodoi{10.1093/mnras/176.1.31}

\bibitem[{{Lee} {et~al.}(2011){Lee}, {Beers}, {An}, {Ivezi{\'c}}, {Just},
  {Rockosi}, {Morrison}, {Johnson}, {Sch{\"o}nrich}, {Bird}, {Yanny},
  {Harding}, \& {Rocha-Pinto}}]{Lee2011}
{Lee}, Y.~S., {Beers}, T.~C., {An}, D., {et~al.} 2011, \apj, 738, 187,
  \dodoi{10.1088/0004-637X/738/2/187}

\bibitem[{Lee {et~al.}(2011)Lee, Beers, Prieto, Lai, Rockosi, Morrison,
  Johnson, An, Sivarani, \& Yanny}]{Lee_2011b}
Lee, Y.~S., Beers, T.~C., Prieto, C.~A., {et~al.} 2011, The Astronomical
  Journal, 141, 90, \dodoi{10.1088/0004-6256/141/3/90}

\bibitem[{Lemasle {et~al.}(2018)Lemasle, Hajdu, Kovtyukh, Inno, Grebel,
  Catelan, Bono, Fran{\c{c} }ois, Kniazev, da~Silva, \& Storm}]{Lemasle2018}
Lemasle, B., Hajdu, G., Kovtyukh, V., {et~al.} 2018, Astronomy {$\&$}
  Astrophysics, 618, A160, \dodoi{10.1051/0004-6361/201834050}

\bibitem[{Leung \& Bovy(2018)}]{Leung_2018}
Leung, H.~W., \& Bovy, J. 2018, Monthly Notices of the Royal Astronomical
  Society, \dodoi{10.1093/mnras/sty3217}

\bibitem[{Lian {et~al.}(2020{\natexlab{a}})Lian, Thomas, Maraston, Zamora,
  Tayar, Pan, Tissera, Fern{\'{a}}ndez-Trincado, \&
  Garcia-Hernandez}]{Lian2020a}
Lian, J., Thomas, D., Maraston, C., {et~al.} 2020{\natexlab{a}}, Monthly
  Notices of the Royal Astronomical Society, 494, 2561,
  \dodoi{10.1093/mnras/staa867}

\bibitem[{Lian {et~al.}(2020{\natexlab{b}})Lian, Thomas, Maraston, Beers,
  Moni Bidin, Fernández-Trincado, García-Hernández, Lane, Munoz, Nitschelm,
  Roman-Lopes, \& Zamora}]{Lian2020b}
---. 2020{\natexlab{b}}, Monthly Notices of the Royal Astronomical Society,
  497, 2371–2384, \dodoi{10.1093/mnras/staa2078}

\bibitem[{Lian {et~al.}(2020{\natexlab{c}})Lian, Zasowski, Hasselquist, Nataf,
  Thomas, Bidin, Fern{\'{a}}ndez-Trincado, Garcia-Hernandez, Lane, Majewski,
  Roman-Lopes, \& Schultheis}]{Lian_2020c}
Lian, J., Zasowski, G., Hasselquist, S., {et~al.} 2020{\natexlab{c}}, Monthly
  Notices of the Royal Astronomical Society, 497, 3557,
  \dodoi{10.1093/mnras/staa2205}

\bibitem[{{Lian} {et~al.}(2021){Lian}, {Zasowski}, {Hasselquist}, {Neumann},
  {Majewski}, {Cohen}, {Fern{\'a}ndez-Trincado}, {Lane}, {Longa-Pe{\~n}a}, \&
  {Roman-Lopes}}]{Lian2021}
{Lian}, J., {Zasowski}, G., {Hasselquist}, S., {et~al.} 2021, \mnras, 500, 282,
  \dodoi{10.1093/mnras/staa3256}

\bibitem[{Lian {et~al.}(2022{\natexlab{a}})Lian, Zasowski, Mackereth, Imig,
  Holtzman, Beaton, Bird, Cunha, Fern{\'{a}}ndez-Trincado, Horta, Lane,
  Masters, Nitschelm, \& Roman-Lopes}]{Lian2022_maps}
Lian, J., Zasowski, G., Mackereth, T., {et~al.} 2022{\natexlab{a}}, Monthly
  Notices of the Royal Astronomical Society, 513, 4130,
  \dodoi{10.1093/mnras/stac1151}

\bibitem[{Lian {et~al.}(2022{\natexlab{b}})Lian, Zasowski, Hasselquist,
  Holtzman, Boardman, Cunha, Fernández-Trincado, Frinchaboy, Garcia-Hernandez,
  Nitschelm, Lane, Thomas, \& Zhang}]{Lian2022_migration}
Lian, J., Zasowski, G., Hasselquist, S., {et~al.} 2022{\natexlab{b}}, Monthly
  Notices of the Royal Astronomical Society, 511, 5639–5655,
  \dodoi{10.1093/mnras/stac479}

\bibitem[{{Lin} {et~al.}(2018){Lin}, {Dotter}, {Ting}, \& {Asplund}}]{Lin2018}
{Lin}, J., {Dotter}, A., {Ting}, Y.-S., \& {Asplund}, M. 2018, \mnras, 477,
  2966, \dodoi{10.1093/mnras/sty709}

\bibitem[{Lin {et~al.}(2019)Lin, Asplund, Ting, Casagrande, Buder,
  Bland-Hawthorn, Casey, Silva, D'Orazi, Freeman, Kos, Lind, Martell, Sharma,
  Simpson, Zwitter, Zucker, Minchev, {\v{C}}otar, Hayden, Horner, Lewis,
  Nordlander, Wyse, \& {\v{Z}}erjal}]{Lin_2019}
Lin, J., Asplund, M., Ting, Y.-S., {et~al.} 2019, Monthly Notices of the Royal
  Astronomical Society, 491, 2043, \dodoi{10.1093/mnras/stz3048}

\bibitem[{{Linden} {et~al.}(2017){Linden}, {Pryal}, {Hayes}, {Troup},
  {Majewski}, {Andrews}, {Beers}, {Carrera}, {Cunha}, {Fern{\'a}ndez-Trincado},
  {Frinchaboy}, {Geisler}, {Lane}, {Nitschelm}, {Pan}, {Allende Prieto},
  {Roman-Lopes}, {Smith}, {Sobeck}, {Tang}, {Villanova}, \&
  {Zasowski}}]{Linden_2017}
{Linden}, S.~T., {Pryal}, M., {Hayes}, C.~R., {et~al.} 2017, \apj, 842, 49,
  \dodoi{10.3847/1538-4357/aa6f17}

\bibitem[{Loebman {et~al.}(2016)Loebman, Debattista, Nidever, Hayden, Holtzman,
  Clarke, Ro{\v{s} }kar, \& Valluri}]{Loebman_2016}
Loebman, S.~R., Debattista, V.~P., Nidever, D.~L., {et~al.} 2016, The
  Astrophysical Journal, 818, L6, \dodoi{10.3847/2041-8205/818/1/l6}

\bibitem[{{Lu} {et~al.}(2022){Lu}, {Ness}, {Buck}, \& {Carr}}]{Lu_2022}
{Lu}, Y.~L., {Ness}, M.~K., {Buck}, T., \& {Carr}, C. 2022, \mnras, 512, 4697,
  \dodoi{10.1093/mnras/stac780}

\bibitem[{Luo {et~al.}(2015)Luo, Zhao, Zhao, Deng, Liu, Jing, Wang, Zhang, Shi,
  Cui, \& et~al.}]{LAMOST_Luo2015}
Luo, A.-L., Zhao, Y.-H., Zhao, G., {et~al.} 2015, Research in Astronomy and
  Astrophysics, 15, 1095–1124, \dodoi{10.1088/1674-4527/15/8/002}

\bibitem[{Mackereth {et~al.}(2018)Mackereth, Crain, Schiavon, Schaye, Theuns,
  \& Schaller}]{Mackereth_2018}
Mackereth, J.~T., Crain, R.~A., Schiavon, R.~P., {et~al.} 2018, Monthly Notices
  of the Royal Astronomical Society, 477, 5072, \dodoi{10.1093/mnras/sty972}

\bibitem[{{Mackereth} {et~al.}(2017){Mackereth}, {Bovy}, {Schiavon},
  {Zasowski}, {Cunha}, {Frinchaboy}, {Garc{\'\i}a Perez}, {Hayden}, {Holtzman},
  {Majewski}, {M{\'e}sz{\'a}ros}, {Nidever}, {Pinsonneault}, \&
  {Shetrone}}]{Mackereth2017}
{Mackereth}, J.~T., {Bovy}, J., {Schiavon}, R.~P., {et~al.} 2017, \mnras, 471,
  3057, \dodoi{10.1093/mnras/stx1774}

\bibitem[{{Mackereth} {et~al.}(2019){Mackereth}, {Bovy}, {Leung}, {Schiavon},
  {Trick}, {Chaplin}, {Cunha}, {Feuillet}, {Majewski}, {Martig}, {Miglio},
  {Nidever}, {Pinsonneault}, {Aguirre}, {Sobeck}, {Tayar}, \&
  {Zasowski}}]{Mackereth2019}
{Mackereth}, J.~T., {Bovy}, J., {Leung}, H.~W., {et~al.} 2019, \mnras, 489,
  176, \dodoi{10.1093/mnras/stz1521}

\bibitem[{Magrini {et~al.}(2016)Magrini, Coccato, Stanghellini, Casasola, \&
  Galli}]{Magrini_2016}
Magrini, L., Coccato, L., Stanghellini, L., Casasola, V., \& Galli, D. 2016,
  Astronomy {$\&$} Astrophysics, 588, A91, \dodoi{10.1051/0004-6361/201527799}

\bibitem[{Magrini {et~al.}(2018)Magrini, Vincenzo, Randich, Pancino, Casali,
  Tautvai{\v{s} }iene{\.}, Drazdauskas, Mikolaitis,
  Minkevi{\v{c}}i{\={u}}te{\.}, Stonkute{\.}, Chorniy, Bagdonas, Kordopatis,
  Friel, Roccatagliata, Jim{\'{e}}nez-Esteban, Gilmore, Vallenari, Bensby,
  Bragaglia, Korn, Lanzafame, Smiljanic, Bayo, Casey, Costado, Franciosini,
  Hourihane, Jofr{\'{e}}, Lewis, Monaco, Morbidelli, Sacco, \&
  Worley}]{Magrini_2018}
Magrini, L., Vincenzo, F., Randich, S., {et~al.} 2018, Astronomy {$\&$}
  Astrophysics, 618, A102, \dodoi{10.1051/0004-6361/201833224}

\bibitem[{{Majewski} {et~al.}(2017){Majewski}, {Schiavon}, {Frinchaboy},
  {Allende Prieto}, {Barkhouser}, {Bizyaev}, {Blank}, {Brunner}, {Burton},
  {Carrera}, {Chojnowski}, {Cunha}, {Epstein}, {Fitzgerald}, {Garc{\'\i}a
  P{\'e}rez}, {Hearty}, {Henderson}, {Holtzman}, {Johnson}, {Lam}, {Lawler},
  {Maseman}, {M{\'e}sz{\'a}ros}, {Nelson}, {Nguyen}, {Nidever}, {Pinsonneault},
  {Shetrone}, {Smee}, {Smith}, {Stolberg}, {Skrutskie}, {Walker}, {Wilson},
  {Zasowski}, {Anders}, {Basu}, {Beland}, {Blanton}, {Bovy}, {Brownstein},
  {Carlberg}, {Chaplin}, {Chiappini}, {Eisenstein}, {Elsworth}, {Feuillet},
  {Fleming}, {Galbraith-Frew}, {Garc{\'\i}a}, {Garc{\'\i}a-Hern{\'a}ndez},
  {Gillespie}, {Girardi}, {Gunn}, {Hasselquist}, {Hayden}, {Hekker}, {Ivans},
  {Kinemuchi}, {Klaene}, {Mahadevan}, {Mathur}, {Mosser}, {Muna}, {Munn},
  {Nichol}, {O'Connell}, {Parejko}, {Robin}, {Rocha-Pinto}, {Schultheis},
  {Serenelli}, {Shane}, {Silva Aguirre}, {Sobeck}, {Thompson}, {Troup},
  {Weinberg}, \& {Zamora}}]{Majewski2017}
{Majewski}, S.~R., {Schiavon}, R.~P., {Frinchaboy}, P.~M., {et~al.} 2017, \aj,
  154, 94, \dodoi{10.3847/1538-3881/aa784d}

\bibitem[{{Marinacci} {et~al.}(2011){Marinacci}, {Fraternali}, {Nipoti},
  {Binney}, {Ciotti}, \& {Londrillo}}]{Marinacci_2011}
{Marinacci}, F., {Fraternali}, F., {Nipoti}, C., {et~al.} 2011, \mnras, 415,
  1534, \dodoi{10.1111/j.1365-2966.2011.18810.x}

\bibitem[{Martig {et~al.}(2016)Martig, Minchev, Ness, Fouesneau, \&
  Rix}]{Martig_2016b}
Martig, M., Minchev, I., Ness, M., Fouesneau, M., \& Rix, H.-W. 2016, The
  Astrophysical Journal, 831, 139, \dodoi{10.3847/0004-637x/831/2/139}

\bibitem[{{Martig} {et~al.}(2016){Martig}, {Fouesneau}, {Rix}, {Ness},
  {M{\'e}sz{\'a}ros}, {Garc{\'\i}a-Hern{\'a}ndez}, {Pinsonneault}, {Serenelli},
  {Silva Aguirre}, \& {Zamora}}]{Martig_2016}
{Martig}, M., {Fouesneau}, M., {Rix}, H.-W., {et~al.} 2016, \mnras, 456, 3655,
  \dodoi{10.1093/mnras/stv2830}

\bibitem[{Martig {et~al.}(2021)Martig, Pinna, Falc{\'{o}}n-Barroso, Gadotti,
  Husemann, Minchev, Neumann, Ruiz-Lara, \& van~de Ven}]{Martig_2021}
Martig, M., Pinna, F., Falc{\'{o}}n-Barroso, J., {et~al.} 2021, Monthly Notices
  of the Royal Astronomical Society, 508, 2458, \dodoi{10.1093/mnras/stab2729}

\bibitem[{{Matteucci} \& {Brocato}(1990)}]{mb1990}
{Matteucci}, F., \& {Brocato}, E. 1990, \apj, 365, 539, \dodoi{10.1086/169508}

\bibitem[{{Matteucci} \& {Francois}(1989)}]{Matteucci1989}
{Matteucci}, F., \& {Francois}, P. 1989, \mnras, 239, 885,
  \dodoi{10.1093/mnras/239.3.885}

\bibitem[{{M{\'e}sz{\'a}ros} {et~al.}(2012){M{\'e}sz{\'a}ros}, {Allende
  Prieto}, {Edvardsson}, {Castelli}, {Garc{\'\i}a P{\'e}rez}, {Gustafsson},
  {Majewski}, {Plez}, {Schiavon}, {Shetrone}, \& {de Vicente}}]{Meszaros2012}
{M{\'e}sz{\'a}ros}, S., {Allende Prieto}, C., {Edvardsson}, B., {et~al.} 2012,
  \aj, 144, 120, \dodoi{10.1088/0004-6256/144/4/120}

\bibitem[{Michtchenko {et~al.}(2016)Michtchenko, Vieira, Barros, \& L{\'{e}
  }pine}]{Michtchenko_2016}
Michtchenko, T.~A., Vieira, R. S.~S., Barros, D.~A., \& L{\'{e} }pine, J. R.~D.
  2016, Astronomy {$\&$} Astrophysics, 597, A39,
  \dodoi{10.1051/0004-6361/201628895}

\bibitem[{Miglio {et~al.}(2021)Miglio, Chiappini, Mackereth, Davies, Brogaard,
  Casagrande, Chaplin, Girardi, Kawata, Khan, Izzard, Montalb{\'{a} }n, Mosser,
  Vincenzo, Bossini, Noels, Rodrigues, Valentini, \& Mandel}]{Miglio_2021}
Miglio, A., Chiappini, C., Mackereth, J.~T., {et~al.} 2021, Astronomy {$\&$}
  Astrophysics, 645, A85, \dodoi{10.1051/0004-6361/202038307}

\bibitem[{Minchev {et~al.}(2013)Minchev, Chiappini, \& Martig}]{Minchev_2013}
Minchev, I., Chiappini, C., \& Martig, M. 2013, Astronomy {$\&$} Astrophysics,
  558, A9, \dodoi{10.1051/0004-6361/201220189}

\bibitem[{Minchev {et~al.}(2014)Minchev, Chiappini, \& Martig}]{Minchev_2014}
---. 2014, Astronomy {$\&$} Astrophysics, 572, A92,
  \dodoi{10.1051/0004-6361/201423487}

\bibitem[{{Minchev} {et~al.}(2015){Minchev}, {Martig}, {Streich},
  {Scannapieco}, {de Jong}, \& {Steinmetz}}]{Minchev2015}
{Minchev}, I., {Martig}, M., {Streich}, D., {et~al.} 2015, \apjl, 804, L9,
  \dodoi{10.1088/2041-8205/804/1/L9}

\bibitem[{Minchev {et~al.}(2015)Minchev, Martig, Streich, Scannapieco, de~Jong,
  \& Steinmetz}]{Minchev_2015}
Minchev, I., Martig, M., Streich, D., {et~al.} 2015, The Astrophysical Journal,
  804, L9, \dodoi{10.1088/2041-8205/804/1/l9}

\bibitem[{{Minchev} {et~al.}(2017){Minchev}, {Steinmetz}, {Chiappini},
  {Martig}, {Anders}, {Matijevic}, \& {de Jong}}]{Minchev2017}
{Minchev}, I., {Steinmetz}, M., {Chiappini}, C., {et~al.} 2017, \apj, 834, 27,
  \dodoi{10.3847/1538-4357/834/1/27}

\bibitem[{{Minchev} {et~al.}(2018){Minchev}, {Anders}, {Recio-Blanco},
  {Chiappini}, {de Laverny}, {Queiroz}, {Steinmetz}, {Adibekyan}, {Carrillo},
  {Cescutti}, {Guiglion}, {Hayden}, {de Jong}, {Kordopatis}, {Majewski},
  {Martig}, \& {Santiago}}]{Minchev_2018}
{Minchev}, I., {Anders}, F., {Recio-Blanco}, A., {et~al.} 2018, \mnras, 481,
  1645, \dodoi{10.1093/mnras/sty2033}

\bibitem[{Moll{\'{a} } {et~al.}(2018)Moll{\'{a} }, D{\'{\i}}az, Cavichia,
  Gibson, Maciel, Costa, Ascasibar, \& Few}]{Molla_2018}
Moll{\'{a} }, M., D{\'{\i}}az, {\'{A}}.~I., Cavichia, O., {et~al.} 2018,
  Monthly Notices of the Royal Astronomical Society,
  \dodoi{10.1093/mnras/sty2877}

\bibitem[{{Mor} {et~al.}(2019){Mor}, {Robin}, {Figueras}, {Roca-F{\`a}brega},
  \& {Luri}}]{Mor_2019}
{Mor}, R., {Robin}, A.~C., {Figueras}, F., {Roca-F{\`a}brega}, S., \& {Luri},
  X. 2019, \aap, 624, L1, \dodoi{10.1051/0004-6361/201935105}

\bibitem[{{Myers} {et~al.}(2022){Myers}, {Donor}, {Spoo}, {Frinchaboy},
  {Cunha}, {Price-Whelan}, {Majewski}, {Beaton}, {Zasowski}, {O'Connell},
  {Ray}, {Bizyaev}, {Chiappini}, {Garc{\'\i}a-Hern{\'a}ndez}, {Geisler},
  {J{\"o}nsson}, {Lane}, {Longa-Pe{\~n}a}, {Minchev}, {Minniti}, {Nitschelm},
  \& {Roman-Lopes}}]{Myers_2022}
{Myers}, N., {Donor}, J., {Spoo}, T., {et~al.} 2022, \aj, 164, 85,
  \dodoi{10.3847/1538-3881/ac7ce5}

\bibitem[{{Ness} {et~al.}(2016){Ness}, {Hogg}, {Rix}, {Martig}, {Pinsonneault},
  \& {Ho}}]{Ness2016}
{Ness}, M., {Hogg}, D.~W., {Rix}, H.~W., {et~al.} 2016, \apj, 823, 114,
  \dodoi{10.3847/0004-637X/823/2/114}

\bibitem[{{Nidever} {et~al.}(2014){Nidever}, {Bovy}, {Bird}, {Andrews},
  {Hayden}, {Holtzman}, {Majewski}, {Smith}, {Robin}, {Garc{\'\i}a P{\'e}rez},
  {Cunha}, {Allende Prieto}, {Zasowski}, {Schiavon}, {Johnson}, {Weinberg},
  {Feuillet}, {Schneider}, {Shetrone}, {Sobeck}, {Garc{\'\i}a-Hern{\'a}ndez},
  {Zamora}, {Rix}, {Beers}, {Wilson}, {O'Connell}, {Minchev}, {Chiappini},
  {Anders}, {Bizyaev}, {Brewington}, {Ebelke}, {Frinchaboy}, {Ge}, {Kinemuchi},
  {Malanushenko}, {Malanushenko}, {Marchante}, {M{\'e}sz{\'a}ros}, {Oravetz},
  {Pan}, {Simmons}, \& {Skrutskie}}]{Nidever2014}
{Nidever}, D.~L., {Bovy}, J., {Bird}, J.~C., {et~al.} 2014, \apj, 796, 38,
  \dodoi{10.1088/0004-637X/796/1/38}

\bibitem[{{Nidever} {et~al.}(2015){Nidever}, {Holtzman}, {Allende Prieto},
  {Beland}, {Bender}, {Bizyaev}, {Burton}, {Desphande}, {Fleming}, {Garc{\'\i}a
  P{\'e}rez}, {Hearty}, {Majewski}, {M{\'e}sz{\'a}ros}, {Muna}, {Nguyen},
  {Schiavon}, {Shetrone}, {Skrutskie}, {Sobeck}, \&
  {Wilson}}]{Nidever_2015_ApogeeDataReduction}
{Nidever}, D.~L., {Holtzman}, J.~A., {Allende Prieto}, C., {et~al.} 2015, \aj,
  150, 173, \dodoi{10.1088/0004-6256/150/6/173}

\bibitem[{{Nissen}(2015)}]{Nissen2015}
{Nissen}, P.~E. 2015, \aap, 579, A52, \dodoi{10.1051/0004-6361/201526269}

\bibitem[{Pilkington \& Gibson(2012)}]{Pilkington_2012}
Pilkington, K., \& Gibson, B.~K. 2012, Metallicity Gradients in Simulated Disk
  Galaxies,  arXiv, \dodoi{10.48550/ARXIV.1201.6417}

\bibitem[{Pinsonneault {et~al.}(2018)Pinsonneault, Elsworth, Tayar, Serenelli,
  Stello, Zinn, Mathur, Garc{\'{\i} }a, Johnson, Hekker, Huber, Kallinger,
  M{\'{e}}sz{\'{a}}ros, Mosser, Stassun, Girardi, Rodrigues, Aguirre, An, Basu,
  Chaplin, Corsaro, Cunha, Garc{\'{\i}}a-Hern{\'{a}}ndez, Holtzman, Jönsson,
  Shetrone, Smith, Sobeck, Stringfellow, Zamora, Beers,
  Fern{\'{a}}ndez-Trincado, Frinchaboy, Hearty, \&
  Nitschelm}]{Pinsonneault_2018}
Pinsonneault, M.~H., Elsworth, Y.~P., Tayar, J., {et~al.} 2018, The
  Astrophysical Journal Supplement Series, 239, 32,
  \dodoi{10.3847/1538-4365/aaebfd}

\bibitem[{Poggio {et~al.}(2022)Poggio, Recio-Blanco, Palicio, Fiorentin,
  de~Laverny, Drimmel, Kordopatis, Lattanzi, Schultheis, Spagna, \&
  Spitoni}]{Poggio2022}
Poggio, E., Recio-Blanco, A., Palicio, P.~A., {et~al.} 2022,
  \dodoi{10.48550/ARXIV.2206.14849}

\bibitem[{Price-Jones {et~al.}(2020)Price-Jones, Bovy, Webb, Prieto, Beaton,
  Brownstein, Cohen, Cunha, Donor, Frinchaboy, Garc{\'{\i} }a-Hern{\'{a}}ndez,
  Lane, Majewski, Nidever, \& Roman-Lopes}]{Price_Jones_2020}
Price-Jones, N., Bovy, J., Webb, J.~J., {et~al.} 2020, Monthly Notices of the
  Royal Astronomical Society, 496, 5101, \dodoi{10.1093/mnras/staa1905}

\bibitem[{{Queiroz} {et~al.}(2021){Queiroz}, {Chiappini}, {Perez-Villegas},
  {Khalatyan}, {Anders}, {Barbuy}, {Santiago}, {Steinmetz}, {Cunha},
  {Schultheis}, {Majewski}, {Minchev}, {Minniti}, {Beaton}, {Cohen}, {da
  Costa}, {Fern{\'a}ndez-Trincado}, {Garcia-Hern{\'a}ndez}, {Geisler},
  {Hasselquist}, {Lane}, {Nitschelm}, {Rojas-Arriagada}, {Roman-Lopes},
  {Smith}, \& {Zasowski}}]{Queiroz_2021}
{Queiroz}, A.~B.~A., {Chiappini}, C., {Perez-Villegas}, A., {et~al.} 2021,
  \aap, 656, A156, \dodoi{10.1051/0004-6361/202039030}

\bibitem[{{Quinn} {et~al.}(1993){Quinn}, {Hernquist}, \&
  {Fullagar}}]{Quinn1993}
{Quinn}, P.~J., {Hernquist}, L., \& {Fullagar}, D.~P. 1993, \apj, 403, 74,
  \dodoi{10.1086/172184}

\bibitem[{Rahimi {et~al.}(2014)Rahimi, Carrell, \& Kawata}]{Rahimi_2014}
Rahimi, A., Carrell, K., \& Kawata, D. 2014, Research in Astronomy and
  Astrophysics, 14, 1406, \dodoi{10.1088/1674-4527/14/11/004}

\bibitem[{Reddy {et~al.}(2006)Reddy, Lambert, \& Prieto}]{Reddy2006}
Reddy, B.~E., Lambert, D.~L., \& Prieto, C.~A. 2006, Monthly Notices of the
  Royal Astronomical Society, 367, 1329–1366,
  \dodoi{10.1111/j.1365-2966.2006.10148.x}

\bibitem[{{Reid} {et~al.}(2019){Reid}, {Menten}, {Brunthaler}, {Zheng}, {Dame},
  {Xu}, {Li}, {Sakai}, {Wu}, {Immer}, {Zhang}, {Sanna}, {Moscadelli}, {Rygl},
  {Bartkiewicz}, {Hu}, {Quiroga-Nu{\~n}ez}, \& {van Langevelde}}]{Reid2019}
{Reid}, M.~J., {Menten}, K.~M., {Brunthaler}, A., {et~al.} 2019, \apj, 885,
  131, \dodoi{10.3847/1538-4357/ab4a11}

\bibitem[{Robin {et~al.}(2017)Robin, Bienaym{\'{e} }, Fern{\'{a}}ndez-Trincado,
  \& Reyl{\'{e}}}]{Robin_2017}
Robin, A.~C., Bienaym{\'{e} }, O., Fern{\'{a}}ndez-Trincado, J.~G., \&
  Reyl{\'{e}}, C. 2017, Astronomy {$\&$} Astrophysics, 605, A1,
  \dodoi{10.1051/0004-6361/201630217}

\bibitem[{{Robin} {et~al.}(2022){Robin}, {Bienaym{\'e}}, {Salomon},
  {Reyl{\'e}}, {Lagarde}, {Figueras}, {Mor}, {Fern{\'a}ndez-Trincado}, \&
  {Montillaud}}]{Robin_2022}
{Robin}, A.~C., {Bienaym{\'e}}, O., {Salomon}, J.~B., {et~al.} 2022, arXiv
  e-prints, arXiv:2208.13827.
\newblock \doarXiv{2208.13827}

\bibitem[{Romano \& Starkenburg(2013)}]{Romano_2013}
Romano, D., \& Starkenburg, E. 2013, Monthly Notices of the Royal Astronomical
  Society, 434, 471, \dodoi{10.1093/mnras/stt1033}

\bibitem[{{Ro{\v s}kar} {et~al.}(2008){Ro{\v s}kar}, {Debattista}, {Quinn},
  {Stinson}, \& {Wadsley}}]{Roskar2008}
{Ro{\v s}kar}, R., {Debattista}, V.~P., {Quinn}, T.~R., {Stinson}, G.~S., \&
  {Wadsley}, J. 2008, \apjl, 684, L79, \dodoi{10.1086/592231}

\bibitem[{{Ruiz-Lara} {et~al.}(2020){Ruiz-Lara}, {Gallart}, {Bernard}, \&
  {Cassisi}}]{RuizLara_2020}
{Ruiz-Lara}, T., {Gallart}, C., {Bernard}, E.~J., \& {Cassisi}, S. 2020, Nature
  Astronomy, 4, 965, \dodoi{10.1038/s41550-020-1097-0}

\bibitem[{{Sahlholdt} {et~al.}(2022){Sahlholdt}, {Feltzing}, \&
  {Feuillet}}]{Sahlholdt_2021}
{Sahlholdt}, C.~L., {Feltzing}, S., \& {Feuillet}, D.~K. 2022, \mnras, 510,
  4669, \dodoi{10.1093/mnras/stab3681}

\bibitem[{{Salaris} \& {Cassisi}(2005)}]{Salaris_2005}
{Salaris}, M., \& {Cassisi}, S. 2005, {Evolution of Stars and Stellar
  Populations}

\bibitem[{Santana {et~al.}(2021)Santana, Beaton, Covey, O'Connell,
  Longa-Pe{\~{n}}a, Cohen, Fern{\'{a}}ndez-Trincado, Hayes, Zasowski, Sobeck,
  Majewski, Chojnowski, Lee, Oelkers, Stringfellow, Almeida, Anguiano, Donor,
  Frinchaboy, Hasselquist, Johnson, Kollmeier, Nidever, Price-Whelan,
  Rojas-Arriagada, Schultheis, Shetrone, Simon, Aerts, Borissova, Drout,
  Geisler, Law, Medina, Minniti, Monachesi, Mu{\~{n}}oz, Poleski, Roman-Lopes,
  Schlaufman, Stutz, Teske, Tkachenko, Saders, Weinberger, \&
  Zoccali}]{Santana_2021}
Santana, F.~A., Beaton, R.~L., Covey, K.~R., {et~al.} 2021, The Astronomical
  Journal, 162, 303, \dodoi{10.3847/1538-3881/ac2cbc}

\bibitem[{{Schlesinger} {et~al.}(2014){Schlesinger}, {Johnson}, {Rockosi},
  {Lee}, {Beers}, {Harding}, {Allende Prieto}, {Bird}, {Sch{\"o}nrich},
  {Yanny}, {Schneider}, {Weaver}, \& {Brinkmann}}]{Schlesinger2014}
{Schlesinger}, K.~J., {Johnson}, J.~A., {Rockosi}, C.~M., {et~al.} 2014, \apj,
  791, 112, \dodoi{10.1088/0004-637X/791/2/112}

\bibitem[{{Sch{\"o}nrich} \& {Binney}(2009{\natexlab{a}})}]{Schonrich_2009a}
{Sch{\"o}nrich}, R., \& {Binney}, J. 2009{\natexlab{a}}, \mnras, 396, 203,
  \dodoi{10.1111/j.1365-2966.2009.14750.x}

\bibitem[{{Sch{\"o}nrich} \& {Binney}(2009{\natexlab{b}})}]{Schonrich_2009b}
---. 2009{\natexlab{b}}, Monthly Notices of the Royal Astronomical Society,
  399, 1145, \dodoi{10.1111/j.1365-2966.2009.15365.x}

\bibitem[{{Sellwood} \& {Binney}(2002)}]{Sellwood2002}
{Sellwood}, J.~A., \& {Binney}, J.~J. 2002, \mnras, 336, 785,
  \dodoi{10.1046/j.1365-8711.2002.05806.x}

\bibitem[{{Shapiro} \& {Field}(1976)}]{Shapiro_1976}
{Shapiro}, P.~R., \& {Field}, G.~B. 1976, \apj, 205, 762,
  \dodoi{10.1086/154332}

\bibitem[{Sharma {et~al.}(2021{\natexlab{a}})Sharma, Hayden, \&
  Bland-Hawthorn}]{Sharma_2021b}
Sharma, S., Hayden, M.~R., \& Bland-Hawthorn, J. 2021{\natexlab{a}}, Monthly
  Notices of the Royal Astronomical Society, 507, 5882,
  \dodoi{10.1093/mnras/stab2015}

\bibitem[{Sharma {et~al.}(2021{\natexlab{b}})Sharma, Hayden, Bland-Hawthorn,
  Stello, Buder, Zinn, Spina, Kallinger, Asplund, Silva, D'Orazi, Freeman, Kos,
  Lewis, Lin, Lind, Martell, Schlesinger, Simpson, Zucker, Zwitter, Chen,
  Cotar, Kafle, Khanna, Tepper-Garcia, Wang, \& Wittenmyer}]{Sharma_2021}
Sharma, S., Hayden, M.~R., Bland-Hawthorn, J., {et~al.} 2021{\natexlab{b}},
  Monthly Notices of the Royal Astronomical Society, 510, 734,
  \dodoi{10.1093/mnras/stab3341}

\bibitem[{{Shetrone} {et~al.}(2015){Shetrone}, {Bizyaev}, {Lawler}, {Allende
  Prieto}, {Johnson}, {Smith}, {Cunha}, {Holtzman}, {Garc{\'\i}a P{\'e}rez},
  {M{\'e}sz{\'a}ros}, {Sobeck}, {Zamora}, {Garc{\'\i}a-Hern{\'a}ndez}, {Souto},
  {Chojnowski}, {Koesterke}, {Majewski}, \& {Zasowski}}]{Shetrone_2015}
{Shetrone}, M., {Bizyaev}, D., {Lawler}, J.~E., {et~al.} 2015, \apjs, 221, 24,
  \dodoi{10.1088/0067-0049/221/2/24}

\bibitem[{{Smith} {et~al.}(2021){Smith}, {Bizyaev}, {Cunha}, {Shetrone},
  {Souto}, {Allende Prieto}, {Masseron}, {M{\'e}sz{\'a}ros}, {J{\"o}nsson},
  {Hasselquist}, {Osorio}, {Garc{\'\i}a-Hern{\'a}ndez}, {Plez}, {Beaton},
  {Holtzman}, {Majewski}, {Stringfellow}, \& {Sobeck}}]{Smith_2021}
{Smith}, V.~V., {Bizyaev}, D., {Cunha}, K., {et~al.} 2021, \aj, 161, 254,
  \dodoi{10.3847/1538-3881/abefdc}

\bibitem[{{Soubiran} {et~al.}(2003){Soubiran}, {Bienaym{\'e}}, \&
  {Siebert}}]{Soubiran2003}
{Soubiran}, C., {Bienaym{\'e}}, O., \& {Siebert}, A. 2003, \aap, 398, 141,
  \dodoi{10.1051/0004-6361:20021615}

\bibitem[{Spitoni {et~al.}(2019{\natexlab{a}})Spitoni, Aguirre, Matteucci,
  Calura, \& Grisoni}]{Spitoni_2019}
Spitoni, E., Aguirre, V.~S., Matteucci, F., Calura, F., \& Grisoni, V.
  2019{\natexlab{a}}, Astronomy {$\&$} Astrophysics, 623, A60,
  \dodoi{10.1051/0004-6361/201834188}

\bibitem[{Spitoni {et~al.}(2022{\natexlab{a}})Spitoni, B{\o}rsen-Koch, Verma,
  \& Stokholm}]{Spitoni_2022b}
Spitoni, E., B{\o}rsen-Koch, V.~A., Verma, K., \& Stokholm, A.
  2022{\natexlab{a}}, Astronomy {$\&$} Astrophysics, 663, A174,
  \dodoi{10.1051/0004-6361/202142469}

\bibitem[{Spitoni {et~al.}(2019{\natexlab{b}})Spitoni, Cescutti, Minchev,
  Matteucci, Aguirre, Martig, Bono, \& Chiappini}]{Spitoni_2019b}
Spitoni, E., Cescutti, G., Minchev, I., {et~al.} 2019{\natexlab{b}}, Astronomy
  {$\&$} Astrophysics, 628, A38, \dodoi{10.1051/0004-6361/201834665}

\bibitem[{Spitoni {et~al.}(2020)Spitoni, Verma, Aguirre, \&
  Calura}]{Spitoni_2020}
Spitoni, E., Verma, K., Aguirre, V.~S., \& Calura, F. 2020, Astronomy {$\&$}
  Astrophysics, 635, A58, \dodoi{10.1051/0004-6361/201937275}

\bibitem[{Spitoni {et~al.}(2021)Spitoni, Verma, Aguirre, Vincenzo, Matteucci,
  Vai{\v{c} }ekauskait{\.{e}}, Palla, Grisoni, \& Calura}]{Spitoni_2021}
Spitoni, E., Verma, K., Aguirre, V.~S., {et~al.} 2021, Astronomy {$\&$}
  Astrophysics, 647, A73, \dodoi{10.1051/0004-6361/202039864}

\bibitem[{Spitoni {et~al.}(2022{\natexlab{b}})Spitoni, Recio-Blanco,
  de~Laverny, Palicio, Kordopatis, Schultheis, Contursi, Poggio, Romano, \&
  Matteucci}]{Spitoni_2022}
Spitoni, E., Recio-Blanco, A., de~Laverny, P., {et~al.} 2022{\natexlab{b}},
  Beyond the two-infall model with Gaia DR3 {$\alpha$}-element abundances,
  arXiv, \dodoi{10.48550/ARXIV.2206.12436}

\bibitem[{{Spitzer} \& {Schwarzschild}(1951)}]{Spitzer1951}
{Spitzer}, Lyman, J., \& {Schwarzschild}, M. 1951, \apj, 114, 385,
  \dodoi{10.1086/145478}

\bibitem[{{Steinmetz} {et~al.}(2020){Steinmetz}, {Matijevi{\v{c}}}, {Enke},
  {Zwitter}, {Guiglion}, {McMillan}, {Kordopatis}, {Valentini}, {Chiappini},
  {Casagrande}, {Wojno}, {Anguiano}, {Bienaym{\'e}}, {Bijaoui}, {Binney},
  {Burton}, {Cass}, {de Laverny}, {Fiegert}, {Freeman}, {Fulbright}, {Gibson},
  {Gilmore}, {Grebel}, {Helmi}, {Kunder}, {Munari}, {Navarro}, {Parker},
  {Ruchti}, {Recio-Blanco}, {Reid}, {Seabroke}, {Siviero}, {Siebert}, {Stupar},
  {Watson}, {Williams}, {Wyse}, {Anders}, {Antoja}, {Birko}, {Bland-Hawthorn},
  {Bossini}, {Garc{\'\i}a}, {Carrillo}, {Chaplin}, {Elsworth}, {Famaey},
  {Gerhard}, {Jofre}, {Just}, {Mathur}, {Miglio}, {Minchev}, {Monari},
  {Mosser}, {Ritter}, {Rodrigues}, {Scholz}, {Sharma}, {Sysoliatina}, \& {RAVE
  Collaboration}}]{Steinmetz2020}
{Steinmetz}, M., {Matijevi{\v{c}}}, G., {Enke}, H., {et~al.} 2020, \aj, 160,
  82, \dodoi{10.3847/1538-3881/ab9ab9}

\bibitem[{{Thomas} {et~al.}(2005){Thomas}, {Maraston}, {Bender}, \& {Mendes de
  Oliveira}}]{thomas2005}
{Thomas}, D., {Maraston}, C., {Bender}, R., \& {Mendes de Oliveira}, C. 2005,
  \apj, 621, 673, \dodoi{10.1086/426932}

\bibitem[{Ting {et~al.}(2015)Ting, Conroy, \& Rix}]{Ting_2015}
Ting, Y.-S., Conroy, C., \& Rix, H.-W. 2015, The Astrophysical Journal, 816,
  10, \dodoi{10.3847/0004-637x/816/1/10}

\bibitem[{{Toth} \& {Ostriker}(1992)}]{Toth1992}
{Toth}, G., \& {Ostriker}, J.~P. 1992, \apj, 389, 5, \dodoi{10.1086/171185}

\bibitem[{{Toyouchi} \& {Chiba}(2018)}]{Toyouchi_2018}
{Toyouchi}, D., \& {Chiba}, M. 2018, \apj, 855, 104,
  \dodoi{10.3847/1538-4357/aab044}

\bibitem[{{Twarog}(1980)}]{Twarog1980}
{Twarog}, B.~A. 1980, \apj, 242, 242, \dodoi{10.1086/158460}

\bibitem[{V{\'{a}}zquez {et~al.}(2022)V{\'{a}}zquez, Magrini, Casali,
  Tautvai{\v{s}}ien{\.{e}}, Spina, der Swaelmen, Randich, Bensby, Bragaglia,
  Friel, Feltzing, Sacco, Turchi, Jim{\'{e}}nez-Esteban, D'Orazi, Delgado-Mena,
  Mikolaitis, Drazdauskas, Minkevi{\v{c}}i{\={u}}t{\.{e}}, Stonkut{\.{e}},
  Bagdonas, Montes, Guiglion, Baratella, Tabernero, Gilmore, Alfaro, Francois,
  Korn, Smiljanic, Bergemann, Franciosini, Gonneau, Hourihane, Worley, \&
  Zaggia}]{Vazquez_2022}
V{\'{a}}zquez, C.~V., Magrini, L., Casali, G., {et~al.} 2022, Astronomy {$\&$}
  Astrophysics, 660, A135, \dodoi{10.1051/0004-6361/202142937}

\bibitem[{Vickers {et~al.}(2021)Vickers, Shen, \& Li}]{Vickers_2021}
Vickers, J.~J., Shen, J., \& Li, Z.-Y. 2021, The Astrophysical Journal, 922,
  189, \dodoi{10.3847/1538-4357/ac27a9}

\bibitem[{Vincenzo {et~al.}(2014)Vincenzo, Matteucci, Vattakunnel, \&
  Lanfranchi}]{Vincenzo_2014}
Vincenzo, F., Matteucci, F., Vattakunnel, S., \& Lanfranchi, G.~A. 2014,
  Monthly Notices of the Royal Astronomical Society, 441, 2815,
  \dodoi{10.1093/mnras/stu710}

\bibitem[{Vincenzo {et~al.}(2019)Vincenzo, Spitoni, Calura, Matteucci, Aguirre,
  Miglio, \& Cescutti}]{Vincenzo_2019}
Vincenzo, F., Spitoni, E., Calura, F., {et~al.} 2019, Monthly Notices of the
  Royal Astronomical Society: Letters, 487, L47, \dodoi{10.1093/mnrasl/slz070}

\bibitem[{Vincenzo {et~al.}(2021{\natexlab{a}})Vincenzo, Weinberg, Miglio,
  Lane, \& Roman-Lopes}]{Vincenzo2021}
Vincenzo, F., Weinberg, D.~H., Miglio, A., Lane, R.~R., \& Roman-Lopes, A.
  2021{\natexlab{a}}, Monthly Notices of the Royal Astronomical Society, 508,
  5903–5920, \dodoi{10.1093/mnras/stab2899}

\bibitem[{Vincenzo {et~al.}(2021{\natexlab{b}})Vincenzo, Weinberg, Montalbán,
  Miglio, Khan, Griffith, Hasselquist, Johnson, Johnson, Nitschelm, \&
  Pinsonneault}]{Vincenzo_2021}
Vincenzo, F., Weinberg, D.~H., Montalbán, J., {et~al.} 2021{\natexlab{b}}, CNO
  dredge-up in a sample of APOGEE/Kepler red giants: Tests of stellar models
  and Galactic evolutionary trends of N/O and C/N,  arXiv,
  \dodoi{10.48550/ARXIV.2106.03912}

\bibitem[{Wang \& Zhao(2013)}]{Wang_2013}
Wang, Y., \& Zhao, G. 2013, The Astrophysical Journal, 769, 4,
  \dodoi{10.1088/0004-637x/769/1/4}

\bibitem[{Wegg {et~al.}(2019)Wegg, Rojas-Arriagada, Schultheis, \&
  Gerhard}]{Wegg_2019}
Wegg, C., Rojas-Arriagada, A., Schultheis, M., \& Gerhard, O. 2019, Astronomy
  {$\&$} Astrophysics, 632, A121, \dodoi{10.1051/0004-6361/201936779}

\bibitem[{{Weinberg} {et~al.}(2017){Weinberg}, {Andrews}, \&
  {Freudenburg}}]{Weinberg2017}
{Weinberg}, D.~H., {Andrews}, B.~H., \& {Freudenburg}, J. 2017, \apj, 837, 183,
  \dodoi{10.3847/1538-4357/837/2/183}

\bibitem[{Weinberg {et~al.}(2019)Weinberg, Holtzman, Hasselquist, Bird,
  Johnson, Shetrone, Sobeck, Prieto, Bizyaev, Carrera, Cohen, Cunha, Ebelke,
  Fernandez-Trincado, Garc{\'{\i} }a-Hern{\'{a}}ndez, Hayes, Jönsson, Lane,
  Majewski, Malanushenko, M{\'{e}}sz{\'{a}}ros, Nidever, Nitschelm, Pan, Rix,
  Rybizki, Schiavon, Schneider, Wilson, \& Zamora}]{Weinberg_2019}
Weinberg, D.~H., Holtzman, J.~A., Hasselquist, S., {et~al.} 2019, The
  Astrophysical Journal, 874, 102, \dodoi{10.3847/1538-4357/ab07c7}

\bibitem[{Weinberg {et~al.}(2022)Weinberg, Holtzman, Johnson, Hayes,
  Hasselquist, Shetrone, Ting, Beaton, Beers, Bird, Bizyaev, Blanton, Cunha,
  Fern{\'{a}}ndez-Trincado, Frinchaboy, Garc{\'{\i}}a-Hern{\'{a}}ndez,
  Griffith, Johnson, Jönsson, Lane, Leung, Mackereth, Majewski,
  M{\'{e}}sz{\'{a}}ros, Nitschelm, Pan, Schiavon, Schneider, Schultheis, Smith,
  Sobeck, Stassun, Stringfellow, Vincenzo, Wilson, \& Zasowski}]{Weinberg2021}
Weinberg, D.~H., Holtzman, J.~A., Johnson, J.~A., {et~al.} 2022, The
  Astrophysical Journal Supplement Series, 260, 32,
  \dodoi{10.3847/1538-4365/ac6028}

\bibitem[{{Wenger} {et~al.}(2019){Wenger}, {Balser}, {Anderson}, \&
  {Bania}}]{Wenger_2019}
{Wenger}, T.~V., {Balser}, D.~S., {Anderson}, L.~D., \& {Bania}, T.~M. 2019,
  \apj, 887, 114, \dodoi{10.3847/1538-4357/ab53d3}

\bibitem[{{Wilson} {et~al.}(2019){Wilson}, {Hearty}, {Skrutskie}, {Majewski},
  {Holtzman}, {Eisenstein}, {Gunn}, {Blank}, {Henderson}, {Smee}, {Nelson},
  {Nidever}, {Arns}, {Barkhouser}, {Barr}, {Beland}, {Bershady}, {Blanton},
  {Brunner}, {Burton}, {Carey}, {Carr}, {Colque}, {Crane}, {Damke}, {Davidson},
  {Dean}, {Di Mille}, {Don}, {Ebelke}, {Evans}, {Fitzgerald}, {Gillespie},
  {Hall}, {Harding}, {Harding}, {Hammond}, {Hancock}, {Harrison}, {Hope},
  {Horne}, {Karakla}, {Lam}, {Leger}, {MacDonald}, {Maseman}, {Matsunari},
  {Melton}, {Mitcheltree}, {O'Brien}, {O'Connell}, {Patten}, {Richardson},
  {Rieke}, {Rieke}, {Roman-Lopes}, {Schiavon}, {Sobeck}, {Stolberg}, {Stoll},
  {Tembe}, {Trujillo}, {Uomoto}, {Vernieri}, {Walker}, {Weinberg}, {Young},
  {Anthony-Brumfield}, {Bizyaev}, {Breslauer}, {De Lee}, {Downey}, {Halverson},
  {Huehnerhoff}, {Klaene}, {Leon}, {Long}, {Mahadevan}, {Malanushenko},
  {Nguyen}, {Owen}, {S{\'a}nchez-Gallego}, {Sayres}, {Shane}, {Shectman},
  {Shetrone}, {Skinner}, {Stauffer}, \&
  {Zhao}}]{apogeespectrographs_Wilson2019}
{Wilson}, J.~C., {Hearty}, F.~R., {Skrutskie}, M.~F., {et~al.} 2019, \pasp,
  131, 055001, \dodoi{10.1088/1538-3873/ab0075}

\bibitem[{Wu {et~al.}(2019)Wu, Xiang, Zhao, Bi, Liu, Shi, Huang, Yuan, Wang,
  Chen, Huo, Ren, Tian, Liu, Zhang, Li, \& Zhang}]{Wu_2019}
Wu, Y., Xiang, M., Zhao, G., {et~al.} 2019, Monthly Notices of the Royal
  Astronomical Society, 484, 5315, \dodoi{10.1093/mnras/stz256}

\bibitem[{{Xiang} \& {Rix}(2022)}]{Xiang2022}
{Xiang}, M., \& {Rix}, H.-W. 2022, \nat, 603, 599,
  \dodoi{10.1038/s41586-022-04496-5}

\bibitem[{{Yanny} {et~al.}(2009){Yanny}, {Rockosi}, {Newberg}, {Knapp},
  {Adelman-McCarthy}, {Alcorn}, {Allam}, {Allende Prieto}, {An}, {Anderson},
  {Anderson}, {Bailer-Jones}, {Bastian}, {Beers}, {Bell}, {Belokurov},
  {Bizyaev}, {Blythe}, {Bochanski}, {Boroski}, {Brinchmann}, {Brinkmann},
  {Brewington}, {Carey}, {Cudworth}, {Evans}, {Evans}, {Gates}, {G{\"a}nsicke},
  {Gillespie}, {Gilmore}, {Nebot Gomez-Moran}, {Grebel}, {Greenwell}, {Gunn},
  {Jordan}, {Jordan}, {Harding}, {Harris}, {Hendry}, {Holder}, {Ivans},
  {Ivezi{\v{c}}}, {Jester}, {Johnson}, {Kent}, {Kleinman}, {Kniazev},
  {Krzesinski}, {Kron}, {Kuropatkin}, {Lebedeva}, {Lee}, {French Leger},
  {L{\'e}pine}, {Levine}, {Lin}, {Long}, {Loomis}, {Lupton}, {Malanushenko},
  {Malanushenko}, {Margon}, {Martinez-Delgado}, {McGehee}, {Monet}, {Morrison},
  {Munn}, {Neilsen}, {Nitta}, {Norris}, {Oravetz}, {Owen}, {Padmanabhan},
  {Pan}, {Peterson}, {Pier}, {Platson}, {Re Fiorentin}, {Richards}, {Rix},
  {Schlegel}, {Schneider}, {Schreiber}, {Schwope}, {Sibley}, {Simmons},
  {Snedden}, {Allyn Smith}, {Stark}, {Stauffer}, {Steinmetz}, {Stoughton},
  {SubbaRao}, {Szalay}, {Szkody}, {Thakar}, {Sivarani}, {Tucker}, {Uomoto},
  {Vanden Berk}, {Vidrih}, {Wadadekar}, {Watters}, {Wilhelm}, {Wyse}, {Yarger},
  \& {Zucker}}]{Yanny_2009}
{Yanny}, B., {Rockosi}, C., {Newberg}, H.~J., {et~al.} 2009, \aj, 137, 4377,
  \dodoi{10.1088/0004-6256/137/5/4377}

\bibitem[{{Yoshii}(1982)}]{Yoshii1982}
{Yoshii}, Y. 1982, \pasj, 34, 365

\bibitem[{{Zamora} {et~al.}(2015){Zamora}, {Garc{\'\i}a-Hern{\'a}ndez},
  {Allende Prieto}, {Carrera}, {Koesterke}, {Edvardsson}, {Castelli}, {Plez},
  {Bizyaev}, {Cunha}, {Garc{\'\i}a P{\'e}rez}, {Gustafsson}, {Holtzman},
  {Lawler}, {Majewski}, {Manchado}, {M{\'e}sz{\'a}ros}, {Shane}, {Shetrone},
  {Smith}, \& {Zasowski}}]{Zamora2015}
{Zamora}, O., {Garc{\'\i}a-Hern{\'a}ndez}, D.~A., {Allende Prieto}, C.,
  {et~al.} 2015, \aj, 149, 181, \dodoi{10.1088/0004-6256/149/6/181}

\bibitem[{{Zasowski} {et~al.}(2013){Zasowski}, {Johnson}, {Frinchaboy},
  {Majewski}, {Nidever}, {Rocha Pinto}, {Girardi}, {Andrews}, {Chojnowski},
  {Cudworth}, {Jackson}, {Munn}, {Skrutskie}, {Beaton}, {Blake}, {Covey},
  {Deshpande}, {Epstein}, {Fabbian}, {Fleming}, {Garcia Hernandez}, {Herrero},
  {Mahadevan}, {M{\'e}sz{\'a}ros}, {Schultheis}, {Sellgren}, {Terrien}, {van
  Saders}, {Allende Prieto}, {Bizyaev}, {Burton}, {Cunha}, {da Costa},
  {Hasselquist}, {Hearty}, {Holtzman}, {Garc{\'\i}a P{\'e}rez}, {Maia},
  {O'Connell}, {O'Donnell}, {Pinsonneault}, {Santiago}, {Schiavon}, {Shetrone},
  {Smith}, \& {Wilson}}]{Zasowski2013}
{Zasowski}, G., {Johnson}, J.~A., {Frinchaboy}, P.~M., {et~al.} 2013, \aj, 146,
  81, \dodoi{10.1088/0004-6256/146/4/81}

\bibitem[{{Zasowski} {et~al.}(2017){Zasowski}, {Cohen}, {Chojnowski},
  {Santana}, {Oelkers}, {Andrews}, {Beaton}, {Bender}, {Bird}, {Bovy},
  {Carlberg}, {Covey}, {Cunha}, {Dell'Agli}, {Fleming}, {Frinchaboy},
  {Garc{\'\i}a-Hern{\'a}ndez}, {Harding}, {Holtzman}, {Johnson}, {Kollmeier},
  {Majewski}, {M{\'e}sz{\'a}ros}, {Munn}, {Mu{\~n}oz}, {Ness}, {Nidever},
  {Poleski}, {Rom{\'a}n-Z{\'u}{\~n}iga}, {Shetrone}, {Simon}, {Smith},
  {Sobeck}, {Stringfellow}, {Szigeti{\'a}ros}, {Tayar}, \&
  {Troup}}]{Zasowski2017}
{Zasowski}, G., {Cohen}, R.~E., {Chojnowski}, S.~D., {et~al.} 2017, \aj, 154,
  198, \dodoi{10.3847/1538-3881/aa8df9}

\bibitem[{{Zasowski} {et~al.}(2019){Zasowski}, {Schultheis}, {Hasselquist},
  {Cunha}, {Sobeck}, {Johnson}, {Rojas-Arriagada}, {Majewski}, {Andrews},
  {J{\"o}nsson}, {Beers}, {Chojnowski}, {Frinchaboy}, {Holtzman}, {Minniti},
  {Nidever}, \& {Nitschelm}}]{Zasowski_2019}
{Zasowski}, G., {Schultheis}, M., {Hasselquist}, S., {et~al.} 2019, \apj, 870,
  138, \dodoi{10.3847/1538-4357/aaeff4}

\bibitem[{Zinn {et~al.}(2019)Zinn, Pinsonneault, Huber, Stello, Stassun, \&
  Serenelli}]{Zinn_2019}
Zinn, J.~C., Pinsonneault, M.~H., Huber, D., {et~al.} 2019, The Astrophysical
  Journal, 885, 166, \dodoi{10.3847/1538-4357/ab44a9}

\end{thebibliography}
\bibliographystyle{aasjournal}



\appendix
\counterwithin{figure}{section}

\section{Stellar Ages \& Related Caveats}
\label{sec:app:ages}

As introduced in Section \ref{sec:data:ages}, our age estimates are adopted from the {\texttt{distmass}} value added catalog (Stone-Martinez et al. 2023 submitted), which utilizes a neural network trained on the ASPCAP stellar parameters and APOKASC asteroseismic masses to derive stellar mass for all APOGEE stars. The reported uncertainties, accuracy, and potential caveats related to the {\texttt{distmass}} age estimates are discussed in detail in Stone-Martinez et al. (2023, submitted). Here, we explore and discuss how these uncertainties and our particular sample selection may influence our age-related results. In summary, we find that while the general trends stay the same, both the qualitative and quantitative aspects can differ significantly particularly among the oldest stars. We caution against drawing strong conclusions from stellar ages until more precise ages are available for larger samples of stars. Again, we emphasize that this will only impact our figures that include stellar age.

\subsection{Variation in Spatial Sampling}

\begin{figure}
    \centering
    \includegraphics[width=0.45\textwidth]{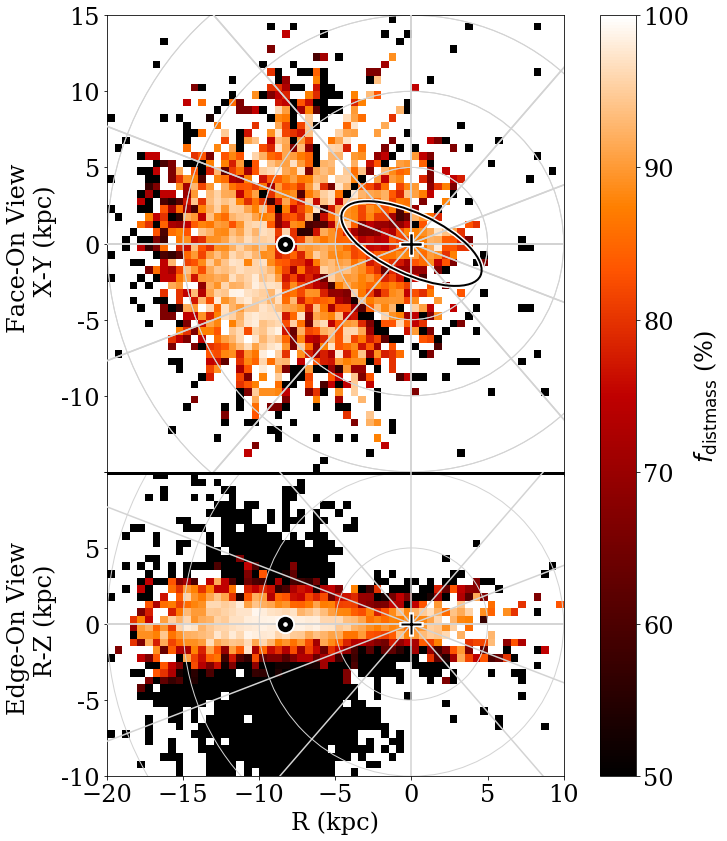}
    \caption{The same as Figure \ref{fig:star_numbers}, but colored by the fraction of stars in our sample that survive the additional distmass quality cuts described in Section \ref{sec:data:ages}. The purpose of this figure is to qualitatively explore how the distmass criterion might bias our results using stellar age estimates.}
    \label{fig:distmass_numbers}
\end{figure}

The {\texttt{distmass}} quality cuts described in \ref{sec:data:ages} notably result in the exclusion of all stars with [Fe/H] $\leq -0.7$ from our sample. In Figure \ref{fig:distmass_numbers}, we qualitatively explore where in the Galaxy this selection criterion may influence our results. This shows the fraction of stars $f_{\rm distmass}$ in every spatial bin ($\Delta X = \Delta Y = \Delta Z = 0.5$ kpc) which survive the additional {\texttt{distmass}} quality cut. Close to the solar neighborhood ($6 \leq R \leq 12$ kpc, $|Z|\leq 1$ kpc), the majority of stars pass this additional selection criterion ($f_{\rm distmass}\geq 97\%$). Near the center of the Galaxy ($R \leq 5$ kpc, $|Z|\leq 1$ kpc), $f_{\rm distmass}$ is generally lower but still exceeds $85\%$. In the outer Galaxy ($R > 15$ kpc) or beyond the plane ($|Z|\geq 1$ kpc), where most of the stars are expected to be metal-poor, there is a sharp drop in $f_{\rm distmass}$, regularly reaching fractions lower than $<80\%$. This shows that our sample with {\texttt{distmass}} ages, while generally a majority, does not equally represent the full sample everywhere in the Galaxy. Particularly beyond $R > 15$ kpc we caution against directly comparing our results including age to those made with the full sample. 

\subsection{Selection of Age Catalog}

\begin{figure*}
    \centering
    \includegraphics[width=\textwidth]{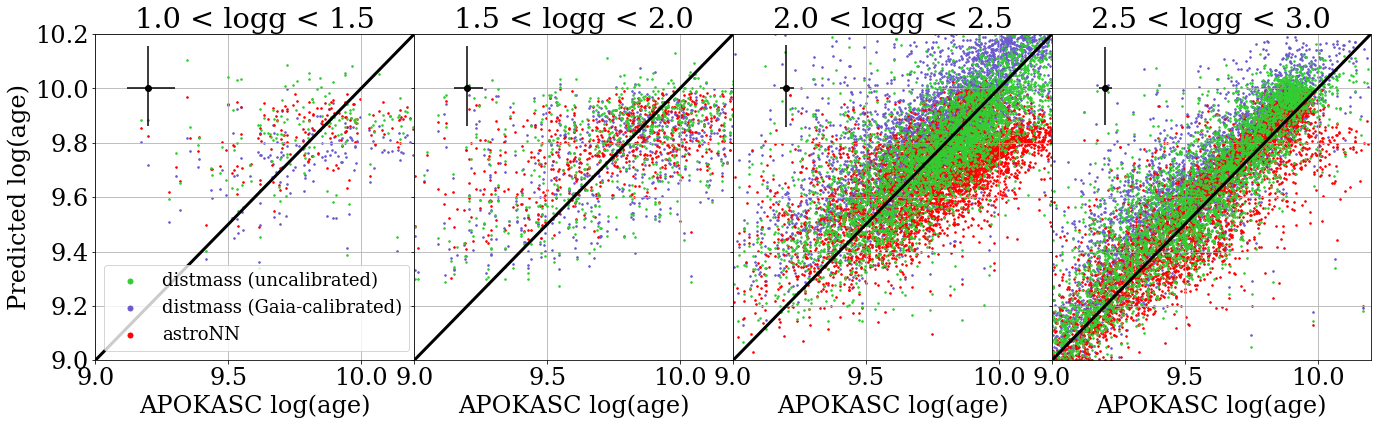}
    \caption{Systematic uncertainties in the stellar age estimates, comparing the distmass (green and blue) and astroNN (red) catalogs across different ranges of stellar $\log g$ (different panels). In each panel, the x-axis is the stellar age from asteroseismology estimates (APOKASC), and the y-axis is the predicted age recovered from the neural network methodology. The typical uncertainty of each point is depicted in black in the upper-left of each panel, with the asteroseismic measurements notably having larger uncertainties at lower $\log g$. In both catalogs and across all $\log g$, young stars are generally recovered too old, and old stars are recovered too young. However, there are higher uncertainties in the asteroseismic measurements for the low-$\log g$ stars, meaning the neural network age estimates may be more precise than the asteroseismic age estimates in this range.}
    \label{fig:age_systematics}
\end{figure*}

\begin{figure*}
    \centering
    \includegraphics[width=\textwidth]{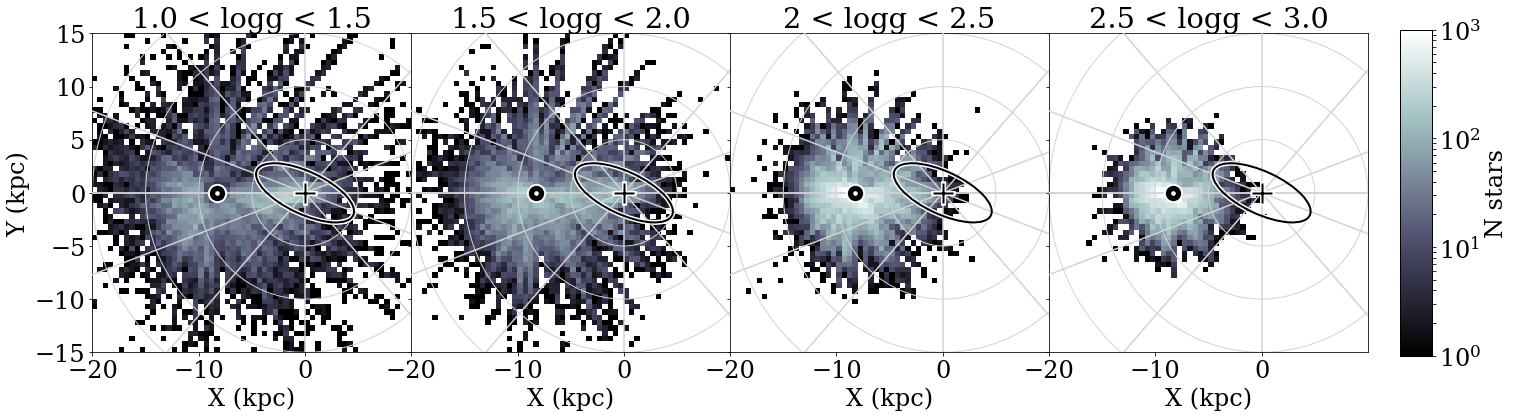}
    \caption{The spatial distribution (X-Y face-on view) of stars in APOGEE for different bins of  $\log g$ (columns). Increasing the $\log g$ range of our sample, which would minimize some of the age-related systematics, would also severely limit the area that our sample can probe.}
    \label{fig:logg_positions}
\end{figure*}

Neural network-derived ages like {\texttt{distmass}} heavily rely on the accuracy of the training set labels which transfer to the model. Small changes the training set can result in differences in the derived ages of the full training set, which contributes to the large uncertainty values around stellar ages.

The {\texttt{distmass}} catalog publishes results from six different neural network models. All six models were trained using the same subset of stars and the same ASPCAP parameters as training labels, with the only difference being the choice of APOKASC asteroseismic masses used in the training set. The APOKASC catalog publishes asteroseismic results from three different groups (initials "SS", "MO" and "TW" - see Pinsonneault et al. 2023 in prep. for details), referred to as the "uncorrected" masses. Each set of masses also has a "corrected" counterpart, using Gaia parallaxes to calibrate the stellar radius derived from asteroseismology \citep[e.g.,][]{Zinn_2019}. The calibrations are dependent on stellar $\log g$ and are less reliable for stars $\log g \leq 2$, which is why we elect to use the uncalibrated age set for our final results. Stone-Martinez et al. 2023 in prep. goes into more detail on this motivation.

Figure \ref{fig:age_systematics} compares different three different age catalogs - the uncalibrated {\texttt{distmass}} ages used in this paper (green points), its calibrated counterpart (blue points), and the {\texttt{astroNN}} APOGEE VAC \citep{Leung_2018,Mackereth2019} (red points) - against the APOKASC derived stellar ages on the x-axis for the subset of stars in the {\texttt{distmass}} training set. The comparison is divided by ranges in stellar $\log g$ (columns, increasing left to right). We find that all three catalogs suffer a "compression" in ages compared to the APOKASC values, with young stars tend to be assigned ages that are too old, and old stars tend to be fit too young. All three sets of ages predict the ages of higher $\log g$ with better accuracy and less scatter. For lower $\log g$ stars, the calibrated {\texttt{distmass}} ages predict the young ages slightly better, but and {\texttt{astroNN}} and the uncalibrated {\texttt{distmass}} predicts the older ages better. For higher $\log g$ stars, {\texttt{astroNN}} predicts the young ages slightly better, and {\texttt{distmass}} predicts the older ages better. This offset is thought to be caused by scatter within the training set that the neural network learns from. The asteroseismic masses are more uncertain for low $\log g$ stars, which results in a larger mismatch between the APOKASC results and the neural network results. The neural network essentially fits for the relation between [C/N] abundance and stellar age, which should be consistent across a broad range of $\log g$. Therefore, we emphasize for low $\log g$ stars, the neural network likely fits more precise ages than available in the pure asteroseismology results used in the training set.


Although the age estimates show less scatter for higher $\log g$ stars, redefining our sample would severely limit the spatial range of the Galaxy we can access. Figure \ref{fig:logg_positions} shows the face-on distribution of stars in the disk for various $\log g$ limits. Above $\log g \geq 2$, the outer disk ($R \geq 20$ kpc) and the inner disk ($R \leq 5$ kpc) disappear from the sample.

Nevertheless, we did test how redefining the sample may influence our results. In general, the observed trends and resulting conclusions do not change, motivating the use of the original sample for the increased spatial range. The radial metallicity gradient as a function of age (Figure \ref{fig:age_metal_gradients}) differs by less than $0.001$ dex kpc$^{-1}$, and the radial age gradient (Figure \ref{fig:radial_age_gradients}) differs by $0.03$ Gyr kpc$^{-1}$ using the higher $\log g$ sample.

\subsection{Alpha-Age relation for different catalogs}

\begin{figure*}
    \centering
    \caption{Reproduction of the age-alpha relation (Figure {\ref{fig:age_alpha_relation}}) using different samples of stellar ages.}
    \label{fig:age_alpha_relation_Appendix}

    \includegraphics[height=0.2\textheight]{figures/age_alpha_contours1.png} \\
   
   {(a) Same as Figure {\ref{fig:age_alpha_relation}}, reproduced here for ease of comparison.}

    \includegraphics[height=0.2\textheight]{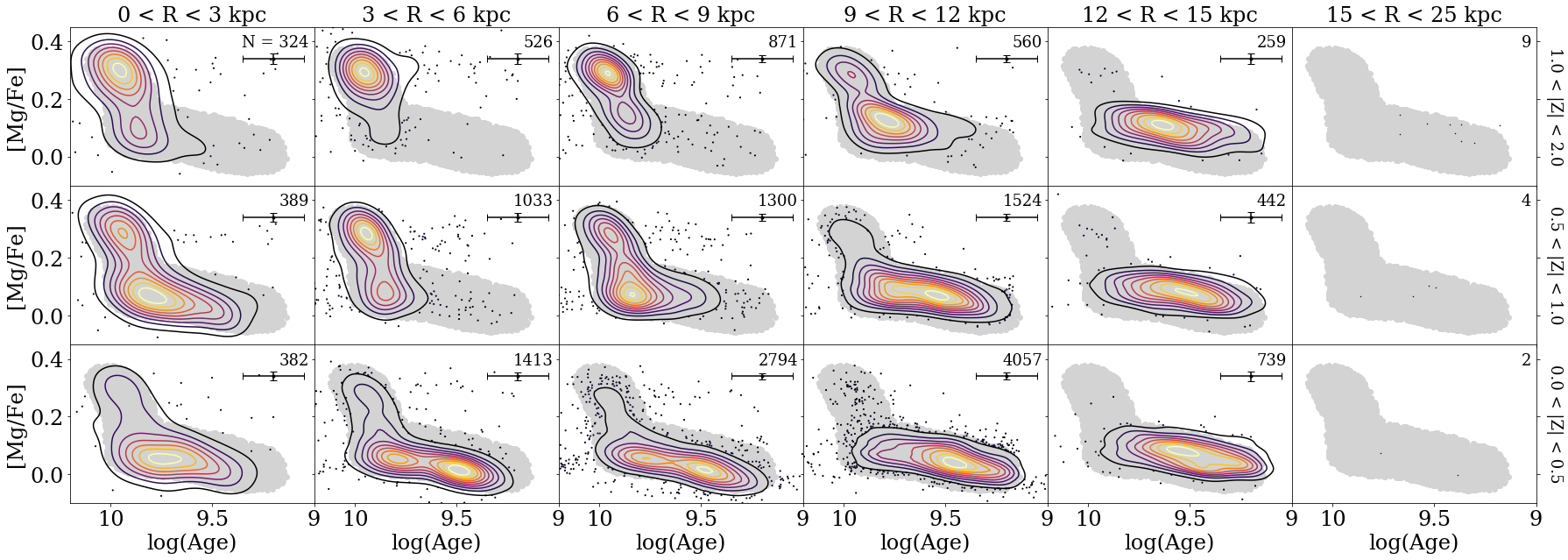} \\
    {(b) Same as (a), but for a sample covering a higher surface gravity range of $2.0 \leq \log g \leq 2.3$ (Test Sample A) \label{subfig:test_A}}
    
    \includegraphics[height=0.2\textheight]{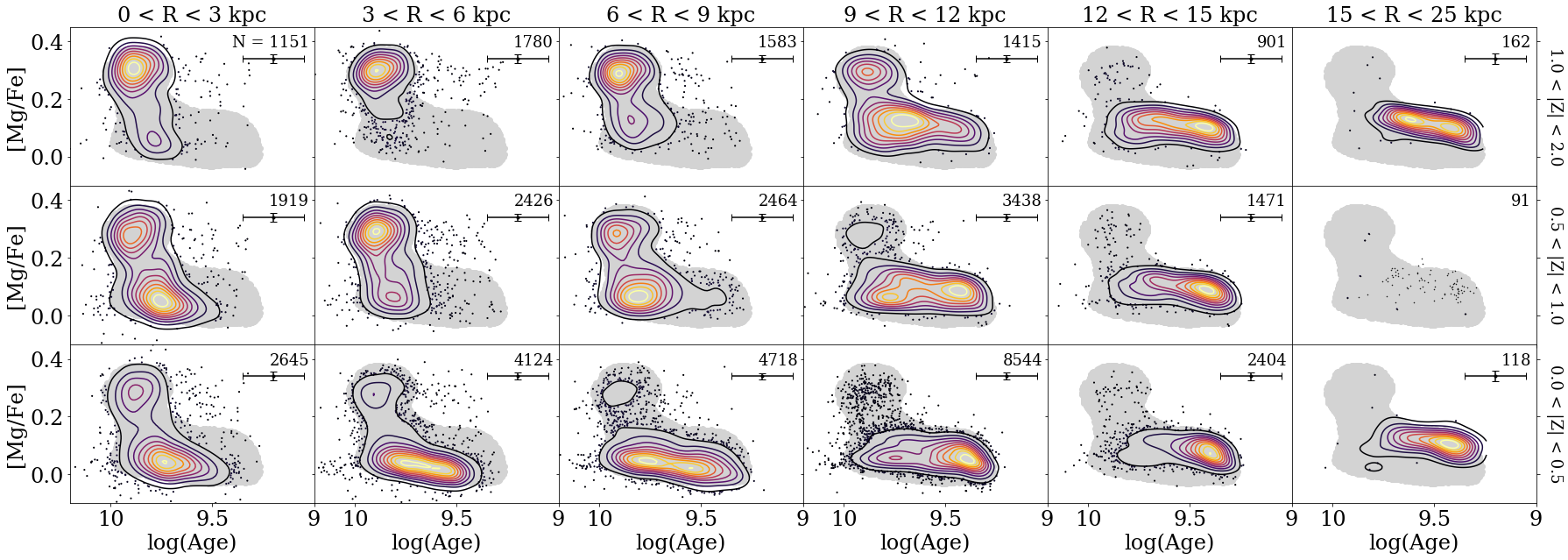} \\
    {(c) Same as (a), but using the "calibrated" ages from distmass trained on the Gaia-corrected APOKASC catalog (Test Sample B). \label{subfig:test_B}}

    \includegraphics[height=0.2\textheight]{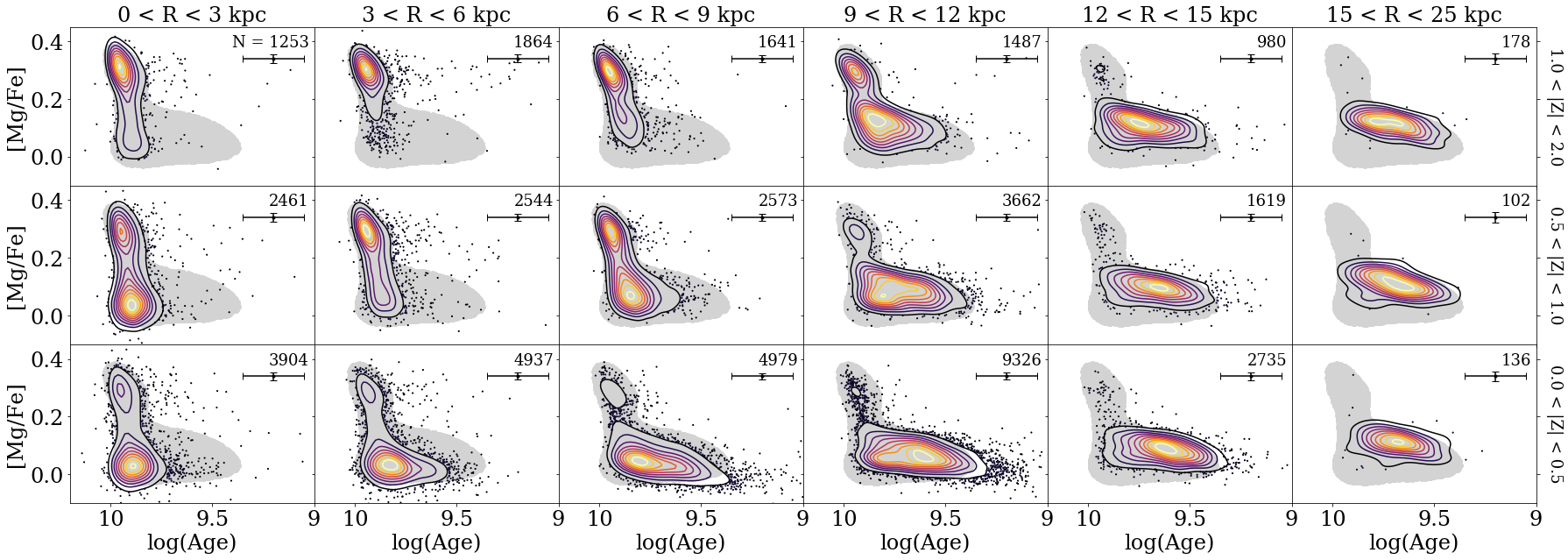} \\
    {(d) Same as (a), but using stellar ages from the astroNN catalog (Test Sample C). \label{subfig:test_C}}

\end{figure*}

To demonstrate how a different selection of age catalog may influence our results, Figure {\ref{fig:age_alpha_relation_Appendix}} shows a recreation of Figure \ref{fig:age_alpha_relation} for three different age samples:

\begin{itemize}
    \item \textbf{Test Sample A ($\log g$ range):} restricting to $2.0 \leq \log g \leq 2.3$: defining a new sample with higher surface gravity
    \item \textbf{Test Sample B (alternate distmass ages):} using the Gaia-corrected APOKASC ages training set from {\texttt{distmass}}, with the original sample of $1.0 \leq \log g \leq 2.0$.
    \item \textbf{Test Sample C (alternate astroNN ages):} using {\texttt{astroNN}} ages instead of the {\texttt{distmass}} catalog, with the original sample of $1.0 \leq \log g \leq 2.0$.
\end{itemize}


Test Sample A (Figure \ref{subfig:test_A}.b) changes our limits to $2.0 \leq \log g \leq 2.3$, shifting our sample to a different $\log g$ range where the age estimates are expected to be more reliable, while purposely avoiding the red clump. However, this suffers from the previously-shown lack of spatial distribution, with significantly fewer stars beyond $R > 12$ kpc and within $R < 6$ kpc than the original sample used in this paper. This test sample does include a larger number of stars near the solar neighborhood, however. The general location and trends across galactic location are similar, but this sample has more scatter which leads to larger contour shapes. The bimodality within the low-$\alpha$ sequence (possible evidence for a "third infall") around $9 < R < 12$ kpc is not obvious in this sample, although is perhaps present at shorter radii ($3 < R < 9$ kpc).

Test Sample B uses the {\texttt{distmass}} ages using the calibrated ASPCAP training set. The Gaia-motivated calibration is dependent on $\log g$ \citep[e.g.,][Pinsonneault et al. 2023 in prep.]{Zinn_2019}, which results in lower $\log g$ stars assigned younger ages in the calibrated catalog and higher $\log g$ stars being assigned systematically older ages and resulting in an overall age compression in the sample obvious is Figure \ref{subfig:test_B}.c. Otherwise, the results are generally similar, with the possible bimodality within the low-$\alpha$ sequence at $9 < R < 12$ kpc still present in this sample. The high-$\alpha$ sequence has more spread in age, possible due to the calibrations propagating uncertainties in the adopted calibrations (from the $T_{\textrm{eff}}$ assumptions for example) through into the final age estimates. 

Test Sample C uses {\texttt{astroNN}} ages instead of {\texttt{distmass}} in Figure \ref{subfig:test_C}.d, which makes obvious the overall "compression" of the ages caused by the systematics discussed earlier; {\texttt{astroNN}} ages do not go as old or as young as the {\texttt{distmass}} ages. The general shapes and trends across the Galaxy otherwise stay the same, and the bimodality within the low-$\alpha$ sequence is not obvious in this sample either.

\begin{figure*}
    \centering
    \caption{Reproduction of the radial age gradients (Figure {\ref{fig:radial_age_gradients}}) using different samples of stellar ages.}
    \label{fig:radial_age_gradients_grav}

    \begin{minipage}{.4\textwidth}
    \includegraphics[width=\textwidth]{figures/radial_age_gradients_RIGHT.png}
    {(a) Same as Figure {\ref{fig:radial_age_gradients}}, reproduced here for ease of comparison.}
    \end{minipage}  
    \begin{minipage}{.4\textwidth}
    \includegraphics[width=\textwidth]{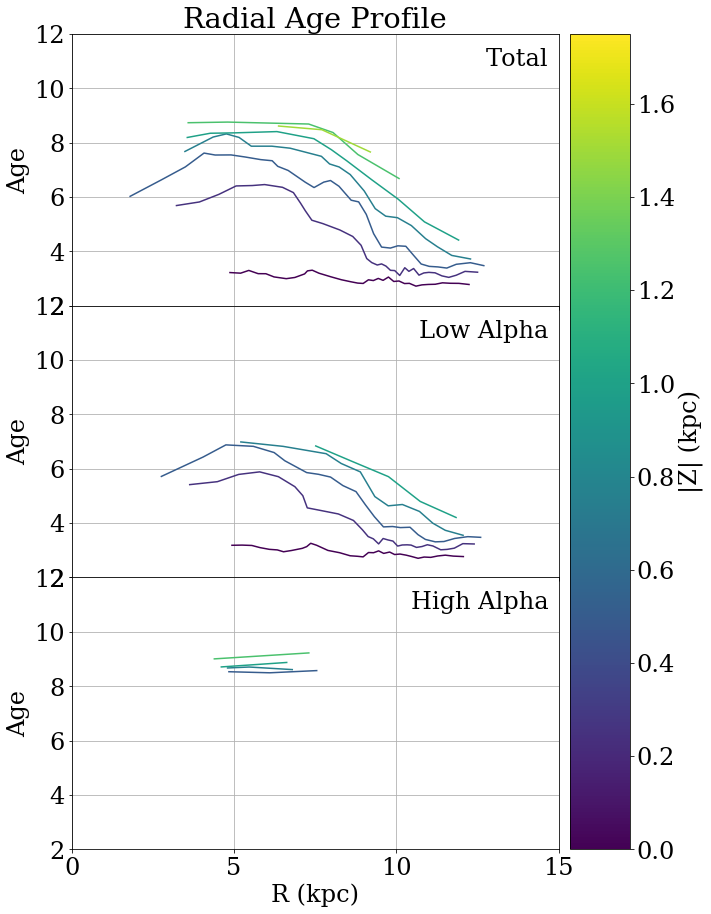}
    {(b) Same as (a), but using a sample covering a higher surface gravity range of $2.0 \leq \log g \leq 2.3$ (Test Sample A)} 
    \end{minipage}
\end{figure*}

\begin{figure*}
    \centering
    \caption{Reproduction of the age-metallicity relation (Figure {{\ref{fig:age_metallicity_relation}}}) using different samples of stellar ages.}
    \label{fig:amr_grav}
    
    \includegraphics[height=0.2\textheight]{figures/age_metalicity_relation.png} \\
    {(a) Same as Figure {\ref{fig:age_metallicity_relation}}, reproduced here for ease of comparison.}

    \includegraphics[height=0.2\textheight]{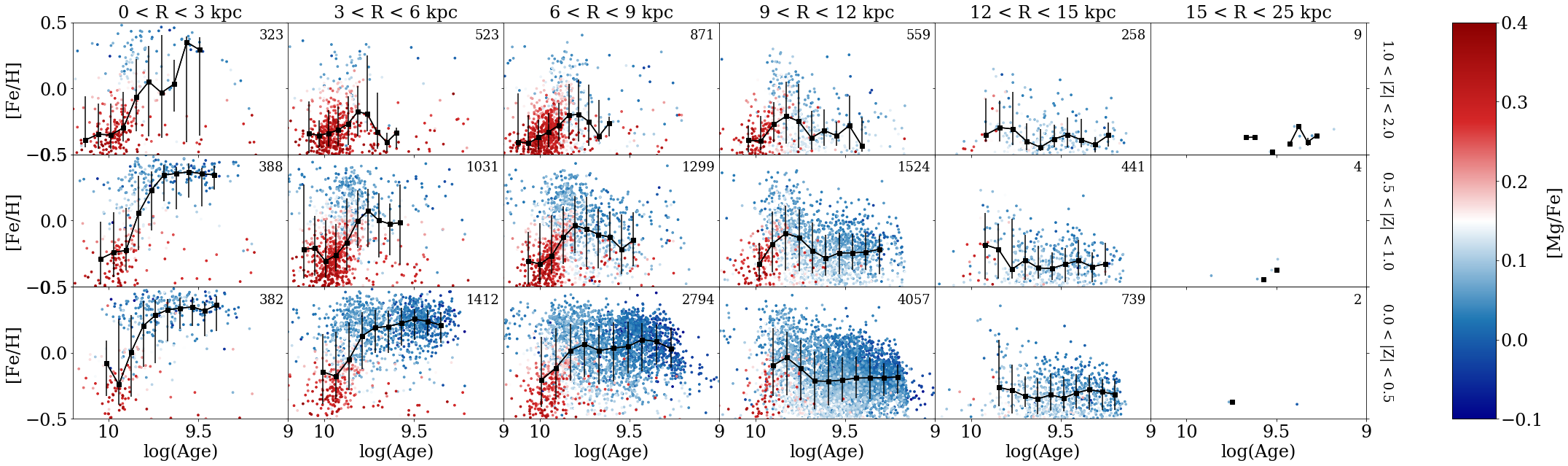} \\
    {(b) Same as (a), but for a sample covering a higher surface gravity range of $2.0 \leq \log g \leq 2.3$ (Test Sample A)}
\end{figure*}

We also include a reproduction of the median radial age profile (Figure {\ref{fig:radial_age_gradients}}) and the age-metallicity relation (Figure {\ref{fig:age_metallicity_relation}}) using the higher-$\log g$ Test Sample A in Figures {\ref{fig:radial_age_gradients_grav}} through {\ref{fig:amr_grav}}. As before, the trends in each plot are similar to their low-$\log g$ counterparts, but with higher scatter, as the range of Galaxy covered by the sample is more limited despite the lower uncertainties in stellar ages.

\subsection{Summary of Age Caveats}

Stellar ages are one of the most powerful quantities available in modern astronomy for revealing the history of the Milky Way, but remain challenging to measure robustly. In this study and others, it is important to consider age-related results with caution and understand the subsequent effects that changes in stellar ages may have on the conclusions. While different age catalogs show generally similar trends for the results of this paper, the quantitative aspects can differ significantly and should be regarded with healthy skepticism until more precise age estimates are available for large samples of stars.


\end{document}